# AN INTRODUCTION TO THE QUANTUM THEORY OF NONLINEAR OPTICS

Mark Hillery[1]

*Department of Physics, Hunter College of CUNY, 695 Park Avenue, New York, NY 10065 USA*

This article is provides an introduction to the quantum theory of optics in nonlinear dielectric media. We begin with a short summary of the classical theory of nonlinear optics, that is nonlinear optics done with classical fields. We then discuss the canonical formalism for fields and its quantization. This is applied to quantizing the electromagnetic field in free space. The definition of a nonclassical state of the electromagnetic field is presented, and several examples are examined. This is followed by a brief introduction to entanglement in the context of field modes. The next task is the quantization of the electromagnetic field in an inhomogeneous, linear dielectric medium. Before going on to field quantization in nonlinear media, we discuss a number of commonly employed phenomenological models for quantum nonlinear optical processes. We then quantize the field in both nondispersive and dispersive nonlinear media. Flaws in the most commonly used methods of accomplishing this task are pointed out and discussed. Once the quantization has been completed, it is used to study a multimode theory of parametric down conversion and the propagation of quantum solitons.

**Contents**

---

[1]E-mail address: mhillery@hunter.cuny.edu

# 1  Introduction

Nonlinear optics is the study of the response of dielectric media to strong optical fields. The fields are sufficiently strong that the response of the medium is nonlinear. That is, the polarization, which is the dipole moment per unit volume in the medium, is not a linear function of the applied electric field. In the equation for the polarization there is a linear term, but, in addition, there are terms containing higher powers of the electric field as well. This leads to significant new types of behavior. One of the most notable is that new frequencies, such as harmonics or subharmonics, can be generated. Linear media do not change the frequency of light incident upon them. In fact, the first observation of a nonlinear optical effect was second-harmonic generation [1]; a laser beam entering a nonlinear medium produced a second beam at twice the frequency of the original. Another type of behavior that becomes possible in nonlinear media is that the index of refraction, rather than being a constant, is a function of the intensity of the light. For a light beam with a nonuniform intensity profile, this can lead to self focussing of the beam.

Most nonlinear optical effects can be described using classical electromagnetic fields, and, in fact, the initial theory of nonlinear optics was done with classical fields [2]. When the fields are quantized, however, a number of new effects emerge. Quantized fields are necessary if we want to describe fields that originate from spontaneous emission. For example, in a process known as spontaneous parametric down conversion, a beam of light at one frequency, the pump, produces a beam at half the original frequency, the signal. and this second beam is a result of spontaneous emission. The quantum properties of the down-converted beam are novel. Its photons are produced in pairs, one pump photon disappearing to produce, simultaneously, two signal photons. This leads to strong correlations between photons in the signal beam, and photons produced in this way can be quantum mechanically entangled. In addition, the signal beam can have smaller phase fluctuations than is possible with classical light. Both of these properties have made light produced by parametric down conversion useful for applications in the field of quantum information.

The quantization of electrodynamics in nonlinear media is not straightforward, and some mistakes were made along the way. The most complete version of a quantum theory of nonlinear optics is the one developed by Peter Drummond and his collaborators. This theory has been used to make detailed comparisons with experiments, with excellent results (for recent work see [3,4]). In many cases, especially if only a few field modes are involved, it is not necessary to use the full theory and phenomenological models are employed. The theory developed by Drummond comes into its own when one wants to describe the propagation of pulses in nonlinear media, and it is essential in order to properly treat quantum solitons in nonlinear fibers.

This review is meant as a pedagogical survey of the field of nonlinear optics with quantized fields. It is hoped that a graduate student or someone not in the field, could use it as an introduction and then have enough preparation to go on and read some of the original literature. This review does not attempt to be comprehensive, and it does have a point of view. Issues involving field quantization are emphasized for a number of reasons. In standard quantum optics texts these issues, to the extent they are discussed at all, are skipped over very quickly. For the most part this is fine, but when quantizing the electromagnetic field in linear or nonlinear media, there are subtleties that are missed by this approach. One result of this is that people working in quantum optics are often not familiar with the canonical formalism for fields and how it is used to quantize them. It is hoped that this review will provide a useful introduction to that subject.

The review begins with a very brief survey of some topics in the classical theory on nonlinear optics. For further reading on this subject, the textbook by Robert Boyd is highly recommended [5]. We then go on to discuss the canonical formalism for fields and how to quantize it. This is then applied to the quantization of the free electromagnetic field. Once we have quantized the field, we go on to discuss nonclassical field states and entanglement. Next, we show how the field can be quantized in the presence of an inhomogeneous linear medium. The next step is to move on to nonlinear media, but before discussing the quantization of the field in such media, we first use phenomenological models to show what kinds of effects can be expected. We then quantize the field in a nonlinear medium, first in a nondispersive medium, then in a dispersive one. Once we have done so, we discuss a multimode theory of spontaneous parametric down conversion. We end by discussing quantum solitons in a nonlinear fiber.

## 2 Classical nonlinear optics

When an electric field is applied to a dielectric medium, a polarization, that is, a dipole moment per unit volume, is created in the medium. If the field is not too strong the response of the medium is linear, which means that the polarization, **P** is linear in the applied field, **E** (we will use S. I. units throughout this paper),

$$P_j = \epsilon_0 \sum_{k=1}^{3} \chi_{jk}^{(1)} E_k. \tag{2.1}$$

In this equation, $\chi^{(1)}$ is the linear susceptibility tensor of the medium. If the field is sufficiently strong, the linear relation breaks down and nonlinear terms must be taken into account, that is

$$P_j = \epsilon_0 [\sum_{k=1}^{3} \chi_{jk}^{(1)} E_k + \sum_{k=1}^{3} \sum_{l=1}^{3} \chi_{jkl}^{(2)} E_k E_l \sum_{k=1}^{3} \sum_{l=1}^{3} \sum_{m=1}^{3} \chi_{jklm}^{(3)} E_k E_l E_m + \ldots]. \tag{2.2}$$

Here, we have kept the first three terms in the power series expansion of the polarization in terms of the field The quantities $\chi^{(2)}$ and $\chi^{(3)}$ are the second and third order nonlinear susceptibilities, respectively. We should also note that Eq. (2.2) is often written as

$$\mathbf{P} = \epsilon_0 [\chi^{(1)} : \mathbf{E} + \chi^{(2)} : \mathbf{EE} + \chi^{(3)} : \mathbf{EEE} + \ldots]. \tag{2.3}$$

The $\chi^{(2)}$ term is only present if the medium is not invariant under spatial inversion ($\mathbf{r} \to -\mathbf{r}$). This follows from the fact that if the medium is invariant under spatial inversion, the $\chi^{(2)}$ for the inverted medium with be the same as that for the original medium, i.e. under spatial inversion we will have $\chi^{(2)} \to \chi^{(2)}$. However, under spatial inversion we also have that $\mathbf{P} \to -\mathbf{P}$ and $\mathbf{E} \to -\mathbf{E}$. Consequently, while $\mathbf{P} \to -\mathbf{P}$ implies that we should have

$$\chi^{(2)} : \mathbf{EE} \to -\chi^{(2)} : \mathbf{EE}, \tag{2.4}$$

the relations $\chi^{(2)} \to \chi^{(2)}$ and $\mathbf{E} \to -\mathbf{E}$ show us that instead we have

$$\chi^{(2)} : \mathbf{EE} \to \chi^{(2)} : \mathbf{EE}. \tag{2.5}$$

The only way these conditions can be consistent is if $\chi^{(2)} = 0$. Therefore, for many materials we do, in fact have $\chi^{(2)} = 0$, and their first nonzero nonlinear susceptibility is $\chi^{(3)}$.

If the nonlinearities in the medium are electronic in origin, then the nonlinear effects should be important when the applied field is of the same order as the electric field in an atom. The field in an atom is of order $E_{atom} \sim e/(4\pi\epsilon_0 a_0^2)$, where $a_0$ is the Bohr radius, which is approximately $5 \times 10^{11}\ V/m$. This can be used to estimate the size of the susceptibilities. For fields of this magnitude, the terms in the expansion for the polarization will be of roughly the same size. Using the fact that $\chi^{(1)}$ is of order one, we then find that

$$\begin{aligned}\chi^{(2)} &\sim \frac{1}{E_{atom}} \simeq 2 \times 10^{-12} \frac{m}{V} \\ \chi^{(3)} &\sim \frac{1}{E_{atom}^2} \simeq 4 \times 10^{-24} \frac{m^2}{V^2}.\end{aligned} \tag{2.6}$$

These estimates, are, in fact, quite good.

We would now like to survey some of the effects to which the nonlinear terms in the series for the polarization give rise. We begin by noting that Maxwell's equations, with the polarization included, give us

$$\nabla \times \nabla \times \mathbf{E} + \frac{1}{c^2}\frac{\partial^2 \mathbf{E}}{\partial t^2} = -\mu_0 \frac{\partial^2 \mathbf{P}}{\partial t^2}. \tag{2.7}$$

This tells us that the polarization acts as a source for the field, and, in particular, if the polarization has terms oscillating at a particular frequency, then those terms will give rise to components of the field oscillating at the same frequency. In our initial survey of nonlinear optical effects, we shall ignore all indices, and treat all quantities as scalars.

Let us first look at second-order nonlinearities. If our applied field oscillates at the frequency $\omega$, $E(t) = E_0 \cos \omega t$, then the nonlinear part of the polarization, $P_{nl}$, will be

$$P_{nl}(t) = \epsilon_0 \chi^{(2)} E(t)^2 = \frac{1}{2}\epsilon_0 \chi^{(2)} E_0^2 (1 + \cos 2\omega t). \tag{2.8}$$

We can see that the polarization has a term oscillating at twice the applied frequency, and this will give rise to a field whose frequency is also $2\omega$. This process is known as second-harmonic generation, and it can be, and is, used to double the frequency of the output of a laser by sending the beam through an appropriate material, that is, one with a nonzero value of $\chi^{(2)}$. As was mentioned, it was the first nonlinear optical effect that was observed [1]. Now suppose our applied field oscillates at two frequencies

$$E(t) = E_1 \cos \omega_1 t + E_2 \cos \omega_2 t. \tag{2.9}$$

The nonlinear polarization is now

$$\begin{aligned} P_{nl}(t) &= \frac{1}{2}\epsilon_0 \chi^{(2)} \{ E_1^2(1 + \cos 2\omega_1 t) + E_2^2(1 + \cos 2\omega_2 t) \\ &+ 2 E_1 E_2 [\cos(\omega_1 + \omega_2)t + \cos(\omega_1 - \omega_2)t] \}. \end{aligned} \tag{2.10}$$

In this case, not only do we have terms oscillating at twice the frequencies of the components of the applied field, we also have terms oscillating at the sum and difference of their frequencies. These processes are called sum and difference frequency generation, respectively.

Now let us move on to a third-order nonlinearity. For an applied field oscillating at a single frequency, $E(t) = E_0 \cos \omega t$, we find that (assuming that $\chi^{(2)} = 0$)

$$P_{nl}(t) = \frac{1}{4}\epsilon_0 \chi^{(3)} E_0^3 [\cos 3\omega t + 3 \cos \omega t]. \tag{2.11}$$

The first term clearly will cause a field at $3\omega$, the third harmonic of the applied field, to be generated. In order to see the effect of the second term, it is useful to combine the linear and nonlinear parts of the polarization to get the total polarization,

$$P(t) = \epsilon_0 \left( \chi^{(1)} + \frac{3}{4}\chi^{(3)} E_0^2 \right) E_0 \cos \omega t + \frac{1}{4}\epsilon_0 \chi^{(3)} E_0^3 \cos 3\omega t. \tag{2.12}$$

When there is no nonlinear polarization, the polarization is proportional to the field, and the constant of proportionality, $\chi^{(1)}$, is directly related to the refractive index of the material. When there is a nonlinearity, we see that the component of the polarization at the same frequency as the applied field is similar to what it is in the linear case, except that

$$\chi^{(1)} \to \chi^{(1)} + \frac{3}{4}\chi^{(3)} E_0^2. \tag{2.13}$$

This results in a refractive index that depends on the intensity of the applied field.

So far we have assumed that the response of the medium to an applied field, that is, the polarization at time $t$, depends only on the electric field at time $t$. This is, of course, an idealization. The response of a medium is not instantaneous, so that the polarization at time $t$ depends on the field at previous times, not just the field at time $t$. In the case of a linear medium, this is expressed as

$$P(t) = \epsilon_0 \int_0^\infty d\tau \tilde{\chi}^{(1)}(\tau) E(t-\tau). \tag{2.14}$$

Taking the Fourier transform of both sides, and defining

$$P(\omega) = \frac{1}{\sqrt{2\pi}} \int_{-\infty}^\infty dt e^{-i\omega t} P(t)$$

$$E(\omega) = \frac{1}{\sqrt{2\pi}} \int_{-\infty}^\infty dt e^{-i\omega t} E(t), \tag{2.15}$$

we find that

$$P(\omega) = \epsilon_0 \chi^{(1)}(\omega) E(\omega), \tag{2.16}$$

where

$$\chi^{(1)}(\omega) = \int_0^\infty d\tau e^{-i\omega \tau} \tilde{\chi}^{(1)}(\tau). \tag{2.17}$$

Therefore, we see that a medium response that is not instantaneous causes the susceptibilities to become frequency dependent. This phenomenon is known as dispersion. It also causes the nonlinear susceptibilities to become frequency dependent. For the second and third order nonlinearities we have

$$\begin{aligned} P^{(2)}(t) &= \epsilon_0 \int_0^\infty d\tau_1 \int_0^\infty d\tau_2 \tilde{\chi}^{(2)}(\tau_1, \tau_2) E(t-\tau_1) E(t-\tau_2) \\ P^{(3)}(t) &= \epsilon_0 \int_0^\infty d\tau_1 \int_0^\infty d\tau_2 \int_0^\infty d\tau_3 \tilde{\chi}^{(3)}(\tau_1, \tau_2, \tau_3) E(t-\tau_1) \\ &\quad E(t-\tau_2) E(t-\tau_3), \end{aligned} \tag{2.18}$$

where $P^{(2)}(t)$ and $P^{(3)}(t)$ are the contributions of the second and third order nonlinearities to the polarization, respectively. If we now take the Fourier transforms of these equations, we find that

$$\begin{aligned} P^{(2)}(\omega) &= \epsilon_0 \int_{-\infty}^\infty d\omega_1 \int_{-\infty}^\infty d\omega_2 \chi^{(2)}(\omega_1, \omega_2) \delta(\omega - \omega_1 - \omega_2) E(\omega_1) E(\omega_2) \\ P^{(3)}(\omega) &= \epsilon_0 \int_{-\infty}^\infty d\omega_1 \ldots \int_{-\infty}^\infty d\omega_3 \chi^{(3)}(\omega_1, \omega_2, \omega_3) \delta(\omega - \omega_1 - \omega_2 - \omega_3) \\ &\quad E(\omega_1) E(\omega_2) E(\omega_3), \end{aligned} \tag{2.19}$$

where

$$\chi^{(2)}(\omega_1,\omega_2) = \frac{1}{\sqrt{2\pi}}\int_0^\infty d\tau_1 \int_0^\infty d\tau_2 e^{-i(\omega_1\tau_1+\omega_2\tau_2)}\tilde{\chi}^{(2)}(\tau_1,\tau_2)$$

$$\chi^{(3)}(\omega_1,\omega_2,\omega_3) = \frac{1}{2\pi}\int_0^\infty d\tau_1 \ldots \int_0^\infty d\tau_3 e^{-i(\omega_1\tau_1+\omega_2\tau_2+\omega_3\tau_3)}$$
$$\tilde{\chi}^{(3)}(\tau_1,\tau_2,\tau_3). \qquad (2.20)$$

Note that because it does not matter which frequency we call $\omega_1$, which we call $\omega_2$, and so on, we can define the frequency-dependent susceptibilities to be invariant under permutations of their arguments, and we shall assume that this is the case.

As an illustration, let us compute the frequency-dependent susceptibilities for a medium consisting of two-level atoms. Let the upper level be $|a\rangle$, with an energy of $\hbar\omega_0$, and let the lower level be $|b\rangle$ with an energy of $0$, and we shall assume that there is one optically active electron. The Hamiltonian describing the interaction of this atom with an incident electromagnetic wave of frequency $\nu$ is

$$H = \hbar\omega_0|a\rangle\langle a| + \hbar g(E(t)e^{-i\nu t}+E^*(t)e^{i\nu t})(\sigma^{(+)}+\sigma^{(-)}) \qquad (2.21)$$

where $\sigma^{(+)}=|a\rangle\langle b|$ and $\sigma^{(-)}=|b\rangle\langle a|$. Here, $\hbar g = \hat{\mathbf{e}}\cdot\mathbf{d}$, where the polarization of the incident wave is in the direction of the unit vector $\hat{\mathbf{e}}$, and the phases of the atomic wave functions are chosen so that dipole matrix element of the transition

$$\mathbf{d} = -e\langle a|\mathbf{r}|b\rangle, \qquad (2.22)$$

is real. In the above equation $-e$ is the charge of the electron, and $\mathbf{r}$ is the position operator of the optically active electron. The electric field amplitude, $E(t)$, has been taken to be time dependent, because we are going to start the atom in its ground state at $t=-\infty$ and adiabatically turn on the field until it reaches a value of $E(t)=E_0$ at $t=0$, after which it remains steady. The reason for doing this is to eliminate transient effects, because we are interested in the steady state response of the system. Note that with this electric field we have that

$$E(\omega) = \frac{1}{\sqrt{2\pi}}\int_{-\infty}^\infty dt e^{-i\omega t}(E_0 e^{-i\nu t}+E_0^* e^{i\nu t})$$
$$= \sqrt{2\pi}[E_0\delta(\omega+\nu)+E_0^*\delta(\omega-\nu)]. \qquad (2.23)$$

We cannot solve the equations of motion resulting from this Hamiltonian in closed form, but this situation changes if we make what is called the rotating-wave approximation. In the limit of no interaction, we have in the Heisenberg picture that $\sigma^{(+)}\sim e^{i\omega_0 t}$ and $\sigma^{(-)}\sim e^{-i\omega_0 t}$. That means that if we are close to resonance, i.e. $\nu$ is close to $\omega_0$, then the terms $E(t)e^{-i\nu t}\sigma^{(+)}$ and $E^*(t)e^{i\nu t}\sigma^{(-)}$ are slowly varying and the terms $E(t)e^{-i\nu t}\sigma^{(-)}$ and $E^*(t)e^{i\nu t}\sigma^{(+)}$ are rapidly varying. Slowly varying terms will have a much larger effect on the dynamics than rapidly varying ones, so we drop the rapidly varying terms to give

$$H = \hbar\omega_0|a\rangle\langle a| + \hbar g(E(t)e^{-i\nu t}\sigma^{(+)}+E^*(t)e^{i\nu t}\sigma^{(-)}) \qquad (2.24)$$

This is the Hamiltonian in the rotating-wave approximation. This approximation is good near resonance, but less good when the detuning between the field frequency and atomic frequency is large. It will suffice for our purposes here.

We now proceed to derive and solve the equations of motion for the atom in the rotating-wave approximation. The quantum state of the atom is given by

$$|\psi(t)\rangle = c_a(t)e^{-i\omega_0 t}|a\rangle + c_b(t)|b\rangle, \tag{2.25}$$

and substituting this into the Schrödinger equation we find (with $E(t) = E_0$)

$$i\frac{d}{dt}\begin{pmatrix} c_a \\ c_b \end{pmatrix} = \begin{pmatrix} 0 & gE_0 e^{i\Delta t} \\ gE_0^* e^{-i\Delta t} & 0 \end{pmatrix}\begin{pmatrix} c_a \\ c_b \end{pmatrix}, \tag{2.26}$$

where $\Delta = \omega_0 - \nu$. Assuming that $c_b = \exp i\mu t$ we find that

$$\mu^2 + \Delta\mu - g^2|E_0|^2 = 0, \tag{2.27}$$

so that we have the following two possible values for $\mu$

$$\mu_\pm = \frac{1}{2}[-\Delta \pm (\Delta^2 + 4g^2|E_0|^2)^{1/2}]. \tag{2.28}$$

The normalized solution that corresponds to $\mu_+$ is

$$\begin{pmatrix} c_a \\ c_b \end{pmatrix} = \frac{1}{N_+}\begin{pmatrix} -(\mu_+/gE_0^*)e^{-i\mu_- t} \\ e^{i\mu_+ t} \end{pmatrix}, \tag{2.29}$$

and the one corresponding to $\mu_-$ is

$$\begin{pmatrix} c_a \\ c_b \end{pmatrix} = \frac{1}{N_-}\begin{pmatrix} -(\mu_-/gE_0^*)e^{-i\mu_+ t} \\ e^{i\mu_- t} \end{pmatrix}. \tag{2.30}$$

The normalization constants in the above equations are

$$N_\pm = \left(1 + \left|\frac{\mu_\pm}{gE_0}\right|^2\right)^{1/2}. \tag{2.31}$$

Examining these solutions in the limit $E_0 \to 0$, we find that it is the solution corresponding to $\mu_+$ that goes to $|b\rangle$, so this is the solution we need.

The polarization of the atom can be computed by taking the expectation value of the dipole-moment operator in the direction of the applied field, $-e\hat{\mathbf{e}} \cdot \mathbf{r}$, which, for our two-level atom, becomes $\hbar g(\sigma^{(+)} + \sigma^{(-)})$ (note that because we are assuming that the atomic states have well-defined parity, $\langle a|\mathbf{r}|a\rangle = \langle b|\mathbf{r}|b\rangle = 0$). If we have a medium consisting of $n$ two-level atoms per unit volume, the the polarization of the medium, $P(t)\hat{\mathbf{e}}$, is just $n$ times the polarization of each atom. Therefore, we find that

$$P(t) = -\frac{n\hbar g^2 \mu_+}{|gE_0|^2 + \mu_+^2}(E_0^* e^{i\nu t} + E_0 e^{-i\nu t}). \tag{2.32}$$

In order to find the different susceptibilities we expand this result in $E_0$. Keeping terms of up to third order, we have

$$P(t) = \left(-\frac{\hbar g^2}{\Delta} + \frac{2\hbar g^4 |E_0|^2}{\Delta^3}\right)(E_0^* e^{i\nu t} + E_0 e^{-i\nu t}). \tag{2.33}$$

From this we find

$$\begin{aligned} P^{(1)}(\omega) &= \sqrt{2\pi}\left(-\frac{\hbar g^2}{\Delta}\right)[E_0\delta(\omega+\nu)+E_0^*\delta(\omega-\nu)] \\ P^{(3)}(\omega) &= \sqrt{2\pi}\left(\frac{2\hbar g^4|E_0|^2}{\Delta^3}\right)[E_0\delta(\omega+\nu)+E_0^*\delta(\omega-\nu)]. \end{aligned} \quad (2.34)$$

We now have to compare these equations to our equations for the susceptibilities. We find that

$$\epsilon_0\chi^{(1)}(\nu) = \epsilon_0\chi^{(1)}(-\nu,) = \frac{-\hbar g^2}{\Delta}, \quad (2.35)$$

and

$$\epsilon_0\chi^{(3)}(-\nu,\nu,\nu) = \epsilon_0\chi^{(3)}(\nu,-\nu,-\nu) = \frac{\hbar g^4}{3\pi\Delta^3}. \quad (2.36)$$

This last equation also holds if we permute the arguments of the third-order nonlinear susceptibilities.

A more general method of calculating susceptibilities is by the use of perturbation theory. One starts with the density matrix equations describing the interaction of the electromagnetic field with a medium. These equations include damping effects. The equations are then solved perturbatively, where the perturbation is the field-matter interaction. The first-order term yields the linear susceptibility, the second-order term yields, $\chi^{(2)}$, and so on. This type of calculation allows us to take into account an arbitrary number of atomic or molecular energy levels, and the effects of the terms that were dropped when we made the rotating-wave approximation. Detailed accounts of these methods can be found in textbooks on nonlinear optics [5].

## 3  Field quantization

In order to proceed, our next step will have to be the quantization of the electromagnetic field. Before doing so, however, it is useful to present the canonical formalism for field quantization. We shall present it for a scalar field, $\phi(\mathbf{r},t)$ and treat the electromagnetic field explicitly in the following section. We start with a Lagrangian, $L$, which can be expressed in terms of a Lagrangian density, $\mathcal{L}$,

$$L = \int d^3r \mathcal{L}, \tag{3.1}$$

and we shall assume for now that $\mathcal{L}$ is a function of $\phi$, $\partial_t \phi$, and $\partial_j \phi$, where $\partial_t$ is a more compact way of writing $\partial/\partial t$ and $\partial_j$, $j=1,2,3$, corresponds to partial spatial derivatives in the $x$, $y$, and $z$ directions. This assumption will be true for most of what we do, but there will be one case, when we discuss fields in dispersive media, for which the Lagrangian density will also depend on mixed space and time derivatives. We define the action to be

$$S = \int_{t_1}^{t_2} dt L. \tag{3.2}$$

When we change the field, $\phi(\mathbf{r},t) \to \phi(\mathbf{r},t) + \delta\phi(\mathbf{r},t)$, where we consider only variations in the field, $\delta\phi(\mathbf{r},t)$, which vanish at $t=t_1$, $t=t_2$, and as $|\mathbf{r}| \to \infty$, then $S$ goes to $S + \delta S$. The equations of motion are determined by the condition that the action is stationary, that is $\delta S = 0$ for any choice of $\delta\phi$ obeying the boundary condtions.

The change in the Lagrangian can be expressed in terms of functional derivatives. These are defined by the equation

$$\delta L = \int d^3r \left[ \frac{\delta L}{\delta \phi} \delta\phi + \frac{\delta L}{\delta(\partial_t \phi)} \partial_t(\delta\phi) \right], \tag{3.3}$$

where only first order terms in $\delta\phi$ and its derivatives have been kept. These functional derivatives can be expressed in terms of partial derivatives of the Lagrangian density. To do so we begin by noting

$$\delta L = \int d^3r \left[ \frac{\partial \mathcal{L}}{\partial \phi} \delta\phi + \sum_{j=1}^{3} \frac{\partial \mathcal{L}}{\partial(\partial_j \phi)} \partial_j(\delta\phi) + \frac{\partial \mathcal{L}}{\partial(\partial_t \phi)} \partial_t(\delta\phi) \right]. \tag{3.4}$$

We can perform a partial integration on the term with the spatial derivatives of $\delta\phi$ yielding

$$\delta L = \int d^3r \left[ \left( \frac{\partial \mathcal{L}}{\partial \phi} - \sum_{j=1}^{3} \partial_j \frac{\partial \mathcal{L}}{\partial(\partial_j \phi)} \right) \delta\phi + \frac{\partial \mathcal{L}}{\partial(\partial_t \phi)} \partial_t(\delta\phi) \right]. \tag{3.5}$$

From this we see that

$$\begin{aligned}
\frac{\delta L}{\delta \phi} &= \frac{\partial \mathcal{L}}{\partial \phi} - \sum_{j=1}^{3} \partial_j \frac{\partial \mathcal{L}}{\partial(\partial_j \phi)} \\
\frac{\delta L}{\delta(\partial_t \phi)} &= \frac{\partial \mathcal{L}}{\partial(\partial_t \phi)}.
\end{aligned} \tag{3.6}$$

We can now go on to examine the variation of the action. We have that

$$\delta S = \int_{t_1}^{t_2} dt \int d^3 r \left[ \frac{\delta L}{\delta \phi} \delta \phi + \frac{\delta L}{\delta(\partial_t \phi)} \partial_t(\delta \phi) \right]. \tag{3.7}$$

We can now perform a partial integration on the time integral to give

$$\delta S = \int_{t_1}^{t_2} dt \int d^3 r \left[ \frac{\delta L}{\delta \phi} - \partial_t \frac{\delta L}{\delta(\partial_t \phi)} \right] \delta \phi. \tag{3.8}$$

If $\delta S = 0$ for any choice of $\delta \phi$, the expression in brackets must vanish, yielding the equation of motion

$$\frac{\delta L}{\delta \phi} - \partial_t \frac{\delta L}{\delta(\partial_t \phi)} = 0, \tag{3.9}$$

or, in terms of the Lagrangian density

$$\frac{\partial \mathcal{L}}{\partial \phi} - \sum_{j=1}^{3} \partial_j \frac{\partial \mathcal{L}}{\partial(\partial_j \phi)} - \partial_t \frac{\partial \mathcal{L}}{\partial(\partial_t \phi)} = 0. \tag{3.10}$$

We can also express the equations of motion in terms of the Hamiltonian. The canonical momentum is just the functional derivative of the Lagrangian with respect to $\partial_t \phi$,

$$\pi = \frac{\delta L}{\delta(\partial_t \phi)}. \tag{3.11}$$

The Hamiltonian is then

$$H = \int d^3 r \pi(\mathbf{r}, t) \partial_t \phi(\mathbf{r}, t) - L, \tag{3.12}$$

and the equation of motion is given by

$$\frac{\delta H}{\delta \phi} = -\partial_t \pi \tag{3.13}$$

The final step in the quantization of the theory is to make all of the quantities above operators, and to impose the canonical equal-time commutation relations between the coordinate, which is the field, and its corresponding momentum

$$[\phi(\mathbf{r}, t), \pi(\mathbf{r}', t)] = i\hbar \delta^{(3)}(\mathbf{r} - \mathbf{r}'), \tag{3.14}$$

as well as the equal-time commutation relations between the field and itself and between the canonical momentum and itself

$$[\phi(\mathbf{r}, t), \phi(\mathbf{r}', t)] = [\pi(\mathbf{r}, t), \pi(\mathbf{r}', t)] = 0. \tag{3.15}$$

These commutation relations, when used in conjunction with the Hamiltonian, can be used to find the Heisenberg equation of motion for the field.

Let us now apply these methods to a simple example, the quantization of the motion of a string of length $l$ with periodic boundary conditions. The string extends in the $x$ direction, from $x = 0$ to $x = l$, and we will assume that its transverse motion is only in the $y$ direction. The motion of the string in the $y$ direction is then described by a field, $\phi(x,t)$, where $\phi(x,t)$ is the $y$ displacement of the string at position $x$ and at time $t$. The equation of motion for $\phi(x,t)$ is

$$v^2 \frac{\partial^2 \phi}{\partial x^2} = \frac{\partial^2 \phi}{\partial t^2}, \tag{3.16}$$

where $v$ is the wave velocity for the string, which depends on its tension and mass density. The Lagrange density for this system is given by

$$\mathcal{L} = \frac{1}{2}\left(\frac{\partial \phi}{\partial t}\right)^2 - \frac{1}{2}v^2\left(\frac{\partial \phi}{\partial x}\right)^2. \tag{3.17}$$

Application of Eq. (3.10) to this Lagrange density yields the equation of motion above. The canonical momentum is found to be

$$\pi(x,t) = \frac{\partial \mathcal{L}}{\partial(\partial_t \phi)} = \frac{\partial \phi}{\partial t}, \tag{3.18}$$

and this gives us the Hamiltonian

$$H = \int_0^l dx \left[\frac{1}{2}\left(\frac{\partial \phi}{\partial t}\right)^2 + \frac{1}{2}v^2\left(\frac{\partial \phi}{\partial x}\right)^2\right]. \tag{3.19}$$

Now that we have the full canonical classical theory, we can quantize it by applying the canonical commutation relation. The equal-time commutator between the field and its canonical momentum is

$$[\phi(x,t), \partial_t \phi(x',t)] = i\hbar \delta(x-x'). \tag{3.20}$$

We can use this to define creation and annihilation operators with the proper commutation relations. The normal modes of the string are characterized by a wave number $k = 2\pi n/l$, where $n$ is an integer, and a frequency $\omega_k = |k|v$. We can define an annihilation operator corresponding the the mode with wave number $k$ by

$$a_k = \frac{1}{\sqrt{2l\hbar}} \int_0^l dx e^{-ikx}\left[\sqrt{\omega_k}\phi(x,t) + \frac{i}{\sqrt{\omega_k}}\partial_t \phi(x,t)\right]. \tag{3.21}$$

With this definition, it is straightforward to verify that $[a_k, a_{k'}^\dagger] = \delta_{k,k'}$. In addition, making use of the fact that

$$\sum_k e^{ik(x-x')} = l\delta(x-x'), \tag{3.22}$$

it is possible to invert the equations for $a_k$ and $a_k^\dagger$ in terms of $\phi$ and $\partial_t \phi$ to find

$$\begin{aligned}\phi(x,t) &= \sum_k \sqrt{\frac{\hbar}{2l\omega_k}}(e^{ikx}a_k + e^{-ikx}a_k^\dagger) \\ \partial_t \phi(x,t) &= i\sum_k \sqrt{\frac{\hbar\omega_k}{2l}}(e^{-ikx}a_k^\dagger - e^{ikx}a_k).\end{aligned} \tag{3.23}$$

These equations can now be inserted into the expression for the Hamiltonian, with the result that

$$H = \frac{1}{2} \sum_k \hbar\omega_k (a_k^\dagger a_k + a_k a_k^\dagger). \tag{3.24}$$

Finally, dropping an (infinite) constant the Hamiltonian becomes

$$H = \sum_k \hbar\omega_k a_k^\dagger a_k. \tag{3.25}$$

The ground state of the string is the vacuum state, $|0\rangle$, the state that is annihilated by all of the annihilation operators, $a_k|0\rangle = 0$. Other states of the string, those containing excitations, are given by applying creation operators to the vacuum state. Not much happens in this theory, because the excitations do not interact; they simply propagate along the string. In order to create an interaction between the excitations, we would have to add terms to the Hamiltonian consisting of products of three or more creation and annihilation operators. These are the types of terms that occur when describing a quantum theory of electrodynamics in nonlinear media.

## 4 The quantized free electromagnetic field

So far we have treated the electromagnetic field classically, but we now need to treat it as a quantum mechanical system. We shall begin with the free field, and we shall treat its quantization in some detail, because when we move to situations involving dielectric media, the quantization becomes more complicated. It will be easier to quantize the field when dielectric media are present, if we have a solid understanding of its quantization in free space. In quantum optics textboooks, the free-space quantization is often accomplished by breaking the fields up into modes and treating each mode as a harmonic oscillator. We shall use a more quantum field theoretic approach, as it will prove useful later. We shall also discuss some properties of the free field, such as squeezing and entanglement, which fields emerging from nonlinear dielectric media frequently possess.

Our ultimate goal is to find a Hamiltonian and commutation relations that lead to the free-space Maxwell equations

$$\nabla \cdot \mathbf{E} = 0 \qquad \nabla \times \mathbf{E} = -\frac{\partial \mathbf{B}}{\partial t}$$
$$\nabla \cdot \mathbf{B} = 0 \qquad \nabla \times \mathbf{B} = \epsilon_0 \mu_0 \frac{\partial \mathbf{E}}{\partial t}. \qquad (4.1)$$

We begin with classical fields, and first find a Lagrangian, and then a Hamiltonian, that leads to the above equations. A Lagrangian requires coordinates, and ours will be the components of the vector potential $A = (A_0, \mathbf{A})$, defined so that

$$\mathbf{E} = -\frac{\partial \mathbf{A}}{\partial t} - \nabla A_0 \qquad \mathbf{B} = \nabla \times \mathbf{A}. \qquad (4.2)$$

Note that with this definition, we automatically satisfy two of the Maxwell equations

$$\nabla \times \mathbf{E} = -\frac{\partial \mathbf{B}}{\partial t} \qquad \nabla \cdot \mathbf{B} = 0. \qquad (4.3)$$

The Lagrangian, which is a function of the vector potential and its time derivative, is expressed as the integral of a Lagrangian density

$$L\left(A_0, \frac{\partial A_0}{\partial t}, \mathbf{A}, \frac{\partial \mathbf{A}}{\partial t}\right) = \int d^3 r \mathcal{L}\left(A_0, \frac{\partial A_0}{\partial t}, \mathbf{A}, \frac{\partial \mathbf{A}}{\partial t}\right). \qquad (4.4)$$

The equations of motion that come from this Lagrangian can be expressed in terms of the Lagrangian density

$$\partial_t \left(\frac{\partial \mathcal{L}}{\partial(\partial_t A_\mu)}\right) + \sum_{j=1}^{3} \partial_j \left(\frac{\partial \mathcal{L}}{\partial(\partial_j A_\mu)}\right) - \frac{\partial \mathcal{L}}{\partial A_\mu} = 0, \qquad (4.5)$$

where $\mu = 0, 1, 2, 3$. With the correct choice of Lagrangian density, these four equations will be the remaining Maxwell equations. We choose

$$\begin{aligned} \mathcal{L} &= \frac{1}{2}\epsilon_0 \mathbf{E}^2 - \frac{1}{2\mu_0} \mathbf{B}^2 \\ &= \frac{1}{2}\epsilon_0 \left(\frac{\partial \mathbf{A}}{\partial t} + \nabla A_0\right)^2 - \frac{1}{2\mu_0}(\nabla \times \mathbf{A})^2, \end{aligned} \qquad (4.6)$$

and now need to confirm that this choice does give us the remaining Maxwell equations. If we now substitute this density into Eq. (4.5) with $\mu = 0$, we obtain Gauss' law, $\nabla \cdot \mathbf{E} = 0$, while doing so with $\mu = 1, 2, 3$ gives us

$$\nabla \times \mathbf{B} = \epsilon_0 \mu_0 \frac{\partial \mathbf{E}}{\partial t}. \tag{4.7}$$

Having found the proper Lagrangian, the next thing to do is to find the corresponding Hamiltonian. In order to do so we first need to find the canonical momentum. Its components are given by

$$\Pi_0 = \frac{\partial \mathcal{L}}{\partial(\partial_t A_0)} = 0, \qquad \Pi_j = \frac{\partial \mathcal{L}}{\partial(\partial_t A_j)} = -\epsilon_0 E_j. \tag{4.8}$$

The vanishing of $\Pi_0$, the momentum canonically conjugate to $A_0$, means we lose $A_0$ as an independent field, and the Hamiltonian will be a function of $\mathbf{A}$ and $\mathbf{\Pi}$ only. The Hamiltonian is expressed in terms of a Hamiltonian density,

$$H = \int d^3 r \, \mathcal{H}(\mathbf{A}, \mathbf{\Pi}), \tag{4.9}$$

which is itself given by

$$\mathcal{H} = \sum_{j=1}^{3} (\partial_t A_j) \Pi_j - \mathcal{L} = \frac{1}{2} \epsilon_0 E^2 + \frac{1}{2\mu_0} B^2 + \epsilon_0 \mathbf{E} \cdot \nabla A_0. \tag{4.10}$$

The last term can be eliminated by noting that when the Hamiltonian density is substituted into the equation for the Hamiltonian, we can integrate it by parts, yielding $-\epsilon_0 \int d^3 r \, A_0 \nabla \cdot \mathbf{E}$, which is zero by Gauss' law. Therefore, the last term in the Hamiltonian density makes no contribution to the Hamiltonian, and it can be dropped. The equation of motion for $\Pi_j$ is given by

$$\partial_t \Pi_j = -\frac{\delta H}{\delta A_j}. \tag{4.11}$$

The left-hand side of this equation is the variational derivative of $H$ with respect to $A_j$, and we can find it in the same way we found the functional derivative of the Lagrangian. Let $A_j(\mathbf{r}, t) \to A_j(\mathbf{r}, t) + \delta A_j(\mathbf{r}, t)$. We then have that

$$H \to H + \int d^3 r \, \frac{\delta H}{\delta A_j} \delta A_j(\mathbf{r}, t) + \ldots \tag{4.12}$$

where the dots indicate terms that are higher order in $\delta A_j(\mathbf{r}, t)$. In our case we have that upon letting $A_j(\mathbf{r}, t) \to A_j(\mathbf{r}, t) + \delta A_j(\mathbf{r}, t)$, we have (keeping in mind that $\mathbf{E} = -\mathbf{\Pi}$ is an independent variable in the Hamiltonian formulation)

$$H \to H + \frac{1}{\mu_0} \int d^3 r \, \delta \mathbf{A} \cdot \nabla \times \mathbf{B} + \ldots, \tag{4.13}$$

after an integration by parts. We therefore find that for the free electromagnetic field

$$\frac{\delta H}{\delta A_j} = \frac{1}{\mu_0} (\nabla \times \mathbf{B})_j, \tag{4.14}$$

and Eq. (4.11) becomes

$$\nabla \times \mathbf{B} = \epsilon_0 \mu_0 \frac{\partial \mathbf{E}}{\partial t}. \tag{4.15}$$

Notice that in the Hamiltonian formulation, we have lost Gauss' law as an equation of motion. This is because $A_0$ is no longer an independent field in this formulation. We will impose it as a constraint on the theory. We will impose an additional constraint by fixing the gauge. The physical fields are invariant under the gauge transformation

$$\begin{aligned} \mathbf{A} &\rightarrow \mathbf{A} - \nabla \xi \\ A_0 &\rightarrow A_0 + \frac{\partial \xi}{\partial t}, \end{aligned} \tag{4.16}$$

where $\xi(\mathbf{r}, t)$ is an arbitrary function of space and time. We shall choose the radiation gauge. This incorporates the Coulomb gauge, which requires that $\mathbf{A}$ be transverse, i.e. $\nabla \cdot \mathbf{A} = 0$, and, in addition requires that $A_0 = 0$. We first show that we can eliminate $A_0$ by a choice of gauge. We start with $(A_0'', \mathbf{A}'')$, and we assume that $A_0'' \neq 0$. If we choose

$$\xi(\mathbf{r}, t) = -\int_{t_0}^{t} dt' \mathbf{A}''(\mathbf{r}, t'), \tag{4.17}$$

then our new vector potential, which we shall call $(A_0', \mathbf{A}')$ does obey the condtion $A_0' = 0$. Now we need to impose the Coulomb gauge condition. Starting from $\mathbf{A}'$, we choose a new function, $\xi(\mathbf{r})$, which is independent of time. This guarantees that the zero component of the new vector potential, which we shall call $\mathbf{A}$ will remain zero. If we choose

$$\xi(\mathbf{r}) = -\int d^3 r' \frac{1}{4\pi |\mathbf{r} - \mathbf{r}'|} \nabla' \cdot \mathbf{A}'(\mathbf{r}, t), \tag{4.18}$$

then we have $\nabla^2 \xi = \nabla \cdot \mathbf{A}'$, which implies that $\nabla \cdot \mathbf{A} = 0$. Note that though it appears that the right-hand side of the above equation depends on time, it does not. Gauss' law along with the fact that $A_0' = 0$ implies that $\partial_t \nabla \cdot \mathbf{A}' = 0$. Finally, at the end of this sequence of gauge transformations, we have a vector potential that satisfies $A_0 = 0$ and $\nabla \cdot \mathbf{A} = 0$.

In order to quantize the theory we would normally impose the equal-time commutation relations

$$\begin{aligned} &[A_j(\mathbf{r}, t), A_l(\mathbf{r}', t)] = 0 \\ &[\Pi_j(\mathbf{r}, t), \Pi_l(\mathbf{r}', t)] = [E_j(\mathbf{r}, t), E_l(\mathbf{r}', t)] = 0 \\ &[A_j(\mathbf{r}, t), \Pi_l(\mathbf{r}', t)] = -[A_j(\mathbf{r}, t), E_l(\mathbf{r}', t)] = i\hbar \delta_{jk} \delta^{(3)}(\mathbf{r} - \mathbf{r}'). \end{aligned} \tag{4.19}$$

However, the last of these commutation relations violates both the Coulomb gauge condition and Gauss' law, so we modify it to reflect the fact that both the vector potential and the electric field are transverse fields, i.e. both have a vanishing divergence. What we need on the right-hand side of the last commutator is a kind of delta function with a vanishing divergence, which acts the same way a standard delta function does for transverse fields. The transverse delta function has exactly these properties.

In order to define the transverse delta function, let us assume that we are quantizing the fields in a box of volume $V$ using periodic boundary conditions. Then we can express the transverse delta function as

$$\delta^{(tr)}_{lm}(\mathbf{r}) = \frac{1}{V}\sum_{\mathbf{k}}(\delta_{lm} - \hat{k}_l\hat{k}_m)e^{i\mathbf{k}\cdot\mathbf{r}}. \tag{4.20}$$

Here, the wave vectors are given by

$$\mathbf{k} = \left(\frac{2\pi n_x}{L_x}, \frac{2\pi n_y}{L_y}, \frac{2\pi n_z}{L_z}\right), \tag{4.21}$$

where $V = L_xL_yL_z$ and $n_x$, $n_y$, and $n_z$ are integers, and $\hat{\mathbf{k}} = \mathbf{k}/|\mathbf{k}|$ is a unit vector. From this definition it is clear that $\sum_{l=1}^{3}\partial_l\delta^{(tr)}_{lm}(\mathbf{r}) = 0$. In addition, the transverse delta function has the property that for any transverse vector field, i.e. one that satisfies $\nabla\cdot\mathbf{V} = 0$, we have

$$V_l(\mathbf{r}) = \sum_{m=1}^{3}\int d^3r'\,\delta^{(tr)}_{lm}(\mathbf{r}-\mathbf{r}')V_m(\mathbf{r}'). \tag{4.22}$$

Finally, we can now modify the commutation relation for the vector potential and the electric field to be

$$[A_j(\mathbf{r},t), E_l(\mathbf{r}',t)] = -i\hbar\delta^{(tr)}_{jl}(\mathbf{r}-\mathbf{r}'). \tag{4.23}$$

The other commutators in Eq. (4) remain the same.

Our next step will be to expand the vector potential and electric field in plane-wave modes, and to define creation and annihilation operators for these modes. The transverse plane-wave modes have the mode functions

$$\mathbf{u}_{\mathbf{k},\alpha}(\mathbf{r}) = \frac{1}{\sqrt{V}}\hat{\mathbf{e}}_{\mathbf{k},\alpha}e^{i\mathbf{k}\cdot\mathbf{r}}, \tag{4.24}$$

where $\alpha = 1,2$. The vectors $\hat{\mathbf{k}}$, $\hat{\mathbf{e}}_{\mathbf{k},1}$, and $\hat{\mathbf{e}}_{\mathbf{k},2}$ form an orthonormal set of vectors, where

$$\begin{aligned}\hat{\mathbf{e}}_{\mathbf{k},1}\times\hat{\mathbf{e}}_{\mathbf{k},2} &= \hat{\mathbf{k}} \quad \hat{\mathbf{e}}_{\mathbf{k},2}\times\hat{\mathbf{k}} = \hat{\mathbf{e}}_{\mathbf{k},1}\\ \hat{\mathbf{k}}\times\hat{\mathbf{e}}_{\mathbf{k},1} &= \hat{\mathbf{e}}_{\mathbf{k},2}.\end{aligned} \tag{4.25}$$

We also choose $\hat{\mathbf{e}}_{-\mathbf{k},\alpha} = -(-1)^{\alpha}\hat{\mathbf{e}}_{\mathbf{k},\alpha}$, which is consistent with the above equations. We now define the operator

$$a_{\mathbf{k},\alpha}(t) = \frac{1}{\sqrt{\hbar}}\int d^3r\,\mathbf{u}^{*}_{\mathbf{k},\alpha}(\mathbf{r})\cdot\left[\sqrt{\frac{\epsilon_0\omega_k}{2}}\mathbf{A}(\mathbf{r},t) - i\sqrt{\frac{\epsilon_0}{2\omega_k}}\mathbf{E}(\mathbf{r},t)\right], \tag{4.26}$$

where $\omega_k = |\mathbf{k}|c$. Making use of the equal-time commutation relations for the field operators, we find that

$$[a_{\mathbf{k},\alpha}, a_{\mathbf{k}',\alpha'}] = 0 \quad [a_{\mathbf{k},\alpha}, a^{\dagger}_{\mathbf{k}',\alpha'}] = \delta_{\mathbf{k},\mathbf{k}'}\delta_{\alpha,\alpha'}, \tag{4.27}$$

so that $a^{\dagger}_{\mathbf{k},\alpha}$ and $a_{\mathbf{k},\alpha}$ can clearly be interpreted as creation and annihilation operators.

It is also possible to invert the relation between the fields and the creation and annihilation operators by making use of the relation

$$\sum_{\mathbf{k},\alpha} u_{\mathbf{k},\alpha}(\mathbf{r})_l u^*_{\mathbf{k},\alpha}(\mathbf{r}')_m = \delta^{(tr)}_{lm}(\mathbf{r}-\mathbf{r}'). \tag{4.28}$$

We find that

$$\begin{aligned}\mathbf{A}(\mathbf{r},t) &= \sum_{\mathbf{k},\alpha} \frac{\hbar}{\sqrt{2\epsilon_0\omega_k V}}\hat{\mathbf{e}}_{\mathbf{k},\alpha}(a_{\mathbf{k},\alpha}e^{i\mathbf{k}\cdot\mathbf{r}} + a^\dagger_{\mathbf{k},\alpha}e^{-i\mathbf{k}\cdot\mathbf{r}}) \\ \mathbf{E}(\mathbf{r},t) &= \sum_{\mathbf{k},\alpha} i\sqrt{\frac{\hbar\omega_k}{2\epsilon_0 V}}\hat{\mathbf{e}}_{\mathbf{k},\alpha}(a_{\mathbf{k},\alpha}e^{i\mathbf{k}\cdot\mathbf{r}} - a^\dagger_{\mathbf{k},\alpha}e^{-i\mathbf{k}\cdot\mathbf{r}}).\end{aligned} \tag{4.29}$$

These expressions can then be substituted into the Hamiltonian to express it in terms of the creation and annihilation operators. The result is

$$H = \frac{1}{2}\sum_{\mathbf{k},\alpha}\hbar\omega_k(a^\dagger_{\mathbf{k},\alpha}a_{\mathbf{k},\alpha} + a_{\mathbf{k},\alpha}a^\dagger_{\mathbf{k},\alpha}). \tag{4.30}$$

The normally ordered form of the Hamiltonian is

$$H = \sum_{\mathbf{k},\alpha}\hbar\omega_k a^\dagger_{\mathbf{k},\alpha}a_{\mathbf{k},\alpha}, \tag{4.31}$$

which is equivalent to the one above it, because the two Hamiltonians differ from each other only by a constant. The ground state of this Hamiltonian is the vacuum state $|0\rangle$, the state that is annihilated by all of the annihilation operators, i.e. $a_{\mathbf{k},\alpha}|0\rangle = 0$ for all $\mathbf{k}$ and $\alpha = 1, 2$.

## 5  Nonclassical states and entanglement

We would like to explore the properties of the quantized electromagnetic field, and, in particular, see what properties a quantized field can have that a classical one cannot. As we shall see, nonlinear optical processes are prime sources of light that cannot be described classically. It is, therefore, useful to start by discussing some of the novel features of quantized light, so that we know what we are looking for when we perform complicated calculations whose goal to find the output of nonlinear optical devices. In order to keep the discussion relatively simple, we shall confine our attention throughout this section to a small number of field modes.

One of the things we can do with light is to count photons. We can open the shutter in front of a photodetector between the times $t$ and $t + \Delta t$ and see how many photons are detected. Photodetection is a probabilistic process, so the number of photons detected will vary from run to run, even if the input fields are identical. Let $p(m; t, t + \Delta t)$ be the probability that $m$ photons are detected in the time interval between $t$ and $t + \Delta t$. The collection of these probabilities, for a fixed time interval but for arbitrary $m$ is called the photocount distribution of the field in that time interval. These probabilities can be calculated by modeling the photodetector as a collection of atoms, which then interact with the electromagnetic field. The probability $p(m; t, t + \Delta t)$ can be found by determining the probability that $m$ of the atoms have absorbed a photon in the interval between $t$ and $t + \Delta t$, which means finding the probability that $m$ atoms are in an excited state at time $t + \Delta t$. The field itself can be modeled as a classical stochastic field or as the quantized electromagnetic field. The theory of photodetection is a well developed part of quantum optics [6, 7].

We would like to know if there are photocount distributions that are possible for quantized fields that are not possible for classical stochastic fields, and how these differences, if they exist, can be observed. The answer, due to R. J. Glauber, is that there are photocount distributions that can only be the result of a quantized field [7]. In order to show what he found, let us consider a single-mode field. A single-mode classical field is characterized by a complex field amplitude, $\alpha$, which contains information about the intensity and phase of the field. If the field is a stochastic one, it is characterized by a probability distribution for this field amplitude, $P_{cl}(\alpha)$. Quantum mechanically, we have a density matrix $\rho$ that describes the state of the field. This is the state of a single field mode, whose creation and annihilation operators are $a^\dagger$ and $a$. If we define a single-mode coherent state, $|\alpha\rangle$, where $\alpha$ is an arbitrary complex number, to be

$$|\alpha\rangle = e^{-|\alpha|^2/2} \sum_{n=0}^{\infty} \frac{(\alpha a^\dagger)^n}{n!} |0\rangle, \tag{5.1}$$

then a short calculation shows that $a|\alpha\rangle = \alpha|\alpha\rangle$, i.e. it is an eigenstate of the annihilation operator with eigenvalue $\alpha$. Any single-mode density matrix can be expressed as

$$\rho = \int d^2\alpha P(\alpha) |\alpha\rangle\langle\alpha|, \tag{5.2}$$

where $P(\alpha)$ is called the P representation of $\rho$, and the integration is over the entire complex plane, $d^2\alpha = d(\text{Re}\,\alpha) d(\text{Im}\,\alpha)$. The P representation is a c-number, not an operator, and is what is known as a quasidistribution function. While it shares some properties with a probability distribution, for example,

$$\int d^2\alpha P(\alpha) = 1 \tag{5.3}$$

it does not share others, for example it need not be positive. In fact, not only need it not be positive, it can be highly singular, containing arbitrarily high derivatives of the delta function. What Glauber found is that the expression for probabilities $p(m; t, t + \delta t)$ when calculated for a quantum field, is almost the same as the the expression calculated for a classical stochastic field, the only difference being the replacement of $P_{cl}(\alpha)$ in the classical expression by $P(\alpha)$ in the quantum. What that means is that if the quantum state $\rho$ has a P representation that has all of the properties of a probability distribution, that is, it is positive and has singularities no worse than a delta function, then there is a classical stochastic field that has the same photocount distribution. However, if $\rho$ has a P representation that does not have all of the properties of a probability distribution, then there is no classical stochastic field with the same photocount distribution. Such states are called nonclassical.

Let us look at some examples of nonclassical states. The number operator for a field mode, which is the observable corresponding to the number of photons in that mode, is given by $\hat{n}_a = a^\dagger a$. The average number of photons in the mode is $\langle \hat{n}_a \rangle$, and the fluctuations in the photon number are characterized by $(\Delta n_a)^2 = \langle (\hat{n}_a)^2 \rangle - \langle \hat{n}_a \rangle^2$. A single-mode state is said to have sub-Poissonian photon statistics if $(\Delta n_a)^2 < \langle \hat{n}_a \rangle$. Such a state is nonclassical. To see this, we first express $\langle \hat{n}_a \rangle$ and $\langle \hat{n}_a^2 \rangle$ in terms of the $P$ representation of the state

$$\begin{aligned} \langle \hat{n}_a \rangle &= \int d^2\alpha P(\alpha) |\alpha|^2 \\ \langle \hat{n}_a^2 \rangle &= \int d^2\alpha P(\alpha) \langle \alpha | (a^\dagger a)^2 | \alpha \rangle = \int d^2\alpha P(\alpha)(|\alpha|^4 + |\alpha|^2). \end{aligned} \tag{5.4}$$

We can now calculate $(\Delta n_a)^2 - \langle \hat{n}_a \rangle$ in terms of the $P$ representation of a state, which gives us

$$(\Delta n_a)^2 - \langle \hat{n}_a \rangle = \int d^2\alpha P(\alpha)(|\alpha|^2 - \langle \hat{n}_a \rangle)^2. \tag{5.5}$$

From this we can see that if the $P$ representation of the state behaves like a probability distribution, then the right-hand side of the above equation is greater than or equal to zero, so that the left-hand side must be as well. Therefore, for classical states, $(\Delta n_a)^2 \geq \langle \hat{n}_a \rangle$, and a state that violates this condition is nonclassical. Photon number states are examples of states that are sub Poissonian. The number state, $|n\rangle$, is an eigenstate of $\hat{n}_a$, i.e. $\hat{n}_a |n\rangle = n|n\rangle$. Since for any photon number state, $(\Delta n_a)^2 = 0$ and $\langle \hat{n}_a \rangle = n$, we see that for $n > 0$ they violate the condition for classical states, which implies that any number state other than the vacuum is nonclassical.

A second nonclassical effect is known as squeezing. Consider the operator

$$X(\phi) = \frac{1}{2}(e^{i\phi} a^\dagger + e^{-i\phi} a). \tag{5.6}$$

This operator and its fluctuations can be measured by means of homodyne detection. In this measurement, the mode to be measured is mixed with a second mode in a strong coherent state at a beam splitter, where $\phi$ is the phase of the coherent state amplitude. The difference in the photon numbers at the two output ports of the beam splitter will be proportional to $X(\phi)$. We also have that $X(0)$ and $X(\pi/2)$ correspond to the real and imaginary parts of the annihilation operator, $a$,

$$a = X(0) + iX(\pi/2). \tag{5.7}$$

Now let us examine the fluctuations in $X(\phi)$. For a state represented in terms of its $P$ representation we have

$$(\Delta X(\phi))^2 = \frac{1}{4} + \frac{1}{4}\int d^2\alpha P(\alpha)[e^{i\phi}(\alpha^* - \langle a^\dagger\rangle) + e^{i\phi}(\alpha - \langle a\rangle)]^2 \tag{5.8}$$

From this equation, we see that for a classical state the second term on the right-hand side is greater than or equal to zero, so that $(\Delta X(\phi))^2 \geq 1/4$. A state with $\Delta X(\phi) < 1/2$ will be nonclassical. As an example of a squeezed state consider

$$|\psi\rangle = c_0|0\rangle - c_2 e^{2i\phi}|2\rangle, \tag{5.9}$$

where $c_0$ and $c_2$ are real and positive, and $|c_0|^2 + |c_2|^2 = 1$. we find that

$$(\Delta X(\phi))^2 = \frac{1}{4} + c_2\left(c_2 - \frac{c_0}{\sqrt{2}}\right). \tag{5.10}$$

This state will be squeezed if $c_0 > \sqrt{2}c_2$.

We can easily generalize the notion of a nonclassical state to more than one mode, and this leads us in a natural way to a discussion of entanglement. As we shall see, nonlinear optical devices are often good sources of entangled light. Let us consider a two-mode state, $\rho_{ab}$. It can be represented in terms of a two-mode $P$ representation as

$$\rho_{ab} = \int d^2\alpha \int d^2\beta P(\alpha,\beta)|\alpha\rangle_a\langle\alpha| \otimes |\beta\rangle_b\langle\beta|, \tag{5.11}$$

and if $P(\alpha,\beta)$ has the properties of a probability distribution, which in general it does not, the state is classical. In order to define entanglement between the two modes, we first have discuss the concept of a separable state. If the two-mode state can be expressed as

$$\rho_{ab} = \sum_j p_j \rho_a^{(j)} \otimes \rho_b^{(j)}, \tag{5.12}$$

where $p_j$ are probabilities whose sum is 1, and $\rho_a^{(j)}$ and $\rho_b^{(j)}$ are density matrices for modes $a$ and $b$, respectively, then $\rho_{ab}$ is said to be separable. A separable state is one in which the correlations between the subsystems, in this case the modes $a$ and $b$, are classical. A separable state can be constructed by two parties, each in possession of one of the subsystems, each acting locally on their subsystem and communicating classically with the other party. A state that is not separable is said to be entangled, and an entangled state possesses quantum correlations between the subsystems. There is clearly a connection between classical states and separable states. In particular, all classical states are separable. Therefore, we can conclude that in order for a two-mode state to be entangled, it must be nonclassical.

Deciding whether a state is entangled or not is, in general, not a simple problem. There are, however, some sufficient conditions. Perhaps the most commonly used one for field modes is the one proved by Simon and by Duan, et al. [8, 9]. Consider a two-mode system in which the annihilation operators for the modes are $a$ and $b$. If a state satisfies the condition

$$[\Delta(x_a + x_b)]^2 + [\Delta(p_a - p_b)]^2 < 2, \tag{5.13}$$

where $x_a = (a^\dagger + a)/\sqrt{2}$, $p_a = i(a^\dagger - a)/\sqrt{2}$ and similarly for $x_b$ and $p_b$, then it is entangled. A related criterion states that a state is entangled if [10]

$$[\Delta(x_a + x_b)]^2 [\Delta(p_a - p_b)]^2 < 1. \tag{5.14}$$

It should be noted that all of the quantities in these inequalities can be measured by using homodyne detection, which means that these conditions can be used to determine whether fields occurring in an experiment are entangled.

Let us prove the first of these conditions. We shall assume that the state is separable, and show that the quantity on the left-hand side must be greater than or equal to 2. Hence, if this quantity is less than 2, the state must be entangled. We start by writing

$$\begin{aligned}[] [\Delta(x_a + x_b)]^2 + [\Delta(p_a - p_b)]^2 &= \sum_j p_j [\langle (x_a + x_b)^2 \rangle_j + \langle (p_a - p_b)^2 \rangle_j] \\ &\quad - \langle x_a + x_b \rangle^2 - \langle p_a - p_b \rangle^2, \end{aligned} \tag{5.15}$$

where expectation values with a subscript $j$ are taken with respect to the density matrix $\rho_a^{(j)} \otimes \rho_b^{(j)}$. This can be expressed as

$$\begin{aligned}[] [\Delta(x_a + x_b)]^2 + [\Delta(p_a - p_b)]^2 &= \sum_j p_j [(\Delta x_a)_j^2 + (\Delta x_b)_j^2 + (\Delta p_a)_j^2 + (\Delta p_b)_j^2] \\ &\quad + \sum_j p_j [(\langle x_a \rangle_j + \langle x_b \rangle_j)^2 + (\langle p_a \rangle_j - \langle p_b \rangle_j)^2] \\ &\quad - \langle x_a + x_b \rangle^2 - \langle p_a - p_b \rangle^2. \end{aligned} \tag{5.16}$$

We now note that because $[x_a, p_a] = 1$ and $[x_b, p_b] = 1$, we have that $\Delta x_a \Delta p_a \geq 1/2$ and $\Delta x_b \Delta p_b \geq 1/2$. These relations imply that

$$\begin{aligned} (\Delta x_a)^2 + (\Delta p_a)^2 &\geq 1 \\ (\Delta x_b)^2 + (\Delta p_b)^2 &\geq 1. \end{aligned} \tag{5.17}$$

We also have that the Schwarz inequality implies that

$$(\sum_j p_j)(\sum_j p_j \langle x_a + x_b \rangle_j^2) \geq (\sum_j p_j \langle x_a + x_b \rangle_j)^2, \tag{5.18}$$

with a similar result for the momenta. These results, when substituted into Eq. (5.16) yield Eq. (5.13).

Now let's use this condition to study the entanglement of the state

$$|\psi\rangle = c_0 |0\rangle + c_1 a^\dagger b^\dagger |0\rangle, \tag{5.19}$$

where $|c_0|^2 + |c_1|^2 = 1$. This is the type of state that is produced, to lowest order in the interaction, by a parametric down converter. The field is either in the vacuum, with amplitude $c_0$, or a pair of photons has been emitted, with amplitude $c_1$. Typically, $|c_1| \ll |c_0|$. This state is entangled as long as neither $c_0 = 0$ or $c_1 = 0$. Substituting this state into the left-hand side of the inequality appearing in Eq. (5.13), we find

$$[\Delta(x_a + x_b)]^2 + [\Delta(p_a - p_b)]^2 = 2 + 2(2|c_1|^2 + c_1^* c_0 + c_0^* c_1). \tag{5.20}$$

The right-hand side will be less than 2 if $2|c_1|^2 < -(c_1^* c_0 + c_0^* c_1)$, and this can certainly happen when $|c_1| \ll |c_0|$, for example when $c_0$ is real and positive, and $c_1$ is real and negative. The smallest value the right-hand side of the above equation can attain is $4 - 2\sqrt{2}$, when $c_0 = \cos(\pi/8)$ and $c_1 = -\sin(\pi/8)$. Therefore, we see that this condition can be used to demonstrate that the light emerging from a parametric down converter is entangled.

## 6 Linear medium without dispersion

We have seen how to quantize the electromagnetic field in free space, so let us now look at its quantization in a linear medium for which there is no dispersion. This was treated for the case of a homogeneous medium in a series of papers by Jauch and Watson [11] and was extended to the case of a single dielectric interface by Carniglia and Mandel [12]. We shall follow the treatment due to Glauber and Lewenstein, who considered a general inhomogeneous medium [13]. The medium is described by a spatially varying dielectric function, $\epsilon(\mathbf{r}) = \epsilon_0(1 + \chi^{(1)}(\mathbf{r}))$, where $\chi^{(1)}(\mathbf{r})$ is the linear polarizability of the medium, which we shall assume to be a scalar, i.e. we are assuming that the medium is isotropic. The equations of motion for the fields are now

$$\nabla \cdot \mathbf{D} = 0 \qquad \nabla \times \mathbf{E} = -\frac{\partial \mathbf{B}}{\partial t}$$
$$\nabla \cdot \mathbf{B} = 0 \qquad \nabla \times \mathbf{B} = \mu_0 \frac{\partial \mathbf{D}}{\partial t}, \tag{6.1}$$

where $\mathbf{D} = \epsilon(\mathbf{r})\mathbf{E}$. As in the case of the free-space theory, the basic field in the theory will be the vector potential, and the electric and magnetic fields will be expressed in terms of it in the same way (see Eq. (4.2)). Our gauge choice will, however, be different. We can still choose $A_0 = 0$, but the Coulomb gauge condition is modified. We now choose

$$\nabla \cdot [\epsilon(\mathbf{r})\mathbf{A}] = 0. \tag{6.2}$$

These gauge conditions and the definitions of the electric and magnetic fields in terms of the vector potential guarantee that the first three of the above equations are satisfied. The remaining equation implies that the vector potential satisfies

$$\mu_0 \epsilon(\mathbf{r}) \frac{\partial^2 \mathbf{A}}{\partial t^2} + \nabla \times (\nabla \times \mathbf{A}) = 0. \tag{6.3}$$

Our next task is to find the Lagrangian and Hamiltonian for this theory. The Lagrangian density is given by replacing $\epsilon_0$ by $\epsilon(\mathbf{r})$ in the free-field Lagrangian density

$$\mathcal{L} = \frac{1}{2}\epsilon(\mathbf{r})\mathbf{E}^2 - \frac{1}{2\mu_0}\mathbf{B}^2. \tag{6.4}$$

From this we find the canonical momentum

$$\Pi_j = \frac{\partial \mathcal{L}}{\partial(\partial_t A_j)} = -D_j. \tag{6.5}$$

Note that the canonical momentum is not the same as in the free theory. This will have consequences when we get to the quantum theory, because it is the canonical momentum that appears in the canonical commutation relations. Consequently, the free-space theory and the dielectric theory will not have the same commutation relations. The Hamiltonian density is now

$$\begin{aligned}\mathcal{H} &= \sum_{j=1}^{3}(\partial_t A_j)\Pi_j - \mathcal{L} = \frac{1}{2\epsilon(\mathbf{r})}\mathbf{\Pi}^2 + \frac{1}{2\mu_0}(\nabla \times \mathbf{A})^2 \\ &= \frac{1}{2}\epsilon(\mathbf{r})\mathbf{E}^2 + \frac{1}{2\mu_0}\mathbf{B}^2.\end{aligned} \tag{6.6}$$

We shall quantize this theory by decomposing the fields in terms of modes and treating each mode as a harmonic oscillator. The $k$th mode is given by

$$\mathbf{A}(\mathbf{r}, t) = e^{-i\omega_k t} \mathbf{f}_k(\mathbf{r}), \tag{6.7}$$

where $\nabla \cdot [\epsilon(\mathbf{r}) \mathbf{f}_k(\mathbf{r})] = 0$. The label $k$ here does not denote the wave number of a mode as it did in the free-field case; here it simply serves as an index to label the mode. It may be discrete or it may be continuous, depending on the boundary conditions imposed on the mode functions. Here we shall assume it is discrete. The functions $\mathbf{f}_k(\mathbf{r})$ satisfy the equation

$$\mu_0 \epsilon(\mathbf{r}) \omega_k^2 \mathbf{f}_k - \nabla \times (\nabla \times \mathbf{f}_k) = 0, \tag{6.8}$$

which follows from Eq. (6.3). These modes obey an orthogonality relation. To see this we first define

$$\mathbf{g}_k(\mathbf{r}) = \sqrt{\epsilon(\mathbf{r})} \mathbf{f}_k(\mathbf{r}), \tag{6.9}$$

and note that $\mathbf{g}_k$ satisfies the equation

$$\mu_0 \omega_k^2 \mathbf{g}_k - \frac{1}{\sqrt{\epsilon(\mathbf{r})}} \nabla \times \left( \nabla \times \frac{\mathbf{g}_k}{\sqrt{\epsilon(\mathbf{r})}} \right) = 0. \tag{6.10}$$

This implies that $\mathbf{g}_k$ is the eigenfunction of an Hermitian operator, so that these functions, when suitably normalized, satisfy

$$\delta_{k,k'} = \int d^3 r\, \mathbf{g}_k^*(\mathbf{r}) \cdot \mathbf{g}_{k'}(\mathbf{r}) = \int d^3 r\, \epsilon(\mathbf{r}) \mathbf{f}_k^*(\mathbf{r}) \cdot \mathbf{f}_{k'}(\mathbf{r}). \tag{6.11}$$

The functions $\mathbf{g}_k$ are complete in the space of functions satisfying $\nabla \cdot [\sqrt{\epsilon(\mathbf{r})} \mathbf{g}(\mathbf{r})] = 0$. We can use the functions $\mathbf{f}_k(\mathbf{r})$ to define a distribution

$$\delta_{mn}^{(\epsilon)}(\mathbf{r}, \mathbf{r}') = \sum_k f_{km}(\mathbf{r}) f_{kn}^*(\mathbf{r}'), \tag{6.12}$$

which, in the absence of a dielectric medium reduces to the transverse delta function.

We can now start the quantization procedure by expanding the vector potential in terms of the mode functions $\mathbf{f}_k$

$$\mathbf{A}(\mathbf{r}, t) = \sum_k Q_k(t) \mathbf{f}_k(\mathbf{r}), \tag{6.13}$$

where the $Q_k$ are operators. The fact that $\mathbf{A}$ is an hermitian operator implies that

$$\sum_k Q_k(t) \mathbf{f}_k(\mathbf{r}) = \sum_k Q_k^\dagger(t) \mathbf{f}_k^*(\mathbf{r}), \tag{6.14}$$

which, along with the orthogonality condition for the $\mathbf{f}_k$, implies that

$$Q_k = \sum_{k'} Q_{k'}^\dagger U_{k'k}^*, \tag{6.15}$$

where the matrix $U^*_{k'k}$ is defined by

$$U^*_{k'k} = \int d^3 r \epsilon(\mathbf{r}) \mathbf{f}^*_{k'}(\mathbf{r}) \mathbf{f}^*_k(\mathbf{r}). \tag{6.16}$$

This matrix is clearly symmetric, $U^*_{k'k} = U^*_{kk'}$, and also satisfies

$$\sum_{k'} U_{kk'} U^*_{k''k'} = \delta_{k,k''}. \tag{6.17}$$

Finally, $U^*_{k'k} = 0$ unless $k$ and $k'$ correspond to modes of the same energy, because modes corresponding to solutions of Eq. (6.8) with different values of $\omega_k$ are orthogonal. We can also expand the canonical momentum in terms of the modes

$$\mathbf{\Pi}(\mathbf{r},t) = \sum_k P_k(t) \epsilon(\mathbf{r}) \mathbf{f}^*_k(\mathbf{r}), \tag{6.18}$$

where, as before, the $P_k$ are operators. The factor of $\epsilon(\mathbf{r})$ is necessary so that $\mathbf{\Pi}$ satisfies $\nabla \cdot \mathbf{\Pi} = -\nabla \cdot \mathbf{D} = 0$. The fact that $\mathbf{\Pi}$ is hermitian and the orthogonality relation for the $\mathbf{f}_k$ imply that

$$P^\dagger_k = \sum_{k'} P_{k'} U^*_{k'k}. \tag{6.19}$$

We can now insert these expressions into the Hamiltonian. We find, making use of the properties of $U_{k'k}$, that

$$\int d^3 r \frac{1}{2\epsilon(\mathbf{r})} \mathbf{\Pi}^2 = \frac{1}{2} \sum_k P^\dagger_k P_k. \tag{6.20}$$

The second term in the Hamiltonian requires more work. We have

$$\begin{aligned}
\frac{1}{2\mu_0} \int d^3 r (\nabla \times \mathbf{A})^2 &= \frac{1}{2\mu_0} \int d^3 r \sum_k \sum_{k'} Q_k Q_{k'} (\nabla \times \mathbf{f}_k) \cdot (\nabla \times \mathbf{f}_{k'}) \\
&= \frac{1}{2\mu_0} \sum_k \sum_{k'} Q_k Q_{k'} \int d^3 r \mathbf{f}_k \cdot \nabla \times (\nabla \times \mathbf{f}_{k'}) \\
&= \frac{1}{2\mu_0} \sum_k \sum_{k'} \mu_0 \omega^2_{k'} U_{kk'} Q_k Q_{k'},
\end{aligned} \tag{6.21}$$

where use has been made of Eq. (6.8). Finally, we can make use of the properties of $U_{kk'}$ to give

$$\frac{1}{2\mu_0} \int d^3 r (\nabla \times \mathbf{A})^2 = \frac{1}{2} \sum_k \omega^2_k Q^\dagger_k Q_k, \tag{6.22}$$

so that the entire Hamiltonian is

$$H = \frac{1}{2} \sum_k (P^\dagger_k P_k + \omega^2_k Q^\dagger_k Q_k). \tag{6.23}$$

If we now impose the equal-time commutation relations

$$[Q_k, Q_{k'}] = [Q^\dagger_k, Q^\dagger_{k'}] = [Q_k, Q^\dagger_{k'}] = 0 \tag{6.24}$$

$$[P_k, P_{k'}] = [P_k^\dagger, P_{k'}^\dagger] = [P_k, P_{k'}^\dagger] = 0 \tag{6.25}$$

$$[Q_k, P_{k'}] = i\hbar \delta_{k,k'}, \tag{6.26}$$

then the resulting Heisenberg equations of motion are identical to the Maxwell equations. If we express $P_k^\dagger$ in terms of $P_k$, we obtain one final commutation relation

$$[Q_k, P_{k'}^\dagger] = i\hbar U_{kk'}^*. \tag{6.27}$$

These commutation relations then give us the equal-time commutation relation between the vector potential and its canonical momentum

$$[A_m(\mathbf{r}, t), \Pi_n(\mathbf{r}', t)] = -[A_m(\mathbf{r}, t), D_n(\mathbf{r}', t)] = i\hbar \delta_{mn}^{(\epsilon)}(\mathbf{r}, \mathbf{r}'). \tag{6.28}$$

Note that these are not the free-space commutation relations.

Next, we would like to define annihilation and creation operators for the modes, $\mathbf{f}_k(\mathbf{r})$, and then express the fields in terms of them. These operators should satisfy the commutation relations

$$[a_k, a_{k'}^\dagger] = \delta_{k,k'} \quad [a_k, a_{k'}] = 0. \tag{6.29}$$

We begin by assuming that

$$a_k = \alpha_k Q_k + \beta_k P_k^\dagger, \tag{6.30}$$

where $\alpha_k$ and $\beta_k$ are constants to be determined. We will assume, however, that they only depend on $k$ through the frequecy, $\omega_k$, that is, if $\omega_k = \omega_{k'}$, then $\alpha_k = \alpha_{k'}$ and $\beta_k = \beta_{k'}$. By making use of the commutation relations for $Q_k$, $P_k$, and their adjoints, we then find

$$[a_k, a_{k'}^\dagger] = i\hbar(\alpha_k \beta_k^* - \alpha_k^* \beta_k)\delta_{k,k'} \tag{6.31}$$

so that we must have

$$i\hbar(\alpha_k \beta_k^* - \alpha_k^* \beta_k) = 1. \tag{6.32}$$

We also find that

$$[a_k, a_{k'}] = i\hbar(\alpha_k \beta_{k'} U_{kk'}^* - \alpha_{k'} \beta_k U_{kk'}^*). \tag{6.33}$$

Because $U_{kk'}$ vanishes unless $\omega_k = \omega_{k'}$, and $\alpha_k$ and $\beta_k$ only depend on $k$ through $\omega_k$, the right-hand side of the above equation vanishes. Our next step is to express $Q_k$ and $P_k$ in terms of the creation and annihilation operators. If we take the adjoint of Eq. (6.30), multiply by $U_{k'k}^*$ and sum over $k'$, we find

$$\sum_{k'} U_{k'k}^* a_{k'}^\dagger = \sum_{k'} U_{k'k}^* (\alpha_{k'}^* Q_{k'}^\dagger + \beta_{k'}^* P_{k'})$$
$$= \alpha_k^* Q_k + \beta_k^* P_k^\dagger, \tag{6.34}$$

where we have again made use of the fact that $U_{kk'}$ vanishes unless $\omega_k = \omega_{k'}$, and $\alpha_k$ and $\beta_k$ only depend on $k$ through $\omega_k$. We can solve the above equation and Eq. (6.30) for $Q_k$ and $P_k^\dagger$, and then take the adjoint to find $P_k$. This gives us

$$Q_k = i\hbar(\beta_k^* a_k - \beta_k \sum_{k'} U_{k'k}^* a_{k'}^\dagger)$$
$$P_k = i\hbar(\alpha_k a_k^\dagger - \alpha_k^* \sum_{k'} U_{k'k} a_{k'}). \tag{6.35}$$

Linear medium without dispersion					29These expressions can now be inserted into the equations for **A** and **D**. Choosing

$$\alpha_k = \left(\frac{\omega_k}{2\hbar}\right)^{1/2} \quad \beta_k = i\left(\frac{1}{2\hbar\omega_k}\right)^{1/2}, \tag{6.36}$$

we find that

$$\begin{aligned}
\mathbf{A}(\mathbf{r},t) &= \sum_k \left(\frac{\hbar}{2\omega_k}\right)^{1/2} [a_k \mathbf{f}_k(\mathbf{r}) + a_k^\dagger \mathbf{f}_k^*(\mathbf{r})] \\
\mathbf{D}(\mathbf{r},t) &= i\epsilon(\mathbf{r}) \sum_k \left(\frac{\hbar\omega_k}{2}\right)^{1/2} [a_k \mathbf{f}_k(\mathbf{r}) - a_k^\dagger \mathbf{f}_k^*(\mathbf{r})].
\end{aligned} \tag{6.37}$$

Finally, the expressions for $Q_k$ and $P_k$ in terms of the creations and annihilation operators can be inserted into the Hamiltonian, yielding

$$\begin{aligned}
H &= \frac{1}{2} \sum_k \hbar\omega_k (a_k^\dagger a_k + a_k a_k^\dagger) \\
&= \sum_k \hbar\omega_k a_k^\dagger a_k + C(\epsilon),
\end{aligned} \tag{6.38}$$

where $C(\epsilon)$ is a formally infinite constant, which can be dropped.

This theory can be used to study how the quantum properties of a field scattered by a dielectric medium change. We are interested in it here as an example of a situation in which the quantization procedure is different than it is in free space. As we have noted, the canonical momentum for the theory with a dielectric is different from that without one, and this has consequences for the commutation relations of the theory. In this case, the commutation relations depend not just upon the presence of a dielectric medium but on the spatial dependence of the polarizability as well.

### 7   Phenomenological models

Before jumping in and quantizing the electromagnetic field in a nonlinear dielectric, it is useful to study some highly simplified models to get an idea of some of the phenomena that can result. Our models will be described by Hamiltonians that couple several modes either to themselves or to each other. We will not derive these Hamiltonians in this section, but merely give plausibility arguments as to why they look as they do. These Hamiltonians underlie most of the work that has been done on the quantum theory of nonlinear optics.

#### 7.1   $\chi^{(2)}$ interactions

Let us start with a two-mode interaction. One mode, whose creation and annihilation operators are $a^\dagger$ and $a$, has a frequency of $\omega$, and the second, whose creation and annihilation operators are $b^\dagger$ and $b$,, has a frequency of $2\omega$. The $\chi^{(2)}$ term in the series expansion of the polarization of the nonlinear medium yields a polarization that is quadratic in the electric field, and this polarization couples back to the electric field, giving an interaction that is cubic in the field. This suggests a Hamiltonian of the form

$$H = \hbar\omega a^\dagger a + 2\hbar\omega b^\dagger b + \hbar\kappa[(a^\dagger)^2 b + a^2 b^\dagger]. \tag{7.1}$$

The first two terms comprise the free-field Hamiltonian of the two modes, and the term proportional to $\kappa$, which is itself proportional to $\chi^{(2)}$, describes the interaction. We have only kept slowly-varying terms in the interaction. In the absence of an interaction, we have that $a(t) = e^{-i\omega t} a(0)$ and $b(t) = e^{-2i\omega t} b(0)$. Because the interaction is weak, we expect that it will still be the case that, even in the presence of the interaction, $a(t)$ will be approximately proportional to $e^{-i\omega t}$ and $b(t)$ will be approximately proportional to $e^{-2i\omega t}$. If we now consider terms that are cubic in the creation and annihilation operators of the two modes, the only slowly varying ones are the terms we have kept in the interaction part of the Hamiltonian. The slowly-varying terms should give the dominant behavior in the dynamics; the remaining terms are rapidly oscillating and their effect will wash out.

The interaction in the Hamiltonian can do two things. It can combine two $a$ photons into a $b$ photon, and it can split one $b$ photon into two $a$ photons. Consequently, it can describe two processes. The first is parametric down conversion, In which a pump at frequency $2\omega$, produces a subharmonic (also called the signal) at frequency $\omega$. The second is second harmonic generation, in which a strong field a frequency $\omega$ produces its second harmonic at frequency $2\omega$.

Let us look at parametric down conversion first. In doing so, we are immediately faced with a problem, we cannot solve the Heisenberg equations of motion resulting from the above Hamiltonian. In order to get around this problem, an additional approximation is introduced, the parametric approximation. We assume that the pump field is in a large-amplitude coherent state, in particular the state $|\beta\rangle$, and that it can be replaced by a time-dependent c-number. This gives us the single-mode Hamiltonian

$$H = \hbar\omega a^\dagger a + \hbar\kappa[\beta e^{-2i\omega t}(a^\dagger)^2 + \beta^* e^{2i\omega t} a^2]. \tag{7.2}$$

The solution of the Heisenberg equations for this Hamiltonian is straightforward. We find that

$$\frac{da}{dt} = \frac{i}{\hbar}[H, a] = -i(\omega a + 2\kappa\beta e^{-2i\omega t} a^\dagger). \tag{7.3}$$

Setting $a(t) = A(t)e^{-i\omega t}$, we obtain the simpler equation

$$\frac{dA}{dt} = -2i\kappa\beta A^\dagger, \tag{7.4}$$

which has the solution

$$A(t) = A(0)\cosh(2|\beta|\kappa t) - ie^{i\phi}A^\dagger(0)\sinh(2|\beta|\kappa t), \tag{7.5}$$

where we have expressed $\beta$ as $\beta = |\beta|e^{i\phi}$.

Let us now consider the case $\phi = \pi/2$ and define the field quadrature components

$$X_1(t) = \frac{1}{2}(A(t) + A^\dagger(t)) \qquad X_2(t) = \frac{i}{2}(A^\dagger(t) - A(t)). \tag{7.6}$$

We then find that

$$X_1(t) = e^{2|\beta|\kappa t}X_1(0) \qquad X_2(t) = e^{-2|\beta|\kappa t}X_2(0). \tag{7.7}$$

Therefore, the parametric down converter amplifies the $X_1$ quadrature, and it attenuates the $X_2$ quadrature. If the input state to the $a$ mode is a coherent state, $|\alpha\rangle$, then if $\alpha$ is real then $\langle A(t)\rangle = \alpha\exp(2|\beta|\kappa t)$, and the input is amplified. If, however, $\alpha$ is imaginary, then $\langle A(t)\rangle = \alpha\exp(-2|\beta|\kappa t)$, and the input is attenuated. This is an example of phase-sensitive amplification.

We can also look at the fluctuations in the output of the down converter. The uncertainties in $X_1$ and $X_2$ obey the uncertainty relation

$$(\Delta X_1)(\Delta X_2) \geq \frac{1}{4}. \tag{7.8}$$

A state that satisfies the the uncertainty relation as an equality is called a minimum uncertainty state. Suppose we start the down-converter in the vacuum state. We then find that $\Delta X_1(0) = \Delta X_2(0) = 1/2$ and

$$\Delta X_1(t) = \frac{1}{2}e^{2|\beta|\kappa t} \qquad \Delta X_2(t) = \frac{1}{2}e^{-2|\beta|\kappa t}. \tag{7.9}$$

Therefore, the state that results from the vacuum state is a minimum uncertainty state. If the uncertainty in one of the quadratures of a state is less than $1/2$, it is called a squeezed state, and its fluctuations in the squeezed quadrature are less than the fluctuations in either quadrature in the vacuum state. What comes out of the output of a down converter when the input is the vacuum state, then, is a squeezed, minimum-uncertainty state, and this state is known as a squeezed vacuum state. This was first pointed out by David Stoler [14].

Squeezed states have a number of uses. They can be used for highly accurate measurements. For example, we might wish to detect a signal by observing the shift in the amplitude of a coherent state, i.e. $|\alpha\rangle \to |\alpha + \delta\alpha\rangle$. This would be the case if we were trying to detect a phase shift in an interferometer using a coherent state as an input state. For a coherent state, as for the vacuum, we have $\Delta X_1(0) = \Delta X_2(0) = 1/2$, and $X_1(0)$ corresponds to the real part of $\alpha$ and $X_2(0)$ to its imaginary part. Consequently, we can only determine $\delta\alpha$ to an accuracy of $1/2$. If we use a squeezed state, however, and we are only interested in determining the change in the $X_2$ component, then we can measure it to an accuracy of $(1/2)\exp(-2|\beta|\kappa t)$. It has been shown that the use of squeezed states in an interferometer can significantly improve its ability to detect small phase changes. Squeezed states have, more recently, found a number of applications in the field of quantum information.

Now let us return to the Hamiltonian in Eq. (7.1). We shall have more to say about the parametric approximation shorlty. While we cannot solve the equations of motion resulting from this Hamiltonian, we can make some statements about the states that are produced. This is because the Hamiltonian obeys a conservation law, in particular, it commutes with the operator

$$M = \hat{n}_a + 2\hat{n}_b, \tag{7.10}$$

so that $M$ is a conserved quantity. One immediate consequence of this is a simple relation between the number of photons in the $a$ mode and the number in the $b$ mode

$$\langle M(0) \rangle = 2\langle \hat{n}_b(t) \rangle + \langle \hat{n}_a(t) \rangle, \tag{7.11}$$

where, it should be noted, the left-hand side requires only knowledge of the initial state of the system.

If we work harder we can find a relation between the photon number fluctuations in the two modes [15]. Using Eq. (7.10) and its square we find

$$\begin{aligned}[] [\Delta M(0)]^2 &= 4[\Delta n_b(t)]^2 + [\Delta n_a(t)]^2 \\ &+ 4[\langle \hat{n}_a(t)\hat{n}_b(t)\rangle - \langle \hat{n}_a(t)\rangle\langle \hat{n}_b(t)\rangle]. \end{aligned} \tag{7.12}$$

The Schwarz inequality implies that

$$|\langle (\hat{n}_a - \langle \hat{n}_a\rangle)(\hat{n}_b - \langle \hat{n}_b\rangle)\rangle| \leq \Delta n_a \Delta n_b, \tag{7.13}$$

where we have dropped the time argument for simplicity. Substituting the previous equation into this one gives

$$-4\Delta n_a \Delta n_b \leq (\Delta M)^2 - 4(\Delta n_b)^2 - (\Delta n_a)^2 \leq 4\Delta n_a \Delta n_b, \tag{7.14}$$

which implies the two inequalities

$$\begin{aligned} \Delta M(0) &\leq 2\Delta n_b(t) + \Delta n_a(t) \\ \Delta M(0) &\geq |2\Delta n_b(t) - \Delta n_a(t)|. \end{aligned} \tag{7.15}$$

Finally, these can be combined to give

$$2\Delta n_b(t) + \Delta M(0) \geq \Delta n_a(t) \geq |2\Delta n_b(t) - \Delta M(0)|, \tag{7.16}$$

which gives us the desired relation between the number fluctuations.

This relation is easiest to interpret if both the $a$ and $b$ modes are initially in number states. This implies that $\Delta M(0) = 0$ which in turn tells us that $\Delta n_a(t) = 2\Delta n_b(t)$. If $\Delta M(0)$ is small but not zero, for example if the signal mode is initially in the vacuum and the photon statistics of the pump are highly sub-Poissonian, then this relation will be approximately true. Under these conditions, even without solving for the detailed dynamics, we can conclude that the number fluctuations in the signal are roughly twice those in the pump.

We can also use the conservation law to determine whether the signal-mode state is nonclassical [16]. As was discussed earlier, any single-mode state, pure or mixed, can be represented in the form

$$\rho = \int d^2\alpha P(\alpha)|\alpha\rangle\langle\alpha|, \tag{7.17}$$

where $P(\alpha)$ is the Glauber-Sudarshan P representation. If $P(\alpha)$ has the properties of a probability distribution, i.e. it is positive and while it may contain delta functions, it does not contain their derivatives, then we call the resulting state classical. A state that is not classical is called nonclassical.

Now let us return to the examination of the states produced by our Hamiltonian, and first give a short argument as to why the states it produces are nonclassical. More detail will be provided shortly. Consider the case in which at $t = 0$ the pump mode is in an arbitrary state and the signal mode is in the vacuum. Because every photon which disappears from the pump produces two photons in the signal, only signal-mode number states with even photon numbers are populated. Such a state cannot be classical unless it is the vacuum state. As a result, if the signal mode state has photons in it, then it is nonclassical.

Now let us look at this argument in more detail. First we need to show that a state with only even photon numbers is either the vacuum or nonclassical. To that end define the projection operators

$$\begin{aligned} Q_o &= \sum_{n=0}^{\infty} |2n+1\rangle\langle 2n+1| \\ Q'_e &= \sum_{n=1}^{\infty} |2n\rangle\langle 2n|. \end{aligned} \tag{7.18}$$

The operator $Q_o$ projects onto the space of odd photon number states and $Q'_e$ projects onto the space which includes all even photon number states except the vacuum. Representing the state of the field in terms of a P representation, we find

$$\langle Q_o \rangle - \langle Q'_e \rangle = \int d^2\alpha P(\alpha) e^{-|\alpha|^2}(1 + \sinh|\alpha|^2 - \cosh|\alpha|^2). \tag{7.19}$$

Now if the state is classical, i. e. $P(\alpha) \geq 0$, then the right-hand side of Eq. (7.19) is nonnegative and

$$\langle Q_o \rangle \geq \langle Q'_e \rangle. \tag{7.20}$$

If this condition is violated the state is nonclassical. If a state contains only even photon numbers then $\langle Q_o \rangle = 0$ so that it is nonclassical if $\langle Q'_e \rangle > 0$. Therefore, if a state contains only even photon numbers it is either the vacuum ($\langle Q'_e \rangle = 0$) or it is nonclassical.

We now need to show that a signal-mode state containing only even photon numbers is produced. In order to do this we need to show that ${}_a\langle n_a|\rho_a|n_a\rangle_a$ is zero unless $n_a$ is even, where $\rho_a$ is the reduced density matrix of the signal mode. We begin by noting that if

$$|\Psi(t)\rangle = U(t)|0\rangle_a \otimes |\psi_b\rangle_b, \tag{7.21}$$

where $U(t)$ is the time development transformation, then, because $[M, H] = 0$, we have

$$e^{i\pi M}|\Psi(t)\rangle = U(t)e^{i\pi M}|0\rangle_a \otimes |\psi_b\rangle_b = U(t)|0\rangle_a \otimes e^{2\pi i \hat{n}_b}|\psi_b\rangle_b. \tag{7.22}$$

The operator $\exp(2\pi i \hat{n}_b)$ is just the identity operator on the $b$ mode as can be seen immediately by considering its action on number states. Eq. (7.22) can now be expressed as

$$e^{i\pi \hat{n}_a}|\Psi(t)\rangle = |\Psi(t)\rangle, \tag{7.23}$$

which implies that

$$e^{i\pi \hat{n}_a}\rho_a(t) = \rho_a(t). \tag{7.24}$$

Taking the expectation of both sides of this equation in the number state $|n_a\rangle_a$ gives

$$(-1)^{n_a}{}_a\langle n_a|\rho_a(t)|n_a\rangle_a = {}_a\langle n_a|\rho_a(t)|n_a\rangle_a. \tag{7.25}$$

This implies that ${}_a\langle n_a|\rho_a(t)|n_a\rangle_a = 0$ if $n_a$ is odd, that is only even photon numbers are present in the signal-mode state. This completes our proof, and shows that if the number of photons in the signal is greater than zero the signal is in a nonclassical state.

It is also possible to show that second harmonic generation, which is described by the same Hamitonian, will produce nonclassical states [16]. In this case one starts from the state $|\alpha\rangle_a|0\rangle_b$, i. e. a coherent state in the $a$ mode and the vacuum in the $b$ mode. The basic idea is that it takes two $a$-mode photons to produce one $b$-mode photon. Consequently, the number distribution of the $b$ mode is more concentrated near zero than that of the $a$ mode. It is sufficiently concentrated, in fact, to be nonlcassical. The argument which shows this in detail will not be reproduced here, but it leads to the result that the $b$ mode is either in the vacuum state or nonclassical.

A final property which can be deduced from the conservation law is the relation between the rotational symmetries in phase space of the initial pump state and the signal-mode state. As we shall see the signal-mode state has twice the rotational symmetry of the pump state. This is perhaps best illustrated by means of an example. Suppose that the pump is initially in a coherent state, $|\beta\rangle_b$ and the signal is in the vacuum state. The pump state can be represented as a point in the complex plane (the $b$-mode phase space) located at $\beta$ surrounded by a circle of radius $1/2$. The circle represents the fluctuations in the real and imaginary parts of the field amplitude, which correspond to the operators $X_{1b} = (b^\dagger + b)/2$ and $X_{2b} = i(b^\dagger - b)/2$, respectively. For a coherent state the fluctuations are the same in both (and, in fact, all) directions. If this state is rotated about the origin by an angle of $2\pi$ it is mapped back into itself (as is any state). On the other hand, the signal-mode state, because it has twice the symmetry of the pump state, is invariant under a rotation by $\pi$. This is consistent with our knowledge of the results from the parametric approximation. In that case a pump mode is a strong coherent state produces a signal mode in a squeezed vacuum state. This state is represented in our two-dimensional phase space as an ellipse centered at the origin; a state which is invariant under a rotation by $\pi$.

In order to prove these assertions let us first define the operators

$$U_a(\theta) = e^{i\theta \hat{n}_a} \qquad U_b(\theta) = e^{i\theta \hat{n}_b} \qquad U_M(\theta) = e^{i\theta M/2} = U_a(\theta/2)U_b(\theta). \tag{7.26}$$

The initial state of the system is assumed to be $|\Psi\rangle = |0\rangle_a|\psi_b\rangle_b$ where

$$U_b(2\pi/n)|\psi_b\rangle_b = e^{i\phi}|\psi_b\rangle_b, \tag{7.27}$$

that is, a rotation by $2\pi/n$ maps $|\psi_b\rangle_b$ into itself multiplied by a phase factor. This implies that

$$U_M(2\pi/n)U(t)|\Psi\rangle = U(t)U_M(2\pi/n)|\Psi\rangle = e^{i\phi}U(t)|\Psi\rangle. \tag{7.28}$$

From this we find

$$\begin{aligned}
{}_b\langle n_b|U(t)|\Psi\rangle\langle\Psi|U^{-1}(t)|n_b\rangle_b &= {}_b\langle n_b|U_M(2\pi/n)U(t)|\Psi\rangle \\
&\quad \langle\Psi|U^{-1}(t)U_M^{-1}(2\pi/n)|n_b\rangle_b \\
&= U_a(\pi/n)\,{}_b\langle n_b|U(t)|\Psi\rangle \\
&\quad \langle\Psi|U^{-1}(t)|n_b\rangle_b U_a^{-1}(\pi/n).
\end{aligned} \tag{7.29}$$

Summing both sides over $n_b$ we obtain

$$\rho_a(t) = U_a(\pi/n)\rho_a(t)U_a^{-1}(\pi/n), \tag{7.30}$$

i. e. the $a$-mode state is invariant under a rotation by $\pi/n$. Summarizing we can say that if at $t = 0$ the signal mode is in the vacuum state and the pump-mode state is invariant under a rotation by $2\pi/n$, then at any time the signal-mode state is invariant under a rotation by $\pi/n$.

This result implies that if the pump mode is initially in a squeezed vacuum state (invariant under a rotation by $\pi$), then the signal will be invariant under a rotation by $\pi/2$. This suggests that in phase space the signal mode state will have a 4-pronged structure. This has been verified by numerical calculations of the Wigner function of the signal-mode state which clearly shows 4 prongs and a 4-fold rotational symmetry [17].

There is also a three-mode version of the process we have been considering. Its Hamiltonian is given by

$$H = \hbar\omega c^\dagger c + \hbar\omega_1 a^\dagger a + \hbar\omega_2 b^\dagger b + \hbar\kappa(c^\dagger ab + ca^\dagger b^\dagger), \tag{7.31}$$

where $c$, $a$, and $b$ are the annihilation operators of the pump, signal and idler modes, respectively, and $\omega = \omega_1 + \omega_2$. This Hamiltonian can describe parametric amplification or sum-frequency generation. In parametric amplification the pump mode is in a large-amplitude coherent state (the parametric approximation is usually employed) and strong correlations are produced between the signal and idler modes. These highly correlated two-mode states have found application in quantum information. For sum-frequency generation, the $a$ and $b$ modes are initially excited, and they give rise to photons in the $c$ mode whose frequency is the sum of those in the $a$ and $b$ modes. In this case the two operators

$$M_1 = \hat{n}_a - \hat{n}_b \qquad M_2 = 2\hat{n}_c + \hat{n}_a + \hat{n}_b, \tag{7.32}$$

or any linear combination of them, commute with $H$. As in the degenerate case we can find a relation between the number fluctuations in the pump and those in either of the other two modes [18]. We find that

$$\begin{aligned}
\Delta n_c(t) + \Delta K_1(0) &\geq \Delta n_a(t) \geq |\Delta n_c(t) - \Delta K_1(0)| \\
\Delta n_c(t) + \Delta K_2(0) &\geq \Delta n_b(t) \geq |\Delta n_c(t) - \Delta K_2(0)|,
\end{aligned} \tag{7.33}$$

where

$$\begin{aligned}
K_1 &= \frac{1}{2}(M_1 + M_2) = \hat{n}_c + \hat{n}_a \\
K_2 &= \frac{1}{2}(M_2 - M_1) = \hat{n}_c + \hat{n}_b.
\end{aligned} \tag{7.34}$$

These relations are most useful when $\Delta K_1(0)$ and $\Delta K_2(0)$ are small. For example, this will occur if the signal and idler modes are initially in their vacuum states and the pump is in a highly sub-Poissonian state.

In the nondegenerate case it is also possible to find a relation between the number fluctuations of the $a$ and $b$ modes. We have

$$\Delta M_1(0) \geq |\Delta n_a(t) - \Delta n_b(t)|. \tag{7.35}$$

If $\Delta M_1(0)$ is small, then the number fluctuations in the signal and idler are similar. If both modes start in the vacuum state, then $\Delta M_1(0) = 0$ and $\Delta n_a(t) = \Delta n_b(t)$. This conclusion is independent of the initial pump state.

We can also use the fact that $M_1$ is conserved to show that if the signal and idler modes are initially in the vacuum state, then at later times the signal-idler state is either the vacuum or nonclassical. This follows from the fact that a two-mode state is nonclassical if

$$\langle(\hat{n}_a(t) - \hat{n}_b(t))^2\rangle - \langle \hat{n}_a(t) - \hat{n}_b(t)\rangle^2 < \langle \hat{n}_a(t)\rangle + \langle \hat{n}_b(t)\rangle. \tag{7.36}$$

If the $a$ and $b$ modes are originally in the vacuum state the left-hand side of Eq. (7.36) will be zero at $t = 0$, and the fact that $[M_1, H] = 0$ implies that it will be zero for all time. Therefore, if either $\langle \hat{n}_a(t)\rangle$ or $\langle \hat{n}_b(t)\rangle$ is greater than zero the signal-idler state will be nonclassical. The reason for this is that the number of photons in the two modes is highly correlated. For example, if we measure the photon number in one mode we immediately know what it is in the other. Correlations which are this strong are not permitted classically. These correlations have been observed experimentally in both a cw, oscillator configuration [19] and in a pulsed, amplifier configuration [20].

## 7.2 $\chi^{(3)}$ interactions

Perhaps the most common phenomenon to arise out of a $\chi^{(3)}$ nonlinearity is that of the intensity-dependent refractive index. In a medium of this type, the refractive index consists of two terms, the first is constant, and is just the usual linear index of refraction, while the second is proportional to the intensity of the field. In the case of a single mode, this leads to the quantum mechanical Hamiltonian

$$H_1 = \hbar\omega a^\dagger a + \frac{\hbar\kappa}{2}(a^\dagger)^2 a^2, \tag{7.37}$$

and in the case of two modes,

$$H_2 = \hbar\omega_1 a^\dagger a + \hbar\omega_2 b^\dagger b + \frac{\hbar\kappa}{2} a^\dagger b^\dagger ab, \tag{7.38}$$

where, in both cases, $\kappa$ is proportional to $\chi^{(3)}$.

In order to see what types of effects these Hamiltonians give rise to, let us first consider the second one and see how an initial state that is a product of a coherent state in the $a$ mode and a number state in the $b$ mode evolves. We have that

$$\begin{aligned} |\psi(t)\rangle &= e^{-itH_2/\hbar}|\alpha\rangle_a|n\rangle_b \\ &= e^{-in\omega_2 t} e^{-it[\omega_1+(n\kappa/2)]a^\dagger a}|\alpha\rangle_a|n\rangle_b \\ &= e^{-in\omega_2 t}|\alpha(t)\rangle_a|n\rangle_b, \end{aligned} \tag{7.39}$$

where $\alpha(t) = \exp\{-it[\omega_1 + (n\kappa/2)]\}\alpha$. Therefore, we see that for times greater than zero, we still have the product of a coherent state and a number state, but the phase of the coherent state has changed, and the amount of the change is proportional to the number of photons initially in the $b$ mode. If we measure the phase shift in the $a$ mode, which could be done by means of an interferometer, then we can determine the number of photons in the $b$ mode. Note also that the state of the $b$ mode is not affected by this measurement. This represents what is known as a quantum nondemolition measurement of photon number [21].

Let us now see what happens when we use the single-mode Hamiltonian to evolve a state that is initially a coherent state. We have that

$$|\psi(t)\rangle = e^{-itH_1/\hbar}|\alpha\rangle = e^{-|\alpha|^2/2} \sum_{n=0}^{\infty} \frac{\alpha^n}{n!} e^{-it(n\omega + n(n-1)(\kappa/2))}|n\rangle. \tag{7.40}$$

We can use this to calculate the expectation value of the annihilation operator, which corresponds to the complex amplitude of the field,

$$\begin{aligned}\langle\psi(t)|a|\psi(t)\rangle &= \alpha e^{-i\omega t}\exp[-|\alpha|^2(1 - e^{-i\kappa t})] \\ &\cong \alpha e^{-it(\omega + \kappa|\alpha|^2)} e^{-(\kappa t|\alpha|)^2/2}.\end{aligned} \tag{7.41}$$

Note that the nonlinear interaction leads to an additional phase, which is proportional to $|\alpha|^2$, and, at longer times, causes the magnitude of the complex amplitude to decay. This is a result of the fact that the different number-state components of the coherent state pick up different phases as the state evolves and this causes the overall phase uncertainty of the state to increase. That means that when adding the contributions of the different number state components to form the complex amplitude, there is some cancellation because of the different phases. Because the number operator commutes with the Hamiltonian, the expectation of the number operator, and of its moments, is not affected by the time evolution.

If we combine the state in Eq. (7.40) with a coherent state at a beam splitter, we can produce another kind of nonclassical field state, one with sub-Poissonian photon statistics [22]. We recall that the photon statistics of a single-mode field are sub-Poissonian if $(\Delta n_a)^2 < \langle \hat{n}_a \rangle$, and that such a state is nonclassical.

The next thing we need to do is describe how a beam splitter works. There are two input modes, with annihilation operators $a_{in}$ and $b_{in}$, and two output modes, with annihilation operators $a_{out}$ and $b_{out}$. The operators are related by

$$\begin{aligned}a_{out} &= \sqrt{T}a_{in} + \sqrt{R}b_{in} \\ b_{out} &= -\sqrt{R}a_{in} + \sqrt{T}b_{in},\end{aligned} \tag{7.42}$$

where $T$ and $R$ are the transmissivity and reflectivity of the beam splitter, respectively. If $T = 1$ the input and output modes are the same, corresponding to complete transmission, and if $T = 0$, the input and output modes are interchanged, with appropriate phase shifts, corresponding to complete reflection. In our case, the input state in the $a$ mode will be the state in Eq. (7.40) and the input state in the $b$ mode will be a coherent state with amplitude $\beta e^{-i\omega t}$. We shall drop the $e^{-i\omega t}$ factors in what follows, because they cancel for the quantities we are calculating. We shall also be interested in the case in which $T$ is close to one and $|\beta|$ is large, with $\sqrt{R}\beta = \xi$. Making these substitutions, we find that in the $a_{out}$ output mode

$$\begin{aligned}(\Delta n_a)_{out}^2 - \langle \hat{n}_{aout}\rangle &= 2[\xi\langle(a^\dagger)^2 a\rangle + \xi^*\langle a^\dagger a^2\rangle - |\alpha|^2(\xi\langle a^\dagger\rangle + \xi^*\langle a\rangle)] \\ &+ \xi^2\langle(a^\dagger)^2\rangle + (\xi^*)^2\langle a^2\rangle - \xi^2\langle a^\dagger\rangle^2 - (\xi^*)^2\langle a\rangle^2 \\ &+ 2|\xi|^2(|\alpha|^2 - |\langle a\rangle|^2),\end{aligned} \quad (7.43)$$

where the expectation values without subscripts, i.e. the ones on the right-hand side of the equation, are expectation values of the *in* operators, that is the operators at the input to the beam splitter. These are just expectation values in the state in Eq. (7.40). Setting $\phi = \kappa t$ and, again neglecting the $e^{-i\omega t}$ factors, we find that

$$\begin{aligned}\langle a^2\rangle &= \alpha^2 e^{-i\phi} e^{-2|\alpha|^2(i\phi+\phi^2)} \\ \langle(a^\dagger)^2 a\rangle &= (\alpha^*)^2\alpha e^{i\phi} e^{|\alpha|^2(i\phi-\phi^2/2)}.\end{aligned} \quad (7.44)$$

Now set $\alpha = |\alpha|e^{i\theta}$, $\xi = re^{i\eta}$, and assume that $\phi \ll 1$, $|\alpha| \gg 1$, and $\phi|\alpha|$ is of order one. Making these substitutions, we find that

$$\begin{aligned}(\Delta n_a)_{out}^2 - \langle\hat{n}_{aout}\rangle &= -4r\phi|\alpha|^3 e^{-(|\alpha|\phi)^2/2}\sin(\eta-\theta+|\alpha|^2\phi) \\ &\quad 2r^2|\alpha|^2(1-e^{-(|\alpha|\phi)^2})\{1 \\ &\quad -e^{-(|\alpha|\phi)^2}\cos[2(\eta-\theta+|\alpha|^2\phi)]\}.\end{aligned} \quad (7.45)$$

Finally, setting $\eta - \theta + |\alpha|^2\phi = \pi/2$ and minimizing with respect to $r$, we find

$$(\Delta n_a)_{out}^2 - \langle\hat{n}_{aout}\rangle = -\frac{2|\alpha|^3\phi e^{-(|\alpha|\phi)^2}}{1-e^{-2(|\alpha|\phi)^2}}. \quad (7.46)$$

We see, then, that using this scheme of mixing the output from a $\chi^{(3)}$ medium with an appropriately chosen coherent state at a beam splitter, we can create a field whose photon statistics are significantly sub-Poissonian. We note that the right-hand side of the above equation is comparable to the expectation value of the photon number in the $a_{out}$ mode, which is just

$$\langle\hat{n}_{aout}\rangle = |\alpha|^2 + \frac{|\alpha|\phi e^{-(|\alpha|\phi)^2/2}}{1-e^{-2(|\alpha|\phi)^2}}. \quad (7.47)$$

This is yet another illustration of the fact that nonlinear media provide good ways to generate nonclassical states of the electromagnetic field.

## 7.3 Parametric Approximation

As we have seen, if one of the coupled modes we are considering is initially in a highly excited coherent state, we can assume that, at least for a certain period of time, it will act like a classical field. To this end, we can replace the operators for this field in the Hamiltonian by c-numbers. This is the parametric approximation. Once this is done we can solve the problem exactly, if the interaction is the result of a $\chi^{(2)}$ nonlinearity, because the equations of motion for the remaining operators are linear. Because this approximation is both important and so often employed, we would now like to enquire into its history and justification.

The parametric approximation was born at the same time as the quantum mechanical study of parametric processes. In their 1961 paper Louisell, Yariv, and Siegman considered a model of a parametric amplifier consisting of two modes of the electromagnetic field (the signal and the idler) coupled by a medium with an oscillating dielectric constant [23]. The modulation of the dielectric constant occurs at a frequency which is the sum of the frequencies of the two modes, and it is assumed that this modulation can be described classically. This led them to the Hamiltonian

$$H = \hbar\omega_1 a^\dagger a + \hbar\omega_2 b^\dagger b + \hbar(\xi^* e^{i\omega t} ab + \xi e^{-i\omega t} a^\dagger b^\dagger), \qquad (7.48)$$

where $a$ and $b$ are the annihilation operators of the two modes, $\omega = \omega_1 + \omega_2$, and $|\xi|$ is proportional to the amplitude of the modulation of the dielectric constant. They used this model to study the quantum noise properties of the parametric amplification process [23, 24].

Mollow and Glauber were the next to analyse the parametric amplifier [25, 26]. They emphasized that the oscillating dielectric constant of Louisell, Yariv, and Siegman is the result of an intense light wave in a nonlinear dielectric. They used the Hamiltonian in the previous paragraph and concentrated their attention on the one-and two-mode P representations of the signal and the idler. They found that if one of the modes starts in a finite-temperature thermal state, then the other mode, no matter what its initial state, will become classical after a sufficient period of time. The situation with the two-mode P function is the opposite; two mode states which are initially classical can remain so for only a finite period of time.

The intense interest in squeezing in the 1980's led to a renewal of work on the parametric amplifier. As we have seen, the degenerate parametric amplifier produces minimum uncertainty squeezed states. In the parametric approximation the amount of squeezing which can be obtained is arbitrarily large; one only need wait long enough. This, however, was clearly an artifact of the parametric approximation. At long enough times the dynamical and quantum mechanical aspects of the pump, both of which are ignored in this approximation, will play a role. It became obvious that it would be necessary to go beyond the parametric approximation to determine how much squeezing is possible.

The first calculation to accomplish this was done by Hillery and Zubairy using a path-integral technique [27, 28]. This was shortly followed by another, done by Scharf and Walls, using an aymptotic method developed by Scharf [29]. The two results coud not be immediately compared because one treated the degenerate parametric amplifier (Hillery and Zubairy) and the other the nondegenerate case (Scharf and Walls). The next step was taken by Crouch and Braunstein who developed two additional techniques for calculating corrections to the parametric approximation and used them to find the maximum amount of squeezing [30]. One method was based on an iterative solution to the Heisenberg equations of motion in which only dominant terms are kept. The solution is expressed as a series in powers of $1/\sqrt{N_p}$, where $N_p$ is the number of photons in the pump. Using numerical techniques they were able to calculate the complete result (dominant terms and corrections to them) for the quadrature variances up to order $1/N_p^2$. The dominant term method has since been extended and refined by Cohen and Braunstein [31]. In addition, Crouch and Braunstein developed a technique, based on the work of Drummond and Gardiner [32], which makes use of a description of quantum systems based on Ito stochastic differential equations. They found the sets of equations for both the degenerate and nondegenerate parametric amplifiers, found approximate solutions to them (again using an iterative approach), and used them to calculate corrections to the parametric approximation.

Crouch and Braunstein also compared the solutions found by different methods. The results from all of them were the same except those found by Scharf and Walls. They concluded that this result is not correct. Subsequent numerical calculations by Kinsler, Fernee, and Drummond [33] and the work of Cohen and Braunstein [31] has lent further support to this conclusion.

More recently the parametric approximation has been extended to cover pumps which are not in coherent states [31, 34, 35]. In particular, pumps which have a large coherent amplitude but are squeezed so that their phase fluctuations are reduced have been treated. In the degenerate parametric amplifier, it was expected that phase squeezing of the pump would enhance the squeezing of the signal. For a coherent state pump, it is the phase fluctuations in the pump that are responsible for the limit on the amount of signal-mode squeezing. If these fluctuations are reduced, the squeezing should increase. Within the parametric approximation, however, it was found that for the pump squeezing to have any effect it had to be so large that it actually produced a degradation of signal- mode squeezing. This paradoxical result is a consequence of the fact that the amplitude fluctuations of the pump become so large that its error ellipse overlaps the origin [31]. This means that if most of the pump has a phase of $\phi$, there is a part of it which has a phase of $\phi + \pi$. This latter part enhances the signal-mode fluctuations in the direction in which the rest of the pump is squeezing them, which reduces the overall squeezing effect [31, 34]. If corrections to the parametric approximation are included, then the situation is improved [31]. The maximum amount of signal-mode squeezing is increased by moderate phase squeezing of the pump.

Now let us look at corrections to the parametric approximation for the degenerate parametric amplifier in more detail. Following Crouch and Braunstein we take for the interaction Hamiltonian (this is equivalent to the Hamiltonian we have been using with a redefinition of the coupling constant and pump phase)

$$H = i\frac{\hbar \kappa}{2}[b(a^\dagger)^2 - b^\dagger a^2]. \tag{7.49}$$

Note that by using the interaction part of the Hamiltonian as the full Hamiltonian, we are neglecting the part of the evolution due to the free field part of the full Hamiltonian. We will be using this Hamiltonian to find equations of motion for the creation and annihilation operators, and neglecting the free field Hamiltonian simply means that we are solving for the slowly varying behavior of these operators, that is, in the notation of the previous subsection, we are solving for $A(t)$ instead of $a(t)$. If the pump mode is in a large-amplitude coherent state with amplitude $\beta = \sqrt{N_p}\exp(i\phi_p)$ we make the parametric approximation by replacing $b$ and $b^\dagger$ in $H$ by $\beta$ and $\beta^*$, respectively, to give

$$H_p = i\frac{\hbar \kappa}{2}\sqrt{N_p}[e^{i\phi_p}(a^\dagger)^2 - e^{-i\phi_p}a^2]. \tag{7.50}$$

The equations of motion for $a$ and $a^\dagger$ which follow from $H_p$ are

$$\frac{da}{dt} = \kappa\sqrt{N_p}e^{i\phi_p}a^\dagger \qquad \frac{da^\dagger}{dt} = \kappa\sqrt{N_p}e^{-i\phi_p}a. \tag{7.51}$$

and are easily solved to give

$$a(t) = a(0)\cosh u + a^\dagger(0)e^{i\phi_p}\sinh u, \tag{7.52}$$

where $u = \kappa\sqrt{N_p}t$. This solution, and its adjoint, allow us to find the properties of the signal as a function of time. In particular, if we define the quadrature operators

$$X_{1a} = \frac{1}{2}(a^\dagger + a) \qquad X_{2a} = \frac{i}{2}(a^\dagger - a), \tag{7.53}$$

and start at $t = 0$ with the signal mode in the vacuum state, we find

$$(\Delta X_{1a})^2 = \frac{1}{4}e^{2u}\cos^2(\phi_p/2) + \frac{1}{4}e^{-2u}\sin^2(\phi_p/2) \tag{7.54}$$

$$(\Delta X_{2a})^2 = \frac{1}{4}e^{-2u}\cos^2(\phi_p/2) + \frac{1}{4}e^{2u}\sin^2(\phi_p/2). \tag{7.55}$$

Note that if $\phi_p = 0$, then $\Delta X_{1a}(u)$ grows exponentially with time and $\Delta X_{2a}(u)$ decreases exponentially with time thereby becoming squeezed.

At this level of approximation there is no limit to how much squeezing in the $X_{2a}$ direction is possible. There are two effects, however, which have been neglected in the parametric approximation which will enforce a limit. The first is pump depletion. In replacing the pump mode operators by c-numbers the parametric approximation assumes that the amplitude of the pump remains constant. This is not a bad approximation for small times, but it gets worse as time progresses. The second effect which has been ignored is pump fluctuations. Wodkiewicz and Zubairy showed that phase fluctuations in a classical pump can limit the amount of squeezing which will be produced [36]. Even if classical fluctuations can be eliminated, however, there will still be quantum fluctuations in the pump field that will affect signal-mode squeezing.

In order to estimate the the effect of pump phase fluctuations we present an argument due to Crouch and Braunstein [30]. Let us assume that the pump is initially in a coherent state with a large, real amplitude $\sqrt{N_p}$. The mean phase of the pump in this state is zero and

$$(\Delta\phi_p)^2 = \langle\phi_p^2\rangle = \frac{1}{4N_p}. \tag{7.56}$$

If the pump phase were exactly zero, we see from Eq. (7.54) that $\Delta X_{2a}$ would be perfectly squeezed. The pump phase fluctuations, however, cause the squeezing direction to fluctuate with the consequence that some of the amplified noise is mixed into $\Delta X_{2a}$. In order to find out how much we use the fact that the phase fluctuations are small to replace, in Eq. (7.54), $\cos^2(\phi_p)$ by 1 and $\sin^2(\phi_p)$ by $\phi_p^2$. We then use Eq. (7.56) to average over the phase noise in the pump to obtain

$$[\Delta X_{2a}(u)]^2 \cong \frac{1}{4}e^{-2u} + \frac{1}{64N_p}e^{2u}. \tag{7.57}$$

As a function of $u$, $\Delta X_{2a}$, first decreases and then increases. Its minimum value is (corresponding to maximum squeezing)

$$(\Delta X_{2a})^2 \cong \frac{1}{8\sqrt{N_p}}, \tag{7.58}$$

which occurs when $u \cong (1/4)\ln(16N_p)$. This tells us that the maximum squeezing (minimum value of $(\Delta X_{2a})^2$) scales as the square root of the number of photons in the pump.

The result that comes from calculating corrections to the parametric approximation is quite similar. Keeping dominant terms up to order $1/N_p^2$ we find [30],

$$[\Delta X_{2a}(u)]^2 \cong \frac{1}{4}e^{-2u} + \frac{1}{64N_p}e^{2u} - \frac{3}{1024N_p^2}e^{4u}, \tag{7.59}$$

which gives a result essentially identical to that of Eq. (7.58) for the maximum squeezing.

One would like to know for how long the parametric approximation, and corrections to it, provide accurate results. For example, as was pointed out by Kinsler, Fernee, and Drummond, at the point of maximum squeezing the first two terms of Eq. (7.59) are of similar size, i. e. the first order correction is comparable to the zeroth order term [33]. This leads one to question the validity of the approximation scheme for times of this order. On the other hand, the next term in the series, the order $1/N_p^2$ term, is much smaller than the first two terms at the maximum squeezing point which suggests that the approximation is, in fact, still accurate there. The latter conclusion is supported by the numerical work of Kinsler, Fernee, and Drummond who found the behavior of the squeezed quadrature by using stochastic simulations to solve a set of stochastic differential equations that describe the dynamics of the degenerate parametric amplifier [33]. It is further supported by improvements to the dominant term method, and careful checks on the results it provides, by Cohen and Braunstein [31].

The picture that emerges from these calculations is that phase fluctuations in the pump limit the amount of squeezing that a parametric or degenerate parametric amplifier can produce. The parametric approximation, which does not take these fluctuations into account, does not predict a limit on squeezing. Corrections to it, however, do contain the effects of the quantum fluctuations of the pump and can be used to find how much squeezing is possible.

So far we have presented the results of the calculations that found corrections to the parametric approximation, and now we would like to briefly discuss two of the methods that have been employed. The first is the path-integral approach due to Hillery and Zubairy, and the second is the dominant term method of Crouch and Braunstein. We shall describe how each can be used to treat the degenerate parametric amplifier.

In the path-integral approach the basic object is the coherent-state propagator which is given by

$$K(\alpha_f, \beta_f, t_f; \alpha_i, \beta_i, t_i) = \langle \alpha_f, \beta_f | e^{-i(t_f - t_i)H/\hbar} | \alpha_i, \beta_i \rangle, \tag{7.60}$$

where the initial and final states are products of signal ($\alpha$) and pump ($\beta$) coherent states, and the Hamiltonian is

$$H = \hbar\omega a^\dagger a + 2\hbar\omega b^\dagger b + \hbar\kappa[b(a^\dagger)^2 + b^\dagger a^2]. \tag{7.61}$$

The propagator gives the amplitude for the system to go from the two-mode coherent state $|\alpha_i, \beta_i\rangle$ at time $t_i$ to another $|\alpha_f, \beta_f\rangle$ at time $t_f$. Knowledge of the propagator and the state of the field at time $t_i$ allows one to find the properties of the field at time $t_f$. The propagator can be expressed as a path integral (where we have set $t_i = 0$ and $t_f = t$)

$$K(\alpha_f, \beta_f, t; \alpha_i, \beta_i, 0) = \int D[\alpha(\tau)] \int D[\beta(\tau)] e^{iS/\hbar}, \tag{7.62}$$

where the action, $S$, is given by

$$iS/\hbar = \int_0^t d\tau [\frac{1}{2}(\dot{\alpha}^*\alpha - \alpha^*\dot{\alpha}) + \frac{1}{2}(\dot{\beta}^*\beta - \beta^*\dot{\beta}) - (i/\hbar)H(\alpha,\alpha^*,\beta,\beta^*)]. \tag{7.63}$$

The paths $\alpha(\tau)$ and $\beta(\tau)$ are such that $\alpha(0) = \alpha_i$, $\alpha(t) = \alpha_f$, $\beta(0) = \beta_i$, and $\beta(t) = \beta_f$, and $H(\alpha,\alpha^*,\beta,\beta^*)$ is the Hamiltonian with $a$ replaced by $\alpha$, $a^\dagger$ replaced by $\alpha^*$, etc.

In the case of the degenerate parametric amplifier it is possible to do the $\beta$ integral with the result that

$$\begin{aligned}K(\alpha_f,\beta_f,t;\alpha_i,\beta_i,0) &= \exp[-\frac{1}{2}(|\beta_f|^2 + |\beta_i|^2) + \beta_f^*\beta_i e^{-2i\omega t}] \\ &\quad \int D[\alpha(\tau)]e^{i(S_0+S_1)/\hbar},\end{aligned} \tag{7.64}$$

where

$$\left(\frac{i}{\hbar}\right)S_0 = \int_0^t d\tau [\frac{1}{2}(\dot{\alpha}^*\alpha - \alpha^*\dot{\alpha}) - i\omega|\alpha|^2 \\ - i\kappa[(\beta_f^* e^{-2i\omega t})e^{2i\omega\tau}\alpha^2 + \beta_i e^{-2i\omega\tau}(\alpha^*)^2]] \tag{7.65}$$

$$\left(\frac{i}{\hbar}\right)S_1 = -\kappa^2 \int_0^t d\tau_2 \int_0^{\tau_2} d\tau_1 e^{-2i\omega(\tau_2-\tau_1)}[\alpha^*(\tau_2)\alpha(\tau_1)]^2. \tag{7.66}$$

The action $S_0$ is similar to the action one would get from making the parametric approximation. In particular, if we set $\beta_f = \beta_i e^{-2i\omega t}$ in Eq, (19), then $S_0$ is the action for the Hamiltonian

$$H_p = \hbar\omega a^\dagger a + \hbar\kappa[\beta_i^* e^{2i\omega t}a^2 + \beta_i e^{-2i\omega t}(a^\dagger)^2]. \tag{7.67}$$

Note that the exponential factor in Eq. (17) falls off rapidly as $\beta_f$ deviates from $\beta_i e^{-2i\omega t}$. Therefore, as our lowest order of approximation, we can neglect $S_1$ (it is second order in the coupling constant and does not contain the strong pump field) and set $\beta_f = \beta_i e^{-2i\omega t}$ in $S_0$. The result is the parametric approximation.

We can calculate corrections to this approximation by doing two things. First, we include the effects of $S_1$ to lowest order. This is done by replacing $e^{iS_1/\hbar}$ inside the path integral by $1 + (i/\hbar)S_1$ and performing the resulting integral. Second, it is necessary to go beyond setting $\beta_f = \beta_i e^{-2i\omega t}$ and to take the dependence of $K$ on $\beta_f$ into account. The result is a propagator which goes beyond the parametric approximation and can be used to find the squeezed field variance.

In the path-integral approach what makes the parametric approximation possible is the fact that the path integral over the pump can be done. What, in turn, makes this possible is the form of the interaction; it is linear in the pump field. One might wonder whether we could devise an approximation based on replacing the signal mode by a classical field. Something like this might emerge if we could do the $\alpha$ path integral first. It is, however, not possible to perform this integral exactly. Consequently, the path integral formulation strongly suggests that the parametric approximation (replacing operators with c-numbers) is only applicable to fields which appear linearly in the interaction.

The dominant term method is based on the Heisenberg equations of motion for the pump and signal fields [30]. We begin by defining the pump mode quarature operators (the signal mode operators were defined in Eq. (7.53)

$$P_1 = \frac{1}{2}(b^\dagger + b) \qquad P_2 = \frac{i}{2}(b^\dagger - b), \tag{7.68}$$

where $\beta_0$, which we are assuming to be real, is the initial pump amplitude. The equations of motion for the quadrature operators which follow from the Hamiltonian in Eq. (7.49), when written in integral form, are

$$\begin{aligned}
X_{1a}(u) &= e^u X_{1a}(0) + \frac{e^u}{\beta_0} \int_0^u dv\, e^{-v}[X_{1a}(v)P_1(v) + X_{2a}(v)P_2(v)] \\
X_{2a} &= e^{-u} X_{2a}(0) + \frac{e^{-u}}{\beta_0} \int_0^u dv\, e^v [X_{1a}(v)P_2(v) - X_{2a}(v)P_1(v)] \\
P_1(u) &= P_1(0) - \frac{1}{2\beta_0} \int_0^u dv\, [X_{1a}(v)^2 - X_{2a}(v)^2] \\
P_2(u) &= P_2(0) - \frac{1}{2\beta_0} \int_0^u dv\, [X_{1a}(v)X_{2a}(v) + X_{2a}(v)X_{1a}(v)].
\end{aligned} \tag{7.69}$$

In order to find a solution these equations can be iterated. The zeroth order solution for each operator is just the first term on the right-hand side of its corresponding equation, e. g. for $X_{1a}(u)$ it is $e^u X_{1a}(0)$. The first order solution is found by inserting the zeroth order solution into the integrals in the above equations. This procedure can be repeated any number of times and it yields the solution as a power series in $1/\beta_0$. However, the number of terms grows rapidly at each step; the calculation of the order $1/\beta_0^4$ correction requires about 1000 terms. Many of these terms will be small for $u \gg 1$, for example, they may go as $e^{-u}$. If we keep only the terms which are largest for $u \gg 1$, the dominant terms, the solution simplifies drastically. Because the dominant terms of order $n$ produce the dominant terms of order $n+1$, all nondominant terms can be neglected throughout the iteration procedure. Quadrature variances can then be computed by taking expectation values of the resulting operator expressions.

### 7.4  Higher order processes: N-photon down conversion

The degenerate parametric amplifier is an example of a down conversion process. If the initial intensity of the pump mode, at frequency $2\omega$, is large and the signal mode at frequency $\omega$ is in the vacuum state, then at later times the intensity of the signal will have increased at the expense of the pump. Light at frequency $2\omega$ is thereby converted into light at frequency $\omega$.

Higher-order nonlinearities can be employed to generalize this process, though the strength of the interaction decreases as the order of the nonlinearity increases. In particular, we consider the process in which a pump photon at frequency $n\omega$ produces $n$ signal photons at frequency $\omega$. The Hamiltonian describing this interaction is

$$H_n = \hbar\omega a^\dagger a + n\hbar\omega b^\dagger b + \hbar\kappa_n[(a^\dagger)^n b + a^n b^\dagger]. \tag{7.70}$$

If we assume the $b$ mode is in a large amplitude coherent state with amplitude $\beta$ (we shall assume $\beta$ is real for simplicity) and then make the parametric approximation, the Hamiltonian in the interaction picture becomes

$$H_{pn} = \hbar \kappa_n \beta [(a^\dagger)^n + a^n]. \tag{7.71}$$

This last Hamiltonian has led to a controversy over whether it produces a well-defined time development transformation for $n > 2$. This question was first raised by Fisher, Nieto, and Sandberg [37], who noted that if the operator

$$U_{pn}(t) = e^{-itH_{pn}/\hbar} \tag{7.72}$$

is defined through a power series expansion of the exponential, the power series for the matrix element $\langle 0|U_{pn}(t)|0\rangle$ does not converge for $n > 2$. From this they concluded that $U_{pn}(t)$ does not exist for $n > 2$.

Their conclusion was disputed by Braunstein and McLachlan [38]. They pointed out that the fact that the power series expansion of $U_{pn}(t)$ is not well-defined does not imply that the operator does not exist; there are other ways to define the exponential of a self-adjoint operator, for example by means of its spectral representation. By using Padé approximants Braunstein and McLachlan were able to obtain numerically the Q functions for the states $U_{p3}(t)|0\rangle$ and $U_{p4}(t)|0\rangle$ for a limited range of time. The Q function for $U_{p3}(t)|0\rangle$ has a 3-pronged structure and is invariant under rotations by $2\pi/3$, while that of $U_{p4}(t)|0\rangle$ has 4 prongs and is invariant under rotations by $\pi/2$.

Additional problems with the operators $U_{pn}(t)$ for $n > 2$ have surfaced. If one assumes that they do exist one finds that they have a rather unpleasant property: if one starts in the vacuum state the photon number will become infinite in a finite time [39]. This is a consequence of the neglect of pump depletion in the parametric approximation. In the cases $n = 1$ and $n = 2$ a photon number divergence occurs, but only as $t \to \infty$. The existence of a divergence at finite time for $n > 2$ at the very least places strong restrictions on the time intervals for which $U_{pn}(t)$ can be used.

All of these problems disappear if one returns to the two-mode Hamiltonians $H_n$. There are no number divergences at all, because pump depletion is taken into account. In addition it can be proved that the operators $U_n = \exp(-itH_n/\hbar)$ exist and produce a well-defined dynamics for any $n$. These conclusions strongly suggest that when studying $n$-photon down conversion for $n > 2$, it is better to use the two-mode Hamiltonian $H_n$ than the single-mode Hamiltonian $H_{pn}$.

## 7.5 The parametric oscillator

Nonlinear effects can be considerably enhanced if the nonlinear medium sits inside of a cavity with reflecting, or partially reflecting, walls. The field inside the cavity is usually driven by a strong, external field coming into the the cavity through a partially reflecting mirror. This also means that the field can leak out of the cavity, so that the internal cavity field is damped. While the effect of the external field can be incorporated by adding a term to the effective Hamiltonian, the existence of damping requires us to adopt methods developed to describe open systems. Here we shall make use of an operator master equation. We shall present an extended example, which is important in its own right, and also serves to illustrate some of the techniques that are used to solve problems of this type. A much more detailed presentation of methods to analyze nonlinear optical systems in cavities can be found in the textbook by Milburn and Walls [42].

As an example we shall consider the parametric oscillator, which was first analyzed by Drummond, Mc Neil and Walls [40]. Here we shall follow a treatment due to Lugiato and Strini [41]. We begin by considering the interaction picture Hamiltonian

$$H_{int} = i\hbar\frac{\kappa}{2}(b(a^\dagger)^2 - b^\dagger a^2) + i\hbar E_0(b^\dagger - b), \tag{7.73}$$

where the first term describes the interaction with the nonlinear medium and the second term is what describes the effect of the driving field, $E_0$. Note that only the pump mode is driven. The dynamics of the two-mode system is give by the master equation for the two-mode density matrix, $\rho$,

$$\frac{d\rho}{dt} = \frac{-i}{\hbar}[H_{int}, \rho] + \Lambda_a(\rho) + \Lambda_b(\rho), \tag{7.74}$$

where the superoperators $\Lambda_a$ and $\Lambda_b$ describe the damping of the modes due to cavity losses. Their action is given by

$$\Lambda_a(\rho) = \gamma_a(2a\rho a^\dagger - a^\dagger a \rho - \rho a^\dagger a), \tag{7.75}$$

and similarly for $\Lambda_b$. The constants $\gamma_a$ and $\gamma_b$ are the damping rates for modes $a$ and $b$, respectively. The equations of motion for $\langle a \rangle$ and $\langle b \rangle$ that result from this master equation are

$$\begin{aligned}
\frac{d}{dt}\langle a\rangle &= \kappa\langle a^\dagger b\rangle - \gamma_a\langle a\rangle \\
\frac{d}{dt}\langle b\rangle &= -\frac{\kappa}{2}\langle a^2\rangle + E_0 - \gamma_b\langle b\rangle.
\end{aligned} \tag{7.76}$$

We cannot solve the above equations, because they couple the expectation values of $a$ and $b$ to expectation values of products and powers of these operators. We could find equations of motion for these additional expectation values, but we would find that they are coupled to yet more new expectation values. The ultimate result of this process is an infinite set of coupled equations. What we do instead is to linearize about the classical solution. We expect that the dominant part of the solution will be the classical part, and that the quantum fluctuations about the classical solution will be small. This will be true as long as we are not too close to threshold (we shall see what we mean by threshold shortly). We can then solve for the quantum fluctuations by keeping only the leading order terms in the equations.

The first step is to find the classical solutions. To do so, we assume that the expectation values factorize, e.g. $\langle a^\dagger b\rangle = \langle a^\dagger\rangle\langle b\rangle$. In addition, we are interested in the steady-state classical solution, so we set the time derivatives equal to zero. Denoting the steady-state values of $\langle a\rangle$ and $\langle b\rangle$ by $\alpha_0$ and $\beta_0$ respectively, we find that they satisfy the equations

$$\begin{aligned}
0 &= \kappa\alpha_0^*\beta_0 - \gamma_a\alpha_0 \\
0 &= -\frac{\kappa}{2}\alpha_0^2 + E_0 - \gamma_b\beta_0.
\end{aligned} \tag{7.77}$$

There are two types of solutions to this equation. The first is given by

$$\alpha_0 = 0 \qquad \beta_0 = \frac{E_0}{\gamma_b}. \tag{7.78}$$

In this case the amplitude of the signal-mode field is zero, and this situation is known as below threshold. The other solution is given by

$$\alpha_0 = \pm \frac{\sqrt{2}}{\kappa}(E_0\kappa - \gamma_a\gamma_b)^{1/2} \qquad \beta_0 = \frac{\gamma_a}{\kappa}. \tag{7.79}$$

This is the above-threshold solution, and in this case the signal mode has a nonzero amplitude, and it can have one of two possible phases, $0$ or $\pi$. In this solution, $\alpha_0$ is required to be real (if it is not, the above expression is not a solution to the equations), so it only exists if $E_0\kappa \geq \gamma_a\gamma_b$.

We now have the following situation. Below threshold, $E_0\kappa < \gamma_a\gamma_b$, we have only one solution, but above threshold, $E_0\kappa \geq \gamma_a\gamma_b$ we have two types of solution, one with $\alpha_0 = 0$ and one with $\alpha_0 \neq 0$. In order to determine which solution is valid above threshold, we look at the stability of the classical solutions. A physical solution must be stable against small perturbations, that is, if the system is in a state given by a stable solution, when we give it a slight kick, it will return to the state given by the stable solution. If the solution is unstable, any small perturbation will cause the system to deviate from the state described by that solution. Because there are always fluctuations present, the only important solutions are the stable ones; the unstable ones will not be seen.

In order to determine when our solutions are stable, we take the differential equations for the classical solutions

$$\begin{aligned} \frac{d}{dt}\alpha &= \kappa\alpha^*\beta - \gamma_a\alpha \\ \frac{d}{dt}\beta &= -\frac{\kappa}{2}\alpha^2 + E_0 - \gamma_b\beta, \end{aligned} \tag{7.80}$$

and their complex conjugates, set $\alpha = \alpha_0 + \delta\alpha$ and $\beta = \beta_0 + \delta\beta$, and neglect terms that are quadratic or higher in $\delta\alpha$ or $\delta\beta$. We find that

$$\frac{d}{dt}\begin{pmatrix} \delta\alpha \\ \delta\alpha^* \\ \delta\beta \\ \delta\beta^* \end{pmatrix} = \begin{pmatrix} -\gamma_a & \kappa\beta_0 & \kappa\alpha_0^* & 0 \\ \kappa\beta_0^* & -\gamma_a & 0 & \kappa\alpha_0 \\ -\kappa\alpha_0 & 0 & -\gamma_b & 0 \\ 0 & -\kappa\alpha_0^* & 0 & -\gamma_b \end{pmatrix}\begin{pmatrix} \delta\alpha \\ \delta\alpha^* \\ \delta\beta \\ \delta\beta^* \end{pmatrix}. \tag{7.81}$$

Let us call the matrix on the right-hand side of the equation $M$, and note that it depends on the classical solution whose stability we are studying. If the real parts of the eigenvalues of $M$ are negative, then initial fluctuations will decay with time, so the state of the system will return to its steady-state value. This indicates that the steady-state solution is stable. However, if any of the eigenvalues of $M$ has a positive real part, then the state of the system will deviate more and more from its steady-state value, and, consequently, the steady-state solution is unstable.

We now must determine the eigenvalues of $M$. The characteristic equation of $M$ can be expressed as

$$\begin{aligned} 0 &= [(\lambda+\gamma_a)(\lambda+\gamma_b) + \kappa^2|\alpha_0|^2 + \kappa|\beta_0|(\lambda+\gamma_b)] \\ &\quad [(\lambda+\gamma_a)(\lambda+\gamma_b) + \kappa^2|\alpha_0|^2 - \kappa|\beta_0|(\lambda+\gamma_b)], \end{aligned} \tag{7.82}$$

where $\lambda$ is an eigenvalue of $M$. This gives us two quadratic equations that determine the eigenvalues of $M$. For the $\alpha_0 = 0$ solution, we find

$$\lambda = -\gamma_b \qquad \lambda = -\gamma_a \pm \frac{\kappa E_0}{\gamma_b}. \tag{7.83}$$

Note that the eigenvalues will all be negative as long as $\gamma_a \gamma_b > \kappa E_0$. Therefore, this solution is stable only below threshold. For both of the $\alpha_0 \neq 0$ solutions, we find

$$\begin{aligned}
\lambda &= \frac{1}{2}\{-(2\gamma_a + \gamma_b) \pm [(2\gamma_a + \gamma_b)^2 - 8E_0\kappa]^{1/2}\} \\
\lambda &= \frac{1}{2}\{-\gamma_b \pm [\gamma_b^2 - 8(\kappa E_0 - \gamma_a \gamma_b)]^{1/2}\}.
\end{aligned} \quad (7.84)$$

All of these eigenvalues have negative real parts as long as $\gamma_a \gamma_b < \kappa E_0$, so that both of the $\alpha_0 \neq 0$ solutions are stable above threshold. Summarizing, we find that below threshold, we have only the $\alpha_0 = 0$ solution, which is stable, and above threshold, only the $\alpha_0 \neq 0$ solutions are stable.

We would now like to study the squeezing of the field below threshold, which means we need to find the fluctuations of the signal mode field about its steady state value. In order to do this, we first define fluctuation operators, $\Delta a = a - \alpha_0$ and $\Delta b = b - \beta_0$. We next find differential equations for expectation values of products of these operators and their adjoints, and, in doing so, we neglect terms of higher than second order in the fluctuation operators. We can do this, because we are assuming that the fluctuations are small, so that the effect of the terms we drop should be small as well. This assumption is a reasonable one as long as we are not too close to threshold. From the master equation for the density matrix, and keeping only quadratic terms in the fluctuations, we find that

$$\begin{aligned}
\frac{d}{dt}\langle \Delta a^\dagger \Delta a \rangle &= 2\kappa(\alpha_0^* \langle \Delta a^\dagger \delta b \rangle + \alpha_0 \langle \delta b^\dagger \Delta a \rangle) \\
&\quad + \kappa(\beta_0 \langle (\delta a^\dagger)^2 \rangle + \beta_0^* \langle (\Delta a)^2 \rangle) - 2\gamma_a \langle \Delta a^\dagger \Delta a \rangle \\
\frac{d}{dt}\langle (\Delta a)^2 \rangle &= 2\kappa(\alpha_0^* \langle \Delta a \Delta b \rangle + \alpha_0 \langle \Delta a^\dagger \Delta b \rangle \\
&\quad + \beta_0 \langle \Delta a^\dagger \Delta a \rangle) - 2\gamma_a \langle (\Delta a)^2 \rangle + \kappa \beta_0.
\end{aligned} \quad (7.85)$$

We are interested in the steady state situation, so the fluctuations will be time-independent. In addition, we are considering the oscillator below threshold. We therefore, set the time derivatives equal to zero, and substitute the below threshold values for $\alpha_0$ and $\beta_0$. This gives us

$$\begin{aligned}
0 &= \frac{\kappa E_0}{\gamma_b}(\langle (\Delta a^\dagger)^2 \rangle + \langle (\Delta a)^2 \rangle) - 2\gamma_a \langle \Delta a^\dagger \Delta a \rangle \\
0 &= \frac{2\kappa E_0}{\gamma_b}\langle \Delta a^\dagger \Delta a \rangle - 2\gamma_a \langle (\Delta a)^2 \rangle + \frac{\kappa E_0}{\gamma_b}.
\end{aligned} \quad (7.86)$$

These equations can be solved for $\langle \Delta a^\dagger \Delta a \rangle$ and $\langle (\Delta a)^2 \rangle$ yielding

$$\begin{aligned}
\langle \Delta a^\dagger \Delta a \rangle &= \frac{(\kappa E_0)^2}{2[(\gamma_a \gamma_b)^2 - (\kappa E_0)^2]} \\
\langle (\Delta a)^2 \rangle &= \frac{\gamma_a \gamma_b \kappa E_0}{2[(\gamma_a \gamma_b)^2 - (\kappa E_0)^2]}.
\end{aligned} \quad (7.87)$$

Note that these fluctuations are small unless we are close to threshold.

We can now compute the squeezing. Because $\langle a \rangle = 0$ below threshold, we have that

$$(\Delta X_2)^2 = \frac{1}{4}(1 + 2\langle \Delta a^\dagger \Delta a \rangle - \langle (\Delta a)^2 \rangle - \langle (\Delta a^\dagger)^2 \rangle). \quad (7.88)$$

Substituting in the values we found in the previous paragraph we have that

$$(\Delta X_2)^2 = \frac{1}{4} \frac{\gamma_a \gamma_b}{\gamma_a \gamma_b + \kappa E_0}. \qquad (7.89)$$

From this we can conclude that the signal mode is squeezed as long as the driving field is greater than zero. The maximum value of squeezing that can be obtained is by a factor of two, i.e. $(\Delta X_2)^2 \to 1/2$ as threshold is approached. It was originally thought that this would severely limit the squeezing that could be obtained from a degenerate parametric oscillator, but it was realized by Yurke [43] that the squeezing outside the cavity can be much larger than the squeezing inside the cavity. Because it is the squeezing outside the cavity that is important, the degenerate parametric oscillator is, in fact, a very good source of highly squeezed light.

## 8 Quantizing the field in a nonlinear dielectric

After our quick tour of some of the quantum effects we can expect to see in fields emerging from nonlinear media, let us now go back and discuss the quantization of electrodynamics is such media. To do so, we will follow the program that has been laid out earlier in the paper. We will first choose the fundamental field of our theory. This is usually the vector and scalar potentials, but we shall find it convenient to make another choice. We then find a Lagrangian that gives us the equations of motion, which in this case are Maxwell's equations. From the Lagrangian, we first find the canonical momentum, and then find the Hamiltonian. Finally, we impose the canonical commutation relations on the fundamental field and the canonical momentum.

Let us now carry out the steps sketched out in the previous paragraph in detail [44]. The equations of motion for our theory are

$$\nabla \cdot \mathbf{D} = 0 \qquad \nabla \times \mathbf{E} = -\frac{\partial \mathbf{B}}{\partial t}$$
$$\nabla \cdot \mathbf{B} = 0 \qquad \nabla \times \mathbf{B} = \mu_0 \frac{\partial \mathbf{D}}{\partial t}, \tag{8.1}$$

in the absence of external charges and currents. Here $\mathbf{D} = \epsilon_0 \mathbf{E} + \mathbf{P}$ is the displacement field and the polarization $\mathbf{P}$ is given by

$$\mathbf{P} = \epsilon_0 \left[ \chi^{(1)} : \mathbf{E} + \chi^{(2)} : \mathbf{EE} + \chi^{(3)} : \mathbf{EEE} + \ldots \right]. \tag{8.2}$$

We shall assume that the medium is lossless, and nondispersive, but it may be inhomogeneous, i.e. the susceptibilities can be functions of position. We want to find a Langrangian which has Eqs. (8.1) as its equations of motion. Before doing so we need to choose a particular field which is to be the basic dynamical variable in the problem. There are two possibilities. The first is the usual vector potential $A = (A_0, \mathbf{A})$ where

$$\mathbf{E} = -\frac{\partial \mathbf{A}}{\partial t} - \nabla A_0 \qquad \mathbf{B} = \nabla \times \mathbf{A}, \tag{8.3}$$

and the second is the dual potential $\Lambda = (\Lambda_0, \mathbf{\Lambda})$ where

$$\mathbf{B} = \mu_0 \left[ \frac{\partial \mathbf{\Lambda}}{\partial t} + \nabla \Lambda_0 \right] \qquad \mathbf{D} = \nabla \times \mathbf{\Lambda}. \tag{8.4}$$

This potential can only be used if external charges and currents are absent. When this is the case, the fact that $\nabla \cdot \mathbf{D} = 0$ implies that $\mathbf{D}$ can be expressed as the curl of some vector field, and that vector field we call the dual potential, $\mathbf{\Lambda}$. We shall explore quantizing the field with both the standard vector potential and with the dual potential. As we shall see, using the dual potential is much simpler, and it is this method that will be used throughout most of the rest of this paper.

### 8.1 Quantization with the standard vector potential

We will now quantize the field in a homogeneous nonlinear dielectric using the standard vector potential. As we will see shortly, if we use the dual potential we can drop the requirement of homogeneity. An appropriate Lagrangian density for the theory is

$$\begin{aligned}\mathcal{L}(A,\dot{A}) &= \epsilon_0[\frac{1}{2}(\mathbf{E}^2 - c^2\mathbf{B}^2) + \frac{1}{2}\chi^{(1)}_{ij}E_iE_j + \frac{1}{3}\chi^{(2)}_{ijk}E_iE_jE_k \\ &+ \frac{1}{4}\chi^{(3)}_{ijkl}E_iE_jE_kE_l + \ldots],\end{aligned} \quad (8.5)$$

where we are now using the convention that repeated indices are summed over. As in the free-field case, the Lagrange equations are

$$\partial_t\left(\frac{\partial \mathcal{L}}{\partial(\partial_t A_\mu)}\right) + \sum_{j=1}^{3}\partial_j\left(\frac{\partial \mathcal{L}}{\partial(\partial_j A_\mu)}\right) - \frac{\partial \mathcal{L}}{\partial A_\mu} = 0, \quad (8.6)$$

but now the Lagrangian density is different. This equation with $\mu = 0$ gives us $\nabla \cdot \mathbf{D} = 0$, and the three remaining equations give us $\nabla \times \mathbf{B} = \mu_0(\partial \mathbf{D}/\partial t)$. The remaining two Maxwell equations follow from the definition of electric and magnetic fields in terms of the vector potential.

We now want to proceed to the Hamiltonian formalism, and the first thing we need to do is to find the canonical momentum. From the above Lagrangian density, we find that the canonical momentum corresponding to $A$, which we denote by $\Pi = (\Pi_0, \mathbf{\Pi})$, is

$$\Pi_0 = \frac{\partial \mathcal{L}}{\partial(\partial_0 A_0)} = 0 \qquad \Pi_i = \frac{\partial \mathcal{L}}{\partial(\partial_0 A_i)} = -D_i. \quad (8.7)$$

Here we note two things. The first is that as in the case of a linear dielectric, the canonical momentum is different from that in the noninteracting theory where $\Pi_i = -E_i$. This is a consequence of the fact that the interaction depends on $\dot{\mathbf{A}} = \partial_t \mathbf{A}$. The second is that the vanishing of $\Pi_0$ implies that $A_0$ is not an independent field. In the case of free fields, if we choose the Coulomb gauge, it is also possible to choose $A_0 = 0$. This follows from the fact that for the free theory, $A_0 = 0$ and $\nabla \cdot \mathbf{E} = 0$ imply that the time derivative of $\nabla \cdot \mathbf{A}$ is zero, so that if $\nabla \cdot \mathbf{A} = 0$ initially, it will remain zero. Therefore, in this case the Coulomb and temporal ($A_0 = 0$) gauges are consistent. This is, however, no longer true when a nonlinear interaction is present, because now instead of $\nabla \cdot \mathbf{E} = 0$, we have $\nabla \cdot \mathbf{D} = 0$, so that if we choose $A_0 = 0$, the time derivative of $\nabla \cdot \mathbf{A}$ is no longer zero. If we choose the Coulomb gauge, which is what we shall do, then $A_0$ must be determined by solving the equation

$$\nabla^2 A_0 = -\nabla \cdot \mathbf{E}, \quad (8.8)$$

where $\mathbf{E}$ will be expressed in terms of the canonical momentum, $-\mathbf{D}$. In order to facilitate this, we define the tensors $\beta^{(i)}$ by

$$E_i = \beta^{(1)}_{ij}D_j + \beta^{(2)}_{ijk}D_jD_k + \ldots. \quad (8.9)$$

These tensors can be expressed in terms of the susceptibility tensors

$$\begin{aligned}\beta^{(1)} &= [\epsilon_0(1+\chi^{(1)})]^{-1} \\ \beta^{(2)}_{imn} &= -\epsilon_0\beta^{(1)}_{ij}\beta^{(1)}_{km}\beta^{(1)}_{ln}\chi^{(2)}_{jkl}.\end{aligned} \quad (8.10)$$

This result is obtained by solving the equation

$$D_i = \epsilon_0[(I+\chi^{(1)})_{ij}E_j + \chi^{(2)}_{ijk}E_jE_k + \ldots] \quad (8.11)$$

perturbatively. One inserts the expansion for $\mathbf{E}$ in terms of $\mathbf{D}$ into the above equation, assuming that each term in the expansion for $\mathbf{E}$ is of lower order that the one that precedes it, i.e. $\beta^{(1)}$ is order $\chi^{(1)}$, $\beta^{(2)}$ is of order $\chi^{(2)}$, etc. One then equates terms of the same order. This procedure is discussed in more detail in the next subsection.

Another consequence of the fact that $A_0$ is not an independent field in the Hamiltonian formulation is that we lose Gauss' law as an equation of motion. However, the equation $\nabla \times \mathbf{B} = \mu_0 \dot{\mathbf{D}}$, which is a result of the theory, implies that $\nabla \cdot \mathbf{D}$ is time independent, and this allows us to impose Gauss' law as an initial condition.

For the Hamiltonian we have

$$\begin{aligned} H(\mathbf{A}, \mathbf{\Pi}) &= \epsilon_0 \int d^3 r [\frac{1}{2}(\mathbf{E}^2 + c^2 \mathbf{B}^2 + \chi^{(1)}_{ij} E_i E_j) + \frac{2}{3} \chi^{(2)}_{ijk} E_i E_j E_k \\ &+ \frac{3}{4} \chi^{(3)}_{ijkl} E_i E_j E_k E_l] + \int d^3 r \mathbf{D} \cdot \nabla A_0. \end{aligned} \quad (8.12)$$

Performing an integration by parts in the last term and using the initial condition $\nabla \cdot \mathbf{D}$ allows us to eliminate the last term. It is useful to express the Hamiltonian directly in terms of the canonical momenta, $D_i$. Making use of the tensors $\beta^{(j)}$ we find for the Hamiltonian

$$\begin{aligned} H(\mathbf{A}, \mathbf{\Pi}) &= \int d^3 r [\frac{1}{2\mu_0} \mathbf{B}^2 + \frac{1}{2} \beta^{(1)}_{ij} D_i D_j + \frac{1}{3} \beta^{(2)}_{ijk} D_i D_j D_k \\ &+ \frac{1}{4} \beta^{(3)}_{ijkl} D_i D_j D_k D_l]. \end{aligned} \quad (8.13)$$

The theory is quantized by imposing the equal-time commutation relations

$$[A_j(\mathbf{r}, t), \Pi_k(\mathbf{r}', t)] = i\hbar \delta^{(tr)}_{jk}(\mathbf{r} - \mathbf{r}'). \quad (8.14)$$

Here, as in standard QED, we use the transverse delta function in order to be consistent with both the Coulomb gauge condition, $\nabla \cdot \mathbf{A} = 0$, and Gauss' law, $\nabla \cdot \mathbf{D} = 0$. As in the case of free QED it is possible to perform a mode expansion for the field and to define creation and annihilation operators. In particular, for the mode with momentum $\mathbf{k}$ and polarization $\hat{\mathbf{e}}_\alpha(\mathbf{k})$ we have the annihilation operator

$$a_{\mathbf{k},\alpha}(t) = \frac{1}{\sqrt{\hbar V}} \int d^3 r e^{-i\mathbf{k}\cdot\mathbf{r}} \hat{\epsilon}_\lambda(\mathbf{k}) \cdot [\sqrt{\frac{\epsilon_0 \omega_k}{2}} \mathbf{A}(\mathbf{r}, t) - \frac{i}{\sqrt{2\epsilon_0 \omega_k}} \mathbf{D}(\mathbf{r}, t)]. \quad (8.15)$$

Note that because $a_{\mathbf{k},\alpha}$ depends on $\mathbf{D}$, and consequently contains matter degrees of freedom, it is not a pure photon operator. It represents a collective matter-field mode.

## 8.2 Dual potential quantization

As has been mentioned, using the dual potential as the basic field of the theory makes things much simpler, and we shall see that explicitly in this section. Recapitulating, the dual potential, $\Lambda = (\Lambda_0, \mathbf{\Lambda})$ is defined so that

$$\mathbf{B} = \mu_0 \left[ \frac{\partial \mathbf{\Lambda}}{\partial t} + \nabla \Lambda_0 \right] \qquad \mathbf{D} = \nabla \times \mathbf{\Lambda}. \quad (8.16)$$

Note that this definition of **D** and **B** in terms of $\boldsymbol{\Lambda}$ and $\Lambda_0$ already guarantees that two of Maxwell's equations are satisfied, i.e.

$$\nabla \cdot \mathbf{D} = 0 \qquad \nabla \times \mathbf{B} = \mu_0 \frac{\partial \mathbf{D}}{\partial t}. \tag{8.17}$$

Before proceeding, we note that the expression given for the polarization in terms of the electric field, Eq. (8.2), is no longer convenient. The relation between **E** and $(\Lambda_0, \boldsymbol{\Lambda})$ is complicated, while the relation between **D** and $\boldsymbol{\Lambda}$ is relatively simple. Therefore, it is better to express the polarization as an expansion in **D** rather than in **E**,

$$\mathbf{P} = \eta^{(1)} : \mathbf{D} + \eta^{(2)} : \mathbf{DD} + \eta^{(3)} : \mathbf{DDD} + \ldots. \tag{8.18}$$

It is possible to express the tensors $\eta^{(j)}$ in terms of the susceptibilities $\chi^{(j)}$. Let us do this for $\eta^{(1)}$ and $\eta^{(2)}$. Neglecting higher order terms, we have

$$\mathbf{P} = \epsilon_0 \left[ \frac{1}{\epsilon_0} \chi^{(1)} : (\mathbf{D} - \mathbf{P}) + \frac{1}{\epsilon_0^2} \chi^{(2)} : (\mathbf{D} - \mathbf{P})(\mathbf{D} - \mathbf{P}) \right]. \tag{8.19}$$

We will solve this equation for **P** in terms of **D** perturbativley, considering the term proportional to $\chi^{(2)}$ as the perturbation. This is justified, because the size of the nonlinear susceptibilities decreases significantly as their order increases. We find the lowest order solution by setting the $\chi^{(2)}$ term equal to zero, giving us

$$\mathbf{P} = (I + \chi^{(1)})^{-1} \chi^{(1)} : \mathbf{D}. \tag{8.20}$$

This implies that to lowest order

$$\mathbf{D} - \mathbf{P} = (I + \chi^{(1)})^{-1} : \mathbf{D}. \tag{8.21}$$

We now take the lowest order solution for $\mathbf{D} - \mathbf{P}$, insert it into the $\chi^{(2)}$ term in Eq. (8.19), and solve the resulting equation for **P**. We shall write the result in terms of components. Setting $\gamma = (I + \chi^{(1)})^{-1}$, we have

$$P_j = \gamma_{jk} \chi^{(1)}_{kl} D_l + \frac{1}{\epsilon_0} \gamma_{jk} \chi^{(2)}_{klm} \gamma_{ln} \gamma_{mp} D_n D_p. \tag{8.22}$$

Comparing this equation to Eq. (8.18), we see that

$$\eta^{(1)}_{jl} = \gamma_{jk} \chi^{(1)}_{kl} \qquad \eta^{(2)}_{jnp} = \frac{1}{\epsilon_0} \gamma_{jk} \chi^{(2)}_{klm} \gamma_{ln} \gamma_{mp}. \tag{8.23}$$

This perturbative approach can be applied to find expressions for the higher-order $\eta$ tensors. We also note that the tensors $\eta$ are related to the tensors $\beta$ by $\eta^{(1)} = I - \epsilon_0 \beta^{(1)}$ and $\eta^{(j)} = -\epsilon_0 \beta^{(j)}$ for $j \geq 2$.

We now need a Lagrangian, or actually a Lagrangian density, that gives us the remaining two Maxwell equations. If we assume that the tensors $\eta^{(j)}$ are symmetric, it is given by

$$\mathcal{L} = \frac{1}{2} \left( \frac{1}{\mu_0} \mathbf{B}^2 - \frac{1}{\epsilon_0} \mathbf{D}^2 \right) + \frac{1}{\epsilon_0} \left( \frac{1}{2} \mathbf{D} \cdot \eta^{(1)} : \mathbf{D} + \frac{1}{3} \mathbf{D} \cdot \eta^{(2)} : \mathbf{DD} + \ldots \right). \tag{8.24}$$

The equations of motion that come from this Lagrangian density are given by

$$\partial_t \left( \frac{\partial \mathcal{L}}{\partial(\partial_t \Lambda_\mu)} \right) + \sum_{j=1}^{3} \partial_j \left( \frac{\partial \mathcal{L}}{\partial(\partial_j \Lambda_\mu)} \right) - \frac{\partial \mathcal{L}}{\partial \Lambda_\mu} = 0, \tag{8.25}$$

where $\mu = 0, \ldots 3$. Setting $\mu = 0$ we obtain

$$\nabla \cdot \mathbf{B} = 0, \tag{8.26}$$

while the other three equations give us

$$\nabla \times \mathbf{E} = -\frac{\partial \mathbf{B}}{\partial t}. \tag{8.27}$$

Deriving the first of these equations is straightforward, but it is useful to fill in a few steps in the derivation of the second one. We first note that

$$\frac{\partial \mathcal{L}}{\partial(\partial_t \Lambda_k)} = B_k$$

$$\frac{\partial D_l}{\partial(\partial_j \Lambda_k)} = \epsilon_{ljk}, \tag{8.28}$$

where $k \in \{1, 2, 3\}$ and $\epsilon_{ljk}$ is the completely antisymmetric tensor of rank 3. Now, for simplicity, let us look at the case $\eta^{(j)} = 0$ for $j \geq 2$. Making use of the above relations we find that

$$\frac{\partial \mathbf{B}}{\partial t} + \frac{1}{\epsilon_0} \nabla \times (\mathbf{D} - \eta^{(1)} : \mathbf{D}) = 0 \tag{8.29}$$

For the case we are considering, this is just the last equation in the previous paragraph. Therefore, we now have a Lagrangian formulation of the theory.

The next step is to find the Hamiltonian formulation. The canonical momentum is given by

$$\Pi_0 = \frac{\partial \mathcal{L}}{\partial(\partial_t \Lambda_0)} = 0 \quad \Pi_j = \frac{\partial \mathcal{L}}{\partial(\partial_t \Lambda_j)} = B_j. \tag{8.30}$$

The Hamiltonian density is then

$$\begin{aligned}\mathcal{H} &= \sum_{j=1}^{3} \Pi_j(\partial_t \Lambda_j) - \mathcal{L} \\ &= \frac{1}{2}\left(\frac{1}{\mu_0}\mathbf{B}^2 + \frac{1}{\epsilon_0}\mathbf{D}^2\right) - \frac{1}{\epsilon_0}\left(\frac{1}{2}\mathbf{D}\cdot\eta^{(1)}:\mathbf{D} + \frac{1}{3}\mathbf{D}\cdot\eta^{(2)}:\mathbf{DD} + \ldots\right) \\ &\quad - B \cdot \nabla \Lambda_0.\end{aligned} \tag{8.31}$$

At this point, we notice that the Hamiltonian equations of motion,

$$\partial_t \Pi_j = -\frac{\delta H}{\delta \Lambda_k} \tag{8.32}$$

give us only three, instead of four, equations. Because of the vanishing of $\Pi_0$, we have lost $\nabla \cdot B$ as an equation of motion, and, in fact, the vanishing of $\Pi_0$ means that we are dealing with a constrained Hamiltonian system. Dirac developed a theory of the quantization of constrained Hamiltonians [45], but we will be able to proceed here by improvisation. We first note that taking the divergence of both sides of Eq. (8.27) gives us that

$$\frac{\partial}{\partial t}\nabla \cdot \mathbf{B} = 0 \tag{8.33}$$

so that if $\nabla \cdot \mathbf{B} = 0$ is true initially, it will remain true. Therefore, we can recover the equation $\nabla \cdot \mathbf{B} = 0$ if we impose it as an initial condition.

Before quantizing the theory we will fix the gauge. The physical fields are unchanged under the transformation

$$\mathbf{\Lambda} \to \mathbf{\Lambda} + \nabla\Theta \quad \Lambda_0 \to \Lambda_0 - \frac{\partial \Theta}{\partial t}, \tag{8.34}$$

where $\Theta(\mathbf{r}, t)$ is an arbitrary function of space and time. We can eliminate $\Lambda_0$ by choosing $\Theta$ to be a solution to

$$\frac{\partial \Theta}{\partial t} = \Lambda_0, \tag{8.35}$$

which determines $\Theta$ up to an arbitrary function of position, which we shall call $\theta(\mathbf{r})$. Since $\nabla \cdot \mathbf{\Lambda}$ is time independent,

$$\frac{\partial}{\partial t}(\nabla \cdot \mathbf{\Lambda}) = \nabla \cdot \mathbf{B} = 0, \tag{8.36}$$

we can choose $\theta$ so that $\nabla \cdot \mathbf{\Lambda} = 0$. The result is we have a radiation gauge for $\mathbf{\Lambda}$ in which $\Lambda_0 = 0$ and $\nabla \cdot \mathbf{\Lambda} = 0$. Note that this gauge choice eliminates the last term of the Hamiltonian density.

In order to quantize the theory we now impose the canonical equal-time commutation relations

$$[\Lambda_j(\mathbf{r}, t), B_k(\mathbf{r}', t)] = i\hbar \delta_{jk}^{(tr)}(\mathbf{r} - \mathbf{r}'), \tag{8.37}$$

where $\delta^{(tr)}$ is the transverse delta function. We have again used the transverse delta function, because this choice makes the above equation consistent with the fact that both $\nabla \cdot \mathbf{\Lambda} = 0$ and $\nabla \cdot \mathbf{B} = 0$.

It is often useful to express the field in terms of creation and annihilation operators for plane-wave modes. These have the mode functions (see Eq. (4.24))

$$\mathbf{u}_\alpha(\mathbf{k}) = \frac{1}{\sqrt{V}} \hat{\mathbf{e}}_{\mathbf{k},\alpha} e^{i\mathbf{k}\cdot\mathbf{r}}, \tag{8.38}$$

where $\alpha = 1, 2$. The annihilation operators are linear combinations of the fields $\mathbf{\Lambda}$ and $\mathbf{B}$. The exact linear combination will be chosen with two requirements in mind. First, we want to obtain the usual commutation relations between the creation and annihilation operators, so the linear combination should be chosen so that these commutation relations follow from the canonical commutation relations between the fields. If we define

$$a_{\mathbf{k},\alpha} = \int d^3 r \, \mathbf{u}_\alpha^*(\mathbf{k}) \cdot \left( c_\mathbf{k} \mathbf{\Lambda}(\mathbf{r}, t) + \frac{i}{2\hbar c_\mathbf{k}} \mathbf{B}(\mathbf{r}, t) \right), \tag{8.39}$$

where the real numbers $c_{\mathbf{k}}$ are, for the moment, arbitrary, then we indeed find that Eq. (8.37) implies that

$$[a_{\mathbf{k},\alpha}, a^\dagger_{\mathbf{k}',\alpha'}] = i\hbar \delta_{\mathbf{k},\mathbf{k}'}\delta_{\alpha,\alpha'}. \tag{8.40}$$

We can now take Eq. (8.39) and its adjoint, and solve for the fields in terms of the creation and annihilation operators. We find that

$$\begin{aligned}
\mathbf{\Lambda}(\mathbf{r},t) &= \sum_{\mathbf{k},\alpha} \frac{1}{2c_{\mathbf{k}}\sqrt{V}} \hat{\mathbf{e}}_{\mathbf{k},\alpha}(e^{i\mathbf{k}\cdot\mathbf{r}}a_{\mathbf{k},\alpha} + e^{-i\mathbf{k}\cdot\mathbf{r}}a^\dagger_{\mathbf{k},\alpha}) \\
\mathbf{B}(\mathbf{r},t) &= \sum_{\mathbf{k},\alpha} \frac{\hbar c_{\mathbf{k}}}{i\sqrt{V}} \hat{\mathbf{e}}_{\mathbf{k},\alpha}(e^{i\mathbf{k}\cdot\mathbf{r}}a_{\mathbf{k},\alpha} - e^{-i\mathbf{k}\cdot\mathbf{r}}a^\dagger_{\mathbf{k},\alpha}).
\end{aligned} \tag{8.41}$$

We can now substitute these expressions into the Hamiltonian. We choose the numbers $c_{\mathbf{k}}$ so that the free Hamiltonian

$$H_{free} = \frac{1}{2} \int d^3r \left( \frac{1}{\mu_0}\mathbf{B}^2 + \frac{1}{\epsilon_0}\mathbf{D}^2 \right), \tag{8.42}$$

has the form

$$H_{free} = \frac{1}{2} \sum_{\mathbf{k},\alpha} \hbar\omega_{\mathbf{k}}(a^\dagger_{\mathbf{k},\alpha}a_{\mathbf{k},\alpha} + a_{\mathbf{k},\alpha}a^\dagger_{\mathbf{k},\alpha}), \tag{8.43}$$

where $\omega_k = kc$. For a general choice of $c_{\mathbf{k}}$ there will also be terms of the form $a^\dagger_{\mathbf{k},\alpha}a^\dagger_{-\mathbf{k},\alpha}$ and $a_{\mathbf{k},\alpha}a_{-\mathbf{k},\alpha}$ present, but if we choose

$$c_{\mathbf{k}} = \left( \frac{\mu_0 \omega_{\mathbf{k}}}{2\hbar} \right)^{1/2}, \tag{8.44}$$

then all of these terms vanish. Summarizing, then, the fields $\mathbf{\Lambda}$ and $\mathbf{B}$ can be expressed in terms of creation and annihilation operators as

$$\begin{aligned}
\mathbf{\Lambda}(\mathbf{r},t) &= \sum_{\mathbf{k},\alpha} \left( \frac{\hbar}{2\mu_0\omega_{\mathbf{k}}V} \right)^{1/2} \hat{\mathbf{e}}_{\mathbf{k}\alpha}(e^{i\mathbf{k}\cdot\mathbf{r}}a_{\mathbf{k},\alpha} + e^{-i\mathbf{k}\cdot\mathbf{r}}a^\dagger_{\mathbf{k},\alpha}) \\
\mathbf{B}(\mathbf{r},t) &= \sum_{\mathbf{k},\alpha} i \left( \frac{\mu_0\hbar\omega_{\mathbf{k}}}{2V} \right)^{1/2} \hat{\mathbf{e}}_{\mathbf{k}\alpha}(e^{-i\mathbf{k}\cdot\mathbf{r}}a^\dagger_{\mathbf{k},\alpha} - e^{i\mathbf{k}\cdot\mathbf{r}}a_{\mathbf{k},\alpha}).
\end{aligned} \tag{8.45}$$

If we are describing the behavior of the electromagnetic field in a cavity or a waveguide, it is often useful to expand the field operators in terms of the mode functions of the relevant system. The mode functions are determined by the linear part of the polarization. For example, a waveguide can be constructed from a spatially varying dielectric and a cavity by surrounding a region of free space by dielectric sheets or slabs. Therefore, let us suppose that the linear behavior of our system is described by a linear susceptibility $\beta^{(1)}(\mathbf{r})$. Then the mode functions must obey the wave equation

$$\nabla \times [\beta^{(1)}(\mathbf{r})\nabla \times \mathbf{\Lambda}(\mathbf{r},t)] = -\ddot{\mathbf{\Lambda}}(\mathbf{r},t), \tag{8.46}$$

and the proper boundary conditions, which we ususally take to be either periodic or vanishing. The modes have well-defined frequencies, $\omega_n$, so that the mode functions are of the form $\mathbf{\Lambda}(\mathbf{r}, t) = \mathbf{\Lambda}_n(\mathbf{r})e^{-i\omega_n t}$, with the functions $\mathbf{\Lambda}_n(\mathbf{r})$ satisfying the equation

$$\nabla \times [\beta^{(1)}(\mathbf{r})\nabla \times \mathbf{\Lambda}_n(\mathbf{r})] = \omega_n^2 \mathbf{\Lambda}_n(\mathbf{r}). \tag{8.47}$$

Note that because the fields are transverse, we must have $\nabla \cdot \mathbf{\Lambda}_n = 0$. We can show that modes corresponding to different frequencies are orthogonal by taking the scalar product of both sides of this equation with $\mathbf{\Lambda}_{n'}^*$ and integrating over the quantization volume. Applying the vector identity $\nabla \cdot [\mathbf{A} \times \mathbf{A}'] = \mathbf{A}' \cdot [\nabla \times \mathbf{A}] - \mathbf{A} \cdot [\nabla \times \mathbf{A}']$ twice, and assuming vanishing boundary terms, gives us

$$(\omega_n^2 - \omega_{n'}^2) \int d^3r \mathbf{\Lambda}_{n'}^*(\mathbf{r}) \cdot \mathbf{\Lambda}_n(\mathbf{r}) = 0, \tag{8.48}$$

which immediately implies that if $\omega_n \neq \omega_{n'}$, then the integral must vanish. It should also be possible to choose different modes with the same frequency to be orthogonal, and we shall assume that this has been done. In particular, note that $\mathbf{\Lambda}_n$ and $\mathbf{\Lambda}_n^*$ are both solutions of Eq. (8.47). These can always be chosen so that they are orthogonal to each other. Therefore, for the mode functions $\mathbf{\Lambda}_n$ we have

$$\int d^3r \mathbf{\Lambda}_{n'}^*(\mathbf{r}) \cdot \mathbf{\Lambda}_n(\mathbf{r}) = \delta_{n,n'}, \tag{8.49}$$

and we shall assume that the modes form a complete set in the space of transverse functions.

We next define annihilation operators, $a_n$, by

$$a_n(t) = \int d^3r \mathbf{\Lambda}_n^*(\mathbf{r}) \cdot \left[ \sqrt{\frac{\omega_n}{2}} \mathbf{\Lambda}(\mathbf{r}, t) - \frac{i}{\sqrt{2\omega_n}} \dot{\mathbf{\Lambda}}(\mathbf{r}, t) \right]. \tag{8.50}$$

These obey the commutation relations

$$[a_n(t), a_{n'}^\dagger(t)] = \delta_{n,n'}. \tag{8.51}$$

We can also invert Eq. (8.50) to express the fields in terms of the creation and annihilation operators. Making use of Eq. (8.50) and its adjoint we find that

$$\sum_n \sqrt{\frac{2}{\omega_n}} [a_n(t)\mathbf{\Lambda}_n(\mathbf{r}) + a_n^\dagger(t)\mathbf{\Lambda}_n^*(\mathbf{r})] = 2\mathbf{\Lambda}(\mathbf{r}, t)$$

$$+ \int d^3r' \sum_n \frac{i}{\omega_n} [\mathbf{\Lambda}_n^*(\mathbf{r})\mathbf{\Lambda}_n(\mathbf{r'}) - \mathbf{\Lambda}_n(\mathbf{r})\mathbf{\Lambda}_n^*(\mathbf{r'})] \cdot \dot{\mathbf{\Lambda}}(\mathbf{r'}, t). \tag{8.52}$$

The second term on the right-hand side vanishes, because both $\mathbf{\Lambda}_n$ and $\mathbf{\Lambda}_n^*$ are mode functions. In particular, for each $n$ there is an $n'$ such that $\mathbf{\Lambda}_{n'} = \mathbf{\Lambda}_n^*$. This implies that

$$\mathbf{\Lambda}_{n'}(\mathbf{r})\mathbf{\Lambda}_{n'}^*(\mathbf{r'}) = \mathbf{\Lambda}_n^*(\mathbf{r})\mathbf{\Lambda}_n(\mathbf{r'}), \tag{8.53}$$

so that the first term in the brackets for $n$ is cancelled by the second term in the brackets for $n'$. This finally gives us our mode expansion for the field

$$\mathbf{\Lambda}(\mathbf{r}, t) = \sum_n \sqrt{\frac{1}{2\omega_n}} [a_n(t)\mathbf{\Lambda}_n(\mathbf{r}) + a_n^\dagger(t)\mathbf{\Lambda}_n^*(\mathbf{r})]. \tag{8.54}$$

The mode expansion of the **D** field can be obtained from this equation by taking the curl of both sides. Similar reasoning gives

$$\dot{\mathbf{\Lambda}}(\mathbf{r},t) = i\sum_n \sqrt{\frac{\omega_n}{2}}[a_n(t)\mathbf{\Lambda}_n(\mathbf{r}) - a_n^\dagger(t)\mathbf{\Lambda}_n^*(\mathbf{r})], \tag{8.55}$$

which immediately yields the mode expansion for the $B$ field. These expressions can be substituted into $H = \int d^3r \mathcal{H}$, were $\mathcal{H}$ is given by Eq. (8.31), which will give us the Hamiltonian in terms of the mode creation and annihilation operators.

## 9 Other approaches

The method used in the previous section to quantize the electromagnetic field is not the one most commonly used in quantum optics. In this section we will review the other methods and point out problems with them.

Perhaps the first treatment of a quantum theory of nonlinear optics was given in [46]. There an interaction Hamiltonian of the form

$$H_{int} = -\int d^3r \mathbf{E} \cdot \mathbf{P} \tag{9.1}$$

was used, where $\mathbf{P}$ was given by Eq. (2.3). The theory was quantized by substituting for the fields their free-field expressions in terms of creation and annihilation operators

$$\begin{aligned}
\mathbf{E}(\mathbf{r},t) &= i\sum_{\mathbf{k},\alpha}\left(\frac{\hbar\omega_k}{2\epsilon_0 V}\right)^{1/2}\hat{\mathbf{e}}_{\mathbf{k}\alpha}(e^{i\mathbf{k}\cdot\mathbf{r}}a_{\mathbf{k},\alpha} - e^{-i\mathbf{k}\cdot\mathbf{r}}a^\dagger_{\mathbf{k},\alpha}) \\
\mathbf{B}(\mathbf{r},t) &= i\sum_{\mathbf{k},\alpha}\left(\frac{\hbar}{2\epsilon_0\omega_k V}\right)^{1/2}(\mathbf{k}\times\hat{\mathbf{e}}_{\mathbf{k}\alpha})(e^{i\mathbf{k}\cdot\mathbf{r}}a_{\mathbf{k},\alpha} - e^{-i\mathbf{k}\cdot\mathbf{r}}a^\dagger_{\mathbf{k},\alpha}).
\end{aligned} \tag{9.2}$$

This theory runs into immediate problems. While the interaction between an object with a fixed electric dipole moment, $\mathbf{d}$, and the electromagnetic field is given by $-\mathbf{d}\cdot\mathbf{E}$, the situation changes if the dipole is induced as is the case for polarizable media. In fact, for a linear medium with a constant polarizability, the expression for the energy of the electromagnetic field in this medium is given by [47]

$$H = \frac{1}{2}\int d^3r\left(\mathbf{E}\cdot\mathbf{D} + \frac{1}{\mu_0}\mathbf{B}^2\right), \tag{9.3}$$

which is also the Hamiltonian. This implies that the interaction between the medium and the field is $(1/2)\int d^3r \mathbf{E}\cdot\mathbf{P}$ and not $-\int d^3r\mathbf{E}\cdot\mathbf{P}$. This is by no means the only problem, however, Note that the form assumed for the electric field operator implies that $\nabla\cdot\mathbf{E} = 0$. Inside of a polarizable medium, however, the proper Maxwell equation is $\nabla\cdot\mathbf{D} = 0$. Both conditions can only be true if $\mathbf{E}$ and $\mathbf{D}$ are proportional, and this will not be the case for a nonlinear medium. In short, the problem with this theory is that it does not give Maxwell's equations as its equations of motion. which a proper theory should do.

The most commonly used method for quantizing fields in nonlinear dielectrics is a modification of the one just described. As we have seen, the electromagnetic field in a uniform nonlinear dielectric can be quantized using the standard vector potential, and this results in the Hamiltonian

$$\begin{aligned}
H &= \epsilon_0\int d^3r\left[\frac{1}{2}(\mathbf{E}^2 + c^2\mathbf{B}^2 + \mathbf{E}\cdot\chi^{(1)}:\mathbf{E}) + \frac{2}{3}\mathbf{E}\cdot\chi^{(2)}:\mathbf{E}\mathbf{E}\right.\\
&\quad + \left.\frac{3}{4}\mathbf{E}\cdot\chi^{(1)}:\mathbf{E}\mathbf{E}\mathbf{E} + \ldots\right].
\end{aligned} \tag{9.4}$$

The theory is quantized by substituting into the Hamiltonian the expressions in Eq. (9.2), i.e. the free-field expressions for the electric and magnetic fields (see, for example, [6]). This also immediately leads to trouble. First, as before, we have $\nabla\cdot\mathbf{E} = 0$ instead of $\nabla\cdot\mathbf{D} = 0$. We also have incorrect commutation relations. As has been discussed, the canonical momentum corresponding to the above Hamiltonian is $-\mathbf{D}$, which means that in the quantized theory the commutation relations should be

$$[A_j(\mathbf{r},t), D_l(\mathbf{r}',t)] = -i\hbar \delta_{jl}^{(tr)}(\mathbf{r}-\mathbf{r}'), \tag{9.5}$$

whereas in the theory that is actually employed we have instead

$$[A_j(\mathbf{r},t), \epsilon_0 E_l(\mathbf{r}',t)] = -i\hbar \delta_{jl}^{(tr)}(\mathbf{r}-\mathbf{r}'). \tag{9.6}$$

Finally, this theory assumes that $A_0 = 0$, which it, in fact, does not. What this means is that this theory also does not have Maxwell's equations as its equations of motion.

Let us conclude by showing what equations of motion do, in fact, emerge from this incorrect theory. We begin by finding the commutation relations for the electric and magnetic fields that result from Eq. (9.2)

$$[E_j(\mathbf{r},t), E_l(\mathbf{r}',t)] = [B_j(\mathbf{r},t), B_l(\mathbf{r}',t)] = 0 \tag{9.7}$$

and

$$[E_j(\mathbf{r},t), B_l(\mathbf{r}',t)] = \frac{i\hbar}{\epsilon_0}\varepsilon_{lmn}\frac{\partial}{\partial r'_m}\delta_{jn}^{(tr)}(\mathbf{r}'-\mathbf{r}), \tag{9.8}$$

where $\varepsilon_{lmn}$ is the completely antisymmetric tensor of rank three. These can be used to find the Heisenberg equations of motion for the field operators. We have for the magnetic field

$$\frac{\partial B_j}{\partial t} = -\frac{i}{\hbar}[B_j, H]. \tag{9.9}$$

This gives us

$$\begin{aligned}\frac{\partial B_j}{\partial t} &= \frac{-1}{\hbar}\varepsilon_{jmn}\frac{\partial}{\partial r_m}\int d^3r'\{\delta_{an}^{(tr)}(\mathbf{r}-\mathbf{r}')E_a(\mathbf{r}',t) \\ &+ \delta_{an}^{(tr)}(\mathbf{r}-\mathbf{r}')\chi_{ab}^{(1)}E_b(\mathbf{r}',t) + \ldots\},\end{aligned} \tag{9.10}$$

where we have assumed that the susceptibility tensors are symmetric. We now want to evaluate the integral, but we have to take into account that $\chi^{(1)} : \mathbf{E}$, and the nonlinear terms as well, are not necessarily transverse. All of the terms on the right-hand side are of the form

$$\nabla \times \int d^3r' \delta_{an}^{(tr)}(\mathbf{r}-\mathbf{r}')V_a(\mathbf{r}',t), \tag{9.11}$$

where $\mathbf{V}$ is a vector field. This field can be split into transverse and longitudinal parts, $\mathbf{V} = \mathbf{V}^{(tr)} + \mathbf{V}^{(l)}$, where $\nabla \cdot \mathbf{V}^{(tr)} = 0$ and $\nabla \times \mathbf{V}^{(l)} = 0$. The above integral is just equal to $\mathbf{V}^{(tr)}$ so that the whole term is just $\nabla \times \mathbf{V}^{(tr)}$. However, this is also equal to $\nabla \times \mathbf{V}$. Therefore, in evaluating the integrals that arise in the commutator term, we can treat the transverse delta functions as regular three-dimensional delta functions. Doing so gives us

$$\frac{\partial \mathbf{B}}{\partial t} = -\nabla \times \mathbf{E} - \nabla \times (\chi^{(1)} : \mathbf{E} + 2\chi^{(2)} : \mathbf{EE} + \ldots). \tag{9.12}$$

This equation is, of course, not the correct equation for the time derivative of $\mathbf{B}$.

Pushing this a little further, we can also find the equation of motion for the electric field. We find that

$$\frac{\partial E_j}{\partial t} = \frac{-i}{\hbar}[E_j, H] = c^2(\nabla \times \mathbf{B})_j \tag{9.13}$$

This is the correct Maxwell equation for free fields, but not for a field in a dielectric. Finally, let us consider these equations in the situation in which the nonlinear susceptibilities are zero and the linear susceptibility is just a constant, $\chi^{(1)} > 0$. From the above equation, the one in the previous paragraph and the fact that in this theory $\nabla \cdot \mathbf{E} = 0$, we obtain

$$(1 + \chi^{(1)})\nabla^2 \mathbf{E} - \frac{1}{c^2}\frac{\partial \mathbf{E}}{\partial t} = 0. \tag{9.14}$$

This implies that the wave velocity in a dielectric is $\sqrt{1 + \chi^{(1)}}c$, which is greater than the speed of light in the vacuum. That is, rather than predict that light slows down as it goes into a dielectric, as is, of course, the case, this theory predicts that it speeds up. The lesson to be learned is that despite the fact that this theory is often employed, it should not be, because the predictions it makes are nonphysical.

## 10   Multimode treatment of parametric down conversion

As an application of our formalism, let us consider parametric down conversion. With the advent of quantum information, this is a process that is extensively employed and intensively studied. A multimode quantum treatment due to Ou, Wang, and Mandel has been a model for many later treatments [48], and we shall pattern our treatment after theirs. Their approach does employ the usual Hamiltonian for the down-conversion process, which, as has been noted, does not give Maxwell's equations as the equations for the field operators.

Down conversion takes place when a photon is incident on a material with a $\chi^{(2)}$ nonlinearity, and two photons, the sum of whose energies is equal to the energy of the incident photon, are produced. We shall assume that the medium is centered on the origin, and of extent $l_x$, $l_y$, and $l_z$ in the $x$, $y$, and $z$ directions, respectively. In addition, we shall assume the linear electric polarizability is zero, the magnetic susceptibility is zero, and that there are no free charges. We then find for the Hamiltonian

$$H = H_{free} + H_{int}, \tag{10.1}$$

where

$$\begin{aligned} H_{free} &= \int_V d^3r \frac{1}{2}(\mathbf{D}^2 + \mathbf{B}^2) \\ H_{int} &= \int_{V_m} d^3r \frac{1}{3}\mathbf{D} \cdot \eta^{(2)}\mathbf{DD}, \end{aligned} \tag{10.2}$$

where $V$ is the quantization volume and $V_m$ is the volume of the medium. The field $\mathbf{D}$ can now be expanded in terms of plane-wave creation and annihilation operators as

$$\mathbf{D}(\mathbf{r}) = i\sum_{\mathbf{k},\alpha}\left(\frac{\hbar}{2\mu_0\omega_k V}\right)^{1/2}(\mathbf{k}\times\hat{\mathbf{e}}_{\mathbf{k},\alpha})(a_{\mathbf{k},\alpha}e^{i\mathbf{k}\cdot\mathbf{r}} - a^\dagger_{\mathbf{k},\alpha}e^{-i\mathbf{k}\cdot\mathbf{r}}), \tag{10.3}$$

where $\alpha = 1, 2$ and $\mathbf{k}\cdot\hat{\mathbf{e}}_{\mathbf{k},\alpha} = 0$. This expression and the corresponding one for $\mathbf{B}$ (see Eq. (8.45)) can now be inserted into the Hamiltonian. For $H_{free}$ we find, as before,

$$H_{free} = \sum_{\mathbf{k},\alpha}\hbar\omega_k a^\dagger_{\mathbf{k},\alpha}a_{\mathbf{k},\alpha}. \tag{10.4}$$

For $H_{int}$ we shall make several assumptions. We shall assume that $\eta^{(2)}$ is symmetric, that the operators are normally ordered, and, since we are interested only in down conversion, we shall drop the two terms that do not contribute to that process. The result is

$$\begin{aligned} H_{int} &= i\sum_{\mathbf{k}_1,\alpha_1}\sum_{\mathbf{k}_2,\alpha_2}\sum_{\mathbf{k}_3,\alpha_3}\frac{1}{V^{3/2}}F(\mathbf{k}_1,\mathbf{k}_2,\mathbf{k}_3;\alpha_1,\alpha_2,\alpha_3) \\ &\quad h(\mathbf{k}_3 - \mathbf{k}_1 - \mathbf{k}_2)(a^\dagger_{\mathbf{k}_3,\alpha_3}a_{\mathbf{k}_1,\alpha_1}a_{\mathbf{k}_2,\alpha_2} - a^\dagger_{\mathbf{k}_2,\alpha_2}a^\dagger_{\mathbf{k}_1,\alpha_1}a_{\mathbf{k}_3,\alpha_3}), \end{aligned} \tag{10.5}$$

where

$$\begin{aligned} F(\mathbf{k}_1,\mathbf{k}_2,\mathbf{k}_3;\alpha_1,\alpha_2,\alpha_3) &= \sum_{j,k,l=1}^3 \frac{1}{\sqrt{\omega_{\mathbf{k}_1}\omega_{\mathbf{k}_2}\omega_{\mathbf{k}_3}}}\left(\frac{\hbar}{2\mu_0}\right)^{3/2}\eta^{(2)}_{jkl}(\mathbf{k}_1\times\hat{\mathbf{e}}_{\mathbf{k}_1,\alpha_1})_j \\ &\quad (\mathbf{k}_2\times\hat{\mathbf{e}}_{\mathbf{k}_2,\alpha_2})_k(\mathbf{k}_3\times\hat{\mathbf{e}}_{\mathbf{k}_3,\alpha_3})_l, \end{aligned} \tag{10.6}$$

and

$$h(\mathbf{k}) = \prod_{j=x,y,z} \frac{2\sin(k_j l_j/2)}{k_j}. \tag{10.7}$$

Note that $F$ depends only on the direction of the wave vectors and not on their magnitudes.

We would now like to use this Hamiltonian to calculate a two-point correlation function that is related to the probability of detecting a single photon at each point. Suppose we have single-atom detectors at points $\mathbf{r}_1$ and $\mathbf{r}_2$. The detector at $\mathbf{r}_1$ is turned on between time $t_1$ and $t_1 + \Delta t$ and the detector at $\mathbf{r}_2$ is turned on between $t_2$ and $t_2 + \Delta t$, where $t_2 > t_1 + \Delta t$. The initial state of the system is a product of coherent states

$$|\psi_{in}\rangle = \prod_{\mathbf{k}} |\beta_{\mathbf{k}}\rangle, \tag{10.8}$$

where $\beta_{\mathbf{k}}$ is only appreciable for $\mathbf{k}$ in a neighborhood of $k_0 \hat{\mathbf{z}}$. This represents the initial state of the pump, the signal and idler are assumed to be initially in the vacuum state. Let us assume that the detector atoms have ground states $|g_1\rangle$ and $|g_2\rangle$ and excited states $\{|q_{1j}\rangle\}$ and $\{|q_{2j}\rangle\}$ for some range of $j$, and we shall assume that these atoms are identical. Let $\omega_{jg}$ be the frequency difference between the levels $|q_{1j}\rangle$ and $|g_1\rangle$, and the dipole matrix element between these two levels be $\mathbf{d}_{jg}$. We shall assume that the frequencies $\omega_{jg}$ are less than the frequency of the pump mode, and are sufficiently far from the pump frequency that the chance of a pump photon being absorbed by one of the detector atoms is negligible. This implies that if both atoms have absorbed a photon, then the state of the field will just be $|\psi_{in}\rangle$. To lowest order in perturbation theory, where both the detector atoms and $H_{int}$ are treated as perturbations, the amplitude for atom 1 to be in $|q_{1j}\rangle$ and atom 2 to be in $|q_{1j'}\rangle$ after time $t_2 + \Delta t$ is

$$\begin{aligned} B_{j,j'} &= \left(\frac{-i}{\hbar}\right)^3 \int_{t_2}^{t_2+\Delta t} dt_2' \int_{t_1}^{t_1+\Delta t} dt_1' \int_0^{t_1} dt\, e^{i\omega_{jg}t_1'} e^{i\omega_{j'g}t_2'} \\ &\quad \langle \psi_{in} | \mathbf{d}_{j'g} \cdot \mathbf{D}^{(+)}(\mathbf{r}_2, t_2') \mathbf{d}_{jg} \cdot \mathbf{D}^{(+)}(\mathbf{r}_1, t_1') H_{int}(t) | \psi_{in} \rangle. \end{aligned} \tag{10.9}$$

In this equation,

$$\mathbf{D}^{(+)}(\mathbf{r},t) = i \sum_{\mathbf{k},\alpha} \left(\frac{\hbar}{2\mu_0 \omega_k V}\right)^{1/2} (\mathbf{k} \times \hat{\mathbf{e}}_{\mathbf{k},\alpha}) a_{\mathbf{k},\alpha} e^{i(\mathbf{k}\cdot\mathbf{r} - \omega_k t)}, \tag{10.10}$$

and $H_{int}(t)$ is the interaction Hamiltonian in the interaction picture. The probability that each atom will have absorbed a photon is

$$p(\mathbf{r}_1, t_1; \mathbf{r}_2, t_2) = \sum_{j,j'} |B_{j,j'}|^2. \tag{10.11}$$

If we define

$$g_{\mu\nu}(\mathbf{r}_1, t_1'; \mathbf{r}_2, t_2') = \left(\frac{-i}{\hbar}\right)^3 \int_0^{t_1} dt \langle \psi_{in} | D_\mu^{(+)}(\mathbf{r}_2, t_2') D_\nu^{(+)}(\mathbf{r}_1, t_1') H_{int}(t) | \psi_{in} \rangle, \tag{10.12}$$

and

$$S_{\mu\mu'}(t) = \sum_j e^{i\omega_{jg}t} d_{jg,\mu} d^*_{jg,\mu'}, \tag{10.13}$$

then we have that

$$\begin{aligned}
p(\mathbf{r}_1, t_1; \mathbf{r}_2, t_2) &= \int_{t_2}^{t_2+\Delta t} dt'_2 \int_{t_2}^{t_2+\Delta t} dt''_2 \int_{t_1}^{t_1+\Delta t} dt'_1 \int_{t_1}^{t_1+\Delta t} dt''_1 \sum_{\mu,\mu'=1}^{3} \sum_{\nu,\nu'=1}^{3} \\
& \quad S_{\nu\nu'}(t'_1 - t''_1) S_{\mu\mu'}(t'_2 - t''_2) g_{\mu'\nu'}(\mathbf{r}_1, t''_1; \mathbf{r}_2, t''_2)^* \\
& \quad g_{\mu\nu}(\mathbf{r}_1, t'_1; \mathbf{r}_2, t'_2).
\end{aligned} \tag{10.14}$$

Note that the function $S_{\mu\mu'}(t)$ characterizes the frequency bandwidth of the detector.

Finally, in order to obtain an idea of how this probability behaves, we will find an expression for $g_{\mu\nu}(\mathbf{r}_1, t'_1; \mathbf{r}_2, t'_2)$. Converting the sums over momentum into integrals, we find that

$$\begin{aligned}
g_{\mu\nu}(\mathbf{r}_1, t'_1; \mathbf{r}_2, t'_2) &= \frac{-\sqrt{V}}{2(2\pi)^9 \mu_0 \hbar^2} \sum_{\alpha_1, \alpha_2, \alpha_3} \int d^3 k_1 \int d^3 k_2 \int d^3 k_3 \\
& \quad \frac{F}{\sqrt{\omega_1 \omega_2}} M_{\mu\nu}(\mathbf{r}_1, t'_1; \mathbf{r}_2, t'_2) h(\mathbf{k}_3 - \mathbf{k}_1 - \mathbf{k}_2) \\
& \quad \frac{e^{i(\omega_1 + \omega_2 - \omega_3)t_1} - 1}{\omega_1 + \omega_2 - \omega_3} \beta_{\mathbf{k}_3, \alpha_3}
\end{aligned} \tag{10.15}$$

where we have set $\omega_j = \omega_{k_j}$ and $\hat{\mathbf{e}}_{j,\alpha_j} = \hat{\mathbf{e}}_{\mathbf{k}_j, \alpha_j}$ for $j = 1, 2, 3$. We also have that

$$\begin{aligned}
M_{\mu\nu}(\mathbf{r}_1, t'_1; \mathbf{r}_2, t'_2) &= (\mathbf{k}_2 \times \hat{\mathbf{e}}_{2,\alpha_2})_\mu (\mathbf{k}_1 \times \hat{\mathbf{e}}_{1,\alpha_1})_\nu e^{i\mathbf{k}_2 \cdot \mathbf{r}_1} e^{i\mathbf{k}_1 \cdot \mathbf{r}_2} e^{i(\omega_2 t'_1 + \omega_1 t'_2)} \\
& \quad + (\mathbf{k}_1 \times \hat{\mathbf{e}}_{1,\alpha_1})_\mu (\mathbf{k}_2 \times \hat{\mathbf{e}}_{2,\alpha_2})_\nu e^{i\mathbf{k}_1 \cdot \mathbf{r}_1} e^{i\mathbf{k}_2 \cdot \mathbf{r}_2} e^{i(\omega_1 t'_1 + \omega_2 t'_2)}.
\end{aligned} \tag{10.16}$$

We can now make some rather rough approximations in order to make some sense of the above expression. If we assume that the nonlinear crystal is large and that $|\mathbf{r}_1|$ and $|\mathbf{r}_2|$ are comparable in size to $l_z$, then $h(\mathbf{k}_3 - \mathbf{k}_1 - \mathbf{k}_2) \to \delta^{(3)}(\mathbf{k}_3 - \mathbf{k}_1 - \mathbf{k}_2)$. We shall also assume that the interaction time $t_1$ satisfies $ct_1 \gg |\mathbf{r}_1|, |\mathbf{r}_2|$, and in this case we can approximate $[e^{i(\omega_1+\omega_2-\omega_3)t_1} - 1]/(\omega_1 + \omega_2 - \omega_3)$ by $\delta(\omega_1 + \omega_2 - \omega_3)$. We shall also assume the pump mode is specified by $\beta_{\mathbf{k}_3,\alpha_3} = \beta_0 \delta_{\alpha_3,1} \delta^{(3)}(\mathbf{k}_3 - k_0 \hat{\mathbf{z}})$. Finally, because the detectors are not sensitive to frequencies at the pump frequency and above, we shall cut the $\mathbf{k}_1$ and $\mathbf{k}_2$ integrals off at $k_0$. This gives us that

$$\begin{aligned}
g_{\mu\nu}(\mathbf{r}_1, t'_1; \mathbf{r}_2, t'_2) &\sim \int_{|\mathbf{k}_1| < k_0} d^3 k_1 \int_{|\mathbf{k}_2| < k_0} d^3 k_2 \delta(\omega_1 + \omega_2 - \omega_3) \delta^{(3)}(\mathbf{k}_3 - \mathbf{k}_1 - \mathbf{k}_2) \\
& \quad [e^{i\mathbf{k}_2 \cdot \mathbf{r}_1} e^{i\mathbf{k}_1 \cdot \mathbf{r}_2} e^{i(\omega_2 t'_1 + \omega_1 t'_2)} + e^{i\mathbf{k}_1 \cdot \mathbf{r}_1} e^{i\mathbf{k}_2 \cdot \mathbf{r}_2} e^{i(\omega_1 t'_1 + \omega_2 t'_2)}],
\end{aligned} \tag{10.17}$$

where now, $\mathbf{k}_3 = k_0 \hat{\mathbf{z}}$. Performing the integrals, we have, setting $\Delta Z = (z_2 - z_1) - c(t'_2 - t'_1)$

$$\begin{aligned}
g_{\mu\nu}(\mathbf{r}_1, t'_1; \mathbf{r}_2, t'_2) &\sim e^{ik_0(z_1 - ct'_1)} \left[ \frac{2i}{(\Delta Z)^3}(1 - e^{ik_0 \Delta Z}) - \frac{k_0}{(\Delta Z)^2}(1 + e^{ik_0 \Delta Z}) \right] \\
& \quad + (\mathbf{r}_1, t'_1 \leftrightarrow \mathbf{r}_2, t'_2).
\end{aligned} \tag{10.18}$$

This expression tells us the following. Because the photons in down conversion are emitted simultaneously, it is most likely to detect them both at points satisfying the condition $\Delta Z = 0$, and the likelihood decays as approximately $1/(\Delta Z)^2$ as $\Delta Z$ increases.

## 11 Dispersion

A realistic description of the propagation of fields in a nonlinear medium must include the effects of linear dispersion (the effects of nonlinear dispersion are small and can be neglected to lowest order). Dispersion, however, is difficult to incorporate into the standard canonical formulation, because it is an effect which is nonlocal in time. It arises from the fact that the polarization of the medium at time $t$, $\mathbf{P}(t)$, depends not only on the electric field at time $t$, but also on its values at previous times [49]

$$\mathbf{P}(t) = \int_0^\infty d\tau \chi^{(1)}(\tau) : \mathbf{E}(t-\tau). \tag{11.1}$$

There are two known approaches to constructing a quantized theory for nonlinear media that incorporate dispersion. The first, which was pioneered by Drummond [50], has as its basic objects narrow-band fields for which it is possible to derive an approximate theory which is local. In the second, the degrees of freedom of the medium are included in the theory, and the entire theory, fields plus medium, is local. Each approach has its advantages. The second is more fundamental, but requires a model for the medium, which for many systems of interest, will be complicated. The first is more phenomenological, but needs only a set of functions, the polarizabilities, to describe the medium. Both are useful and we shall consider each of them.

In the case of linear theories, a number of other methods of quantizing fields in the presence of dispersive media have been developed. One starts with the equations of motion for the field operators and then introduces frequency dependence into the susceptibilities and noise currents [51]. A second starts from the results of a microscopic model and generalizes them to be able to treat arbitrary frequency-dependent susceptibilities [52]. A third is able to define a Lagrangian and Hamiltonian for fields in a medium with arbitrary frequency response by introducing auxiliary fields [53]. This theory can then be quantized in the usual way. These methods have not yet been used to treat nonlinear media.

Let us first look at the approximate macroscopic theory due to Drummond [50]. The basic field in this theory is the dual potential. It is simplest to start by considering a linear, dispersive medium in which case the electric field is related to the dual potential by

$$E_i(\mathbf{r},t) = \int_0^\infty d\tau \beta_{ij}^{(1)}(\mathbf{r},\tau) D_j(\mathbf{r},t-\tau), \tag{11.2}$$

where $\mathbf{D} = \nabla \times \mathbf{\Lambda}$. As is evident from this equation, $\beta^{(1)}$ is in general a tensor, but we shall assume, for the sake of simplicity, that the medium is isotropic which implies that $\beta^{(1)}$ is a scalar. The relation between $\mathbf{E}$ and $\mathbf{D}$ and Maxwell's equations imply that $\mathbf{\Lambda}(\mathbf{r},t)$ satisfies the equation

$$\nabla \times \int_0^\infty d\tau \beta^{(1)}(\mathbf{r},\tau)[\nabla \times \mathbf{\Lambda}(\mathbf{r},t-\tau)] = -\ddot{\mathbf{\Lambda}}(\mathbf{r},t). \tag{11.3}$$

which is clearly nonlocal in time. Now suppose that $\mathbf{\Lambda}^\nu$ is a narrow-band field with frequency components near $\omega_\nu$ (that is, $\mathbf{\Lambda}^\nu \sim e^{-i\omega_\nu t}$), and that $\mathbf{\Lambda}$ can be expressed as

$$\mathbf{\Lambda} = \mathbf{\Lambda}^\nu + \mathbf{\Lambda}^{-\nu}, \tag{11.4}$$

where $\mathbf{\Lambda}^{-\nu} = (\mathbf{\Lambda}^{\nu})^*$. The field $\mathbf{\Lambda}^{\nu}$ will also satisfy Eq. (11.3). Let us now define

$$\beta^{(1)}(\mathbf{r},\omega) = \int_0^{\infty} d\tau e^{i\omega\tau} \beta^{(1)}(\mathbf{r},\tau), \tag{11.5}$$

where $\beta^{(1)}(\mathbf{r},\omega) = 1/\epsilon(\mathbf{r},\omega)$, and $\epsilon(\mathbf{r},\omega)$ is the usual frequency-dependent dielectric function for the medium. For a nonabsorbing medium, $\epsilon(\mathbf{r},\omega)$ is real and $\epsilon(\mathbf{r},-\omega) = \epsilon(\mathbf{r},\omega)$. Because we are interested in frequencies near $\omega_\nu$, expand this quantity up to second order in $\omega - \omega_\nu$

$$\beta^{(1)}(\mathbf{r},\omega) \cong \beta_\nu(\mathbf{r}) + \omega\beta'_\nu(\mathbf{r}) + \frac{1}{2}\omega^2 \beta''_\nu(\mathbf{r}), \tag{11.6}$$

where

$$\begin{aligned}
\beta_\nu(\mathbf{r}) &\equiv \beta^{(1)}(\mathbf{r},\omega_\nu) - \omega_\nu \frac{d\beta^{(1)}}{d\omega}(\mathbf{r},\omega_\nu) + \frac{1}{2}\omega^2 \frac{d^2\beta^{(1)}}{d\omega^2}(\mathbf{r},\omega_\nu), \\
\beta'_\nu(\mathbf{r}) &\equiv \frac{d\beta^{(1)}}{d\omega}(\mathbf{r},\omega_\nu) - \omega_\nu \frac{d^2\beta^{(1)}}{d\omega^2}(\mathbf{r},\omega_\nu) \\
\beta''_\nu(\mathbf{r}) &\equiv \frac{d^2\beta^{(1)}}{d\omega^2}(\mathbf{r},\omega_\nu)
\end{aligned} \tag{11.7}$$

We now consider the wave equation which $\mathbf{\Lambda}^{\nu}$ satisfies, Eq. (11.3), and make the following approximation. The quantity $e^{-i\omega_\nu\tau}\mathbf{\Lambda}^{\nu}(t-\tau)$ is a slowly varying function of $\tau$, so we expand it in a Taylor series in $\tau$ up to second order. Taylor expansions are often used in this way in classical dispersion theory to simplify wave equations, and this technique was first introduced into macroscopic field quantization by Kennedy and Wright [54]. Doing so we find that (we shall not explicitly indicate the $\mathbf{r}$ dependence of $\mathbf{\Lambda}^{\nu}$ and $\beta^{(1)}$)

$$\int_0^{\infty} d\tau \beta^{(1)}(\tau) e^{i\omega_\nu\tau}[e^{-i\omega_\nu\tau}\nabla \times \mathbf{\Lambda}^{\nu}(t-\tau)]$$

$$\cong \int_0^{\infty} d\tau \beta^{(1)}(\tau) e^{i\omega_\nu\tau} \nabla \times \{\mathbf{\Lambda}^{\nu}(t) - \tau[\dot{\mathbf{\Lambda}}^{\nu}(t) - i\omega_\nu \mathbf{\Lambda}^{\nu}(t)]$$
$$+ \frac{1}{2}\tau^2[\ddot{\mathbf{\Lambda}}^{\nu}(t) + 2i\omega_\nu \dot{\mathbf{\Lambda}}^{\nu}(t) - \omega_\nu^2 \mathbf{\Lambda}^{\nu}(t)]\}$$
$$\cong \beta_\nu \mathbf{\Lambda}^{\nu}(t) + i\beta'_\nu \dot{\mathbf{\Lambda}}^{\nu}(t) - \frac{1}{2}\beta''_\nu \ddot{\mathbf{\Lambda}}^{\nu}(t). \tag{11.8}$$

Substituting this expansion into the wave equation, Eq. (11.3), we find

$$\begin{aligned}
-\ddot{\mathbf{\Lambda}} &= \nabla \times [\beta_\nu \nabla \times \mathbf{\Lambda}^{\nu} + i\beta'_\nu \nabla \times \dot{\mathbf{\Lambda}}^{\nu} \\
&- \frac{1}{2}\beta''_\nu \nabla \times \ddot{\mathbf{\Lambda}}^{\nu}],
\end{aligned} \tag{11.9}$$

which is a local equation for $\mathbf{\Lambda}^{\nu}$, which can, in turn, be derived from a local Lagrangian. The Lagrangian density from which it follows is

$$\begin{aligned}
\mathcal{L} &= \frac{1}{2}[2\mu_0(\dot{\mathbf{\Lambda}}^{-\nu}) \cdot \dot{\mathbf{\Lambda}}^{\nu} - 2(\nabla \times \mathbf{\Lambda}^{-\nu}) \cdot \beta_\nu(\nabla \times \mathbf{\Lambda}^{\nu}) - i(\nabla \times \mathbf{\Lambda}^{-\nu}) \cdot \beta'_\nu(\nabla \times \dot{\mathbf{\Lambda}}^{\nu}) \\
&+ i(\nabla \times \mathbf{\Lambda}^{\nu}) \cdot \beta'_\nu(\nabla \times \dot{\mathbf{\Lambda}}^{-\nu}) - (\nabla \times \dot{\mathbf{\Lambda}}^{-\nu}) \cdot \beta''_\nu(\nabla \times \dot{\mathbf{\Lambda}}^{\nu})].
\end{aligned} \tag{11.10}$$

The coordinates in this Lagrangian density are $\mathbf{\Lambda}^\nu$ and $\mathbf{\Lambda}^{-\nu}$. The equation of motion for $\mathbf{\Lambda}^\nu$, which follows from the condition $\delta \int dt \int d^3r \mathcal{L} = 0$ is

$$
\begin{aligned}
0 &= \frac{\partial \mathcal{L}}{\partial \Lambda_j^\nu} - \sum_{k=1}^{3} \partial_k \left( \frac{\partial \mathcal{L}}{\partial(\partial_k \Lambda_j^\nu)} \right) - \partial_t \left( \frac{\partial \mathcal{L}}{\partial(\partial_t \Lambda_j^\nu)} \right) \\
&+ \sum_{k=1}^{3} \partial_t \partial_k \left( \frac{\partial \mathcal{L}}{\partial(\partial_t \partial_k \Lambda_j^\nu)} \right).
\end{aligned} \quad (11.11)
$$

Insertion of the above Lagrangian density into this equation gives Eq. (11.9). The canonical momentum, $\mathbf{\Pi}^\nu$, is given by

$$\Pi_j^\nu = \frac{\delta L}{\delta \dot{\lambda}_j^\nu} = \frac{\partial \mathcal{L}}{\partial \dot{\Lambda}_j^\nu} - \sum_{k=1}^{3} \partial_k \left( \frac{\partial \mathcal{L}}{\partial(\partial_k \dot{\Lambda}_j^\nu)} \right), \quad (11.12)$$

which gives us that

$$\mathbf{\Pi}^\nu = \mu_0 \dot{\mathbf{\Lambda}}^{-\nu} - \frac{1}{2} \nabla \times [\beta''_\nu (\nabla \times \dot{\mathbf{\Lambda}}^{-\nu}) + i\beta'_\nu (\nabla \times \mathbf{\Lambda}^{-\nu})]. \quad (11.13)$$

Finally, from the Lagrangian and the canonical momentum we can find the Hamiltonian density

$$\mathcal{H} = \mathbf{\Pi}^\nu \cdot \dot{\mathbf{\Lambda}}^\nu + \mathbf{\Pi}^{-\nu} \cdot \dot{\mathbf{\Lambda}}^{-\nu} - \mathcal{L}. \quad (11.14)$$

The Hamiltonian is then found by integrating the Hamiltonian density over the quantization volume. It is possible to simplify the Hamiltonian density by integrating some of the terms by parts and assuming that the boundary terms vanish, in particular, it is useful to make use of the identity

$$\int d^3r \mathbf{V_1} \cdot \nabla \times \mathbf{V_2} = \int d^3r \mathbf{V_2} \cdot \nabla \times \mathbf{V_1}. \quad (11.15)$$

This allows us to combine terms in the Hamiltonian density, and the resulting Hamiltonian is

$$
\begin{aligned}
H &= \int d^3r [\mu_0 \dot{\mathbf{\Lambda}}^{-\nu} \cdot \dot{\mathbf{\Lambda}}^\nu + (\nabla \times \mathbf{\Lambda}^{-\nu}) \cdot \beta_\nu (\nabla \times \mathbf{\Lambda}^\nu) \\
&- \frac{1}{2} (\nabla \times \dot{\mathbf{\Lambda}}^{-\nu}) \beta''_\nu (\nabla \times \dot{\mathbf{\Lambda}}^\nu)]
\end{aligned} \quad (11.16)
$$

where $\dot{\mathbf{\Lambda}}^\nu$ is to be considered a function of $\mathbf{\Pi}^\nu$. As has been shown by Drummond [50], this Hamiltonian is the energy for a classical field in a dispersive dielectric. He has also emphasized that the Langrangian for the theory, which is not unique (it can, for example, be scaled by an arbitrary factor and the equations of motion will be unaffected) should be chosen so that it does give a Hamiltonian which is the classical energy.

If the dispersion in the nonlinear interaction is ignored, so that the interaction is considered local, it can be included in the theory by adding the terms

$$\int d^3r [\frac{1}{3} \mathbf{D} \cdot \beta^{(2)} : \mathbf{DD} + \frac{1}{4} \mathbf{D} \cdot \beta^{(3)} : \mathbf{DDD}], \quad (11.17)$$

to the Hamiltonian in Eq. (11.16). In addition, if fields with several discrete frequencies are present, they can be accomodated by adding additional fields, $\mathbf{\Lambda}^\nu$, centered about these frequencies to the theory. This has the effect of adding summations over $\nu$ to the Langrangian density and Hamiltonian in Eqs. (11.10) and (11.16).

In order to quantize the theory we would like to simply impose the commutation relations

$$[\Lambda_j^\nu(\mathbf{r},t), \Pi_{j'}^\nu(\mathbf{r}',t)] = i\delta_{jj'}^{(tr)}(\mathbf{r}-\mathbf{r}'). \tag{11.18}$$

However, in order for this to be true the fields must have Fourier components of arbitrarily high frequency, while $\mathbf{\Lambda}^\nu$ is limited in bandwidth. An alternative, and in this case better, approach is to expand the field in terms of spatial modes and to use the expansion coefficients as coordinates. One then finds the corresponding canonical momentum for each coordinate and then imposes the usual commutation relations between coordinates and momenta.

To begin let us expand $\Lambda^\nu(\mathbf{r},t)$ in plane wave modes

$$\Lambda^\nu(\mathbf{r},t) = \frac{1}{\sqrt{V}} \sum_{\mathbf{k},\alpha} \lambda_{\mathbf{k},\alpha}^\nu \hat{\mathbf{e}}_{\mathbf{k},\alpha} e^{i\mathbf{k}\cdot\mathbf{r}}, \tag{11.19}$$

where the $\lambda_{\mathbf{k},\alpha}^\nu$ will become our coordinates. The fact that $\mathbf{\Lambda}^{-\nu} = (\mathbf{\Lambda}^\nu)^*$ and that $\hat{\mathbf{e}}_{-\mathbf{k},\alpha} = -(-1)^\alpha \hat{\mathbf{e}}_{\mathbf{k},\alpha}$ imply that

$$\lambda_{\mathbf{k},\alpha}^\nu = -(-1)^\alpha (\lambda_{-\mathbf{k},\alpha}^{-\nu})^*. \tag{11.20}$$

The plane-wave expansion can be inserted into the Lagrangian, yielding

$$\begin{aligned} L &= \sum_{\mathbf{k},\alpha} \sum_{\mathbf{k}',\alpha'} [\dot{\lambda}_{\mathbf{k}',\alpha'}^{-\nu} \dot{\lambda}_{\mathbf{k},\alpha}^\nu M_{(\mathbf{k}',\alpha'),(\mathbf{k},\alpha)}^{(1)} + \lambda_{\mathbf{k}',\alpha'}^{-\nu} \lambda_{\mathbf{k},\alpha}^\nu M_{(\mathbf{k}',\alpha'),(\mathbf{k},\alpha)}^{(2)} \\ &\quad + \lambda_{\mathbf{k}',\alpha'}^{-\nu} \dot{\lambda}_{\mathbf{k},\alpha}^\nu M_{(\mathbf{k}',\alpha'),(\mathbf{k},\alpha)}^{(3)} - \dot{\lambda}_{\mathbf{k}',\alpha'}^{-\nu} \lambda_{\mathbf{k},\alpha}^\nu M_{(\mathbf{k}',\alpha'),(\mathbf{k},\alpha)}^{(3)}]. \end{aligned} \tag{11.21}$$

The matrices $M_{(\mathbf{k}',\alpha'),(\mathbf{k},\alpha)}^{(j)}$, $j=1,2,3$ are given by

$$\begin{aligned} M_{(\mathbf{k}',\alpha'),(\mathbf{k},\alpha)}^{(1)} &= -\mu_0 (-1)^\alpha \delta_{\mathbf{k}',-\mathbf{k}} \delta_{\alpha',\alpha} \\ &\quad + \frac{1}{2V} \int d^3 r\, e^{i(\mathbf{k}+\mathbf{k}')\cdot\mathbf{r}} \beta_\nu''(\mathbf{k}' \times \hat{\mathbf{e}}_{\mathbf{k}',\alpha'}) \cdot (\mathbf{k} \times \hat{\mathbf{e}}_{\mathbf{k},\alpha}) \\ M_{(\mathbf{k}',\alpha'),(\mathbf{k},\alpha)}^{(2)} &= +\frac{1}{V} \int d^3 r\, e^{i(\mathbf{k}+\mathbf{k}')\cdot\mathbf{r}} \beta_\nu(\mathbf{k}' \times \hat{\mathbf{e}}_{\mathbf{k}',\alpha'}) \cdot (\mathbf{k} \times \hat{\mathbf{e}}_{\mathbf{k},\alpha}) \\ M_{(\mathbf{k}',\alpha'),(\mathbf{k},\alpha)}^{(3)} &= +\frac{i}{2V} \int d^3 r\, e^{i(\mathbf{k}+\mathbf{k}')\cdot\mathbf{r}} \beta_\nu'(\mathbf{k}' \times \hat{\mathbf{e}}_{\mathbf{k}',\alpha'}) \cdot (\mathbf{k} \times \hat{\mathbf{e}}_{\mathbf{k},\alpha}) \end{aligned} \tag{11.22}$$

Our next step is to find the canonical momenta. These are given by

$$\pi_{\mathbf{k},\alpha}^\nu = \frac{\partial L}{\partial \dot{\lambda}_{\mathbf{k},\alpha}^\nu} = \sum_{\mathbf{k}',\alpha'} [\dot{\lambda}_{\mathbf{k}',\alpha'}^{-\nu} M_{(\mathbf{k}',\alpha'),(\mathbf{k},\alpha)}^{(1)} + \lambda_{\mathbf{k}',\alpha'}^{-\nu} M_{(\mathbf{k}',\alpha'),(\mathbf{k},\alpha)}^{(3)}] \tag{11.23}$$

and

$$\pi_{\mathbf{k}',\alpha'}^{-\nu} = \frac{\partial L}{\partial \dot{\lambda}_{\mathbf{k}',\alpha'}^{-\nu}} = \sum_{\mathbf{k},\alpha} [\dot{\lambda}_{\mathbf{k},\alpha}^\nu M_{(\mathbf{k}',\alpha'),(\mathbf{k},\alpha)}^{(1)} - \lambda_{\mathbf{k},\alpha}^\nu M_{(\mathbf{k}',\alpha'),(\mathbf{k},\alpha)}^{(3)}]. \tag{11.24}$$

We can now find the Hamiltonian

$$\begin{aligned} H &= \sum_{\mathbf{k},\alpha}(\pi^\nu_{\mathbf{k},\alpha}\dot{\lambda}^\nu_{\mathbf{k},\alpha} + \pi^{-\nu}_{\mathbf{k},\alpha}\dot{\lambda}^{-\nu}_{\mathbf{k},\alpha}) - L \\ &= \sum_{\mathbf{k},\alpha}\sum_{\mathbf{k}',\alpha'}[\dot{\lambda}^{-\nu}_{\mathbf{k}',\alpha'}\dot{\lambda}^{\nu}_{\mathbf{k},\alpha}M^{(1)}_{(\mathbf{k}',\alpha'),(\mathbf{k},\alpha)} - \lambda^{-\nu}_{\mathbf{k}',\alpha'}\lambda^{\nu}_{\mathbf{k},\alpha}M^{(2)}_{(\mathbf{k}',\alpha'),(\mathbf{k},\alpha)}]. \end{aligned} \quad (11.25)$$

The Hamiltonian should be expressed in terms of the coordinates and the canonical momenta, and in order to do so Eqs. (11.23) and (11.24) must be inverted to find expressions for $\dot{\lambda}^\nu_{\mathbf{k},\alpha}$ and $\dot{\lambda}^{-\nu}_{\mathbf{k},\alpha}$ in terms of the canonical momenta and the coordinates.

We shall now consider a simplified version of this theory. We shall suppose that the field consists of plane waves which are polarized in the $y$ direction and propagate in the $x$ direction. In this case the field, $\Lambda^\nu(x,t)$ is a scalar and a function of only one spatial coordinate. This means that $\nabla \times \boldsymbol{\Lambda}^\nu$ becomes $\partial_x \Lambda^\nu \hat{\mathbf{z}}$ and integrals over the quantization volume are replaced by $A \int_0^l dx$, where the quantization volume $V = l^3$, and $A = l^2$. If, in addition, we assume that the only nonlinearity present is described by $\beta^{(3)}$ and that the medium is homogeneous, we find that the Hamiltonian is

$$\begin{aligned} H &= A\int dx \{\mu_0 \dot{\Lambda}^\nu \dot{\Lambda}^{-\nu} + \beta_\nu(\partial_x \Lambda^\nu)(\partial_x \Lambda^{-\nu}) - \frac{1}{2}\beta''_\nu(\partial_x \dot{\Lambda}^\nu)(\partial_x \dot{\Lambda}^{-\nu}) \\ &\quad + \frac{1}{4}\beta^{(3)}[\partial_x(\Lambda^\nu + (\Lambda^\nu)^*)]^4\}. \end{aligned} \quad (11.26)$$

The field $\Lambda(x,t)$ is now given by

$$\Lambda(x,t) = \frac{1}{\sqrt{V}}\sum_k (\lambda^\nu_k e^{ikx} + \lambda^{-\nu}_{-k} e^{-ikx}). \quad (11.27)$$

Note that, because the nonlinear term does not depend on the time derivative of $\Lambda(x,t)$, its addition to the theory does not change the canonical momenta, i.e. the canonical momenta for the theory without the nonlinear term are the same as the canonical momenta for the theory with the nonlinear term.

What we would like to do is to express the Hamiltonian in terms of $\lambda^\nu_k$ and $\lambda^{-\nu}_k$ and their corresponding momenta. Note that we have suppressed the polarization subscript, because it is not necessary for the simplified theory we are considering (there is only one polarization present). For plane-wave modes propagating in the $x$ direction in a homogeneous medium, and assuming the polarization vectors satisfy $\hat{\mathbf{e}}_k = \hat{\mathbf{e}}_{-k}$, we have that

$$\begin{aligned} M^{(1)}_{k,k'} &= (\mu_0 - \frac{1}{2}\beta''_\nu k^2)\delta_{k,-k'} \\ M^{(2)}_{k,k'} &= -k^2 \beta_\nu \delta_{k,-k'} \\ M^{(3)}_{k,k'} &= -\frac{i}{2}k^2 \beta'_\nu \delta_{k,-k'}. \end{aligned} \quad (11.28)$$

From this, for the linear part of the Lagrangian, $L_0$, we obtain

$$\begin{aligned} L_0 &= \sum_k [\dot{\lambda}^{-\nu}_{-k}\dot{\lambda}^{\nu}_k(\mu_0 - \frac{1}{2}\beta''_\nu k^2) - \lambda^{-\nu}_{-k}\lambda^{\nu}_k k^2 \beta_\nu \\ &\quad - \frac{i}{2}k^2 \beta'_\nu(\lambda^{-\nu}_{-k}\dot{\lambda}^{\nu}_k - \dot{\lambda}^{-\nu}_{-k}\lambda^{\nu}_k)], \end{aligned} \quad (11.29)$$

and the canonical momenta

$$\begin{aligned}
\pi_k^\nu &= \dot{\lambda}_{-k}^{-\nu}(\mu_0 - \frac{1}{2}\beta_\nu'' k^2) - \frac{i}{2}k^2\beta_\nu'\lambda_{-k}^{-\nu} \\
\pi_{-k}^{-\nu} &= \dot{\lambda}_k^\nu(\mu_0 - \frac{1}{2}\beta_\nu'' k^2) + \frac{i}{2}k^2\beta_\nu'\lambda_k^\nu.
\end{aligned} \quad (11.30)$$

The linear part of the Hamiltonian, $H_0$, is now given by

$$\begin{aligned}
H_0 &= \sum_k [\dot{\lambda}_{-k}^{-\nu}\dot{\lambda}_k^\nu(\mu_0 - \frac{1}{2}\beta_\nu'' k^2) + \lambda_{-k}^{-\nu}\lambda_k^\nu k^2\beta_\nu] \\
&= \sum_k \left[ \frac{1}{\mu_0 - (\beta_\nu''/2)k^2}(\pi_{-k}^{-\nu} - \frac{i}{2}k^2\beta_\nu'\lambda_k^\nu)(\pi_k^\nu + \frac{i}{2}k^2\beta_\nu'\lambda_{-k}^{-\nu}) \right. \\
&\quad + \left. \lambda_{-k}^{-\nu}\lambda_k^\nu k^2\beta_\nu \right].
\end{aligned} \quad (11.31)$$

Our next step is to quantize the theory. We promote $\lambda_k^\nu$ and $\pi_k^\nu$ to operators and impose the canonical commutation relations

$$[\lambda_k^\nu, \pi_{k'}^\nu] = i\hbar\delta_{k,k'} \qquad [\lambda_k^{-\nu}, \pi_{k'}^{-\nu}] = i\hbar\delta_{k,k'}. \quad (11.32)$$

Note that because $\Lambda = \Lambda^\nu + \Lambda^{-\nu}$ is hermitian, we must have $\lambda_{-k}^{-\nu} = (\lambda_k^\nu)^\dagger$, and because $\dot{\Lambda}$ is hermitian, we have that $\pi_{-k}^{-\nu} = (\pi_k^\nu)^\dagger$. We can now define two sets of creation and annihilation operators

$$\begin{aligned}
a_k &= \frac{1}{\sqrt{2\hbar}}\left[ A_k\lambda_k^\nu + \left(\frac{i}{A_k^*}\right)(\pi_k^\nu)^\dagger \right] \\
b_k^\dagger &= \frac{1}{\sqrt{2\hbar}}\left[ A_k\lambda_k^\nu - \left(\frac{i}{A_k^*}\right)(\pi_k^\nu)^\dagger \right],
\end{aligned} \quad (11.33)$$

where $A_k$ is a c-number yet to be determined. These relations can be inverted to give

$$\begin{aligned}
\lambda_k^\nu &= \frac{1}{A_k}\sqrt{\frac{\hbar}{2}}(a_k + b_k^\dagger) \\
\pi_k^\nu &= iA_k\sqrt{\frac{\hbar}{2}}(a_k^\dagger - b_k),
\end{aligned} \quad (11.34)$$

and the results can be substituted into the Hamiltonian. If $A_k$ is chosen to be

$$A_k = \left[ \frac{1}{4}k^4(\beta_\nu')^2 + (\mu_0 - \frac{1}{2}\beta_\nu'' k^2)k^2\beta_\nu \right]^{1/4}, \quad (11.35)$$

then the terms proportional to $a_k b_k$ and $a_k^\dagger b_k^\dagger$ vanish, with the result that

$$H = \hbar\sum_k (\omega_+(k)a_k^\dagger a_k + \omega_-(k)b_k^\dagger b_k), \quad (11.36)$$

where an overall constant has been dropped. The frequencies $\omega_\pm(k)$ are given by

$$\omega_\pm(k) = \frac{1}{\mu_0 - \beta''_\nu k^2/2} \left\{ \pm \frac{1}{2} k^2 \beta'_{nu} \right.$$
$$\left. + \left[ \frac{1}{4} k^4 (\beta'_\nu)^2 + (\mu_0 - \frac{1}{2} \beta''_\nu k^2) k^2 \beta_\nu \right]^{1/2} \right\}. \qquad (11.37)$$

The frequencies $\omega_\pm(k)$ are solutions of the equation

$$\omega_\pm^2 = k^2 (\beta_\nu \pm \omega_\pm \beta'_\nu + \frac{1}{2} \omega_\pm^2 \beta''_\nu). \qquad (11.38)$$

Note that the expression in parentheses in the above equation is identical to the expansion of $\beta^{(1)}(\omega)$ about $\omega_\nu$ if the plus sign is used. Because $\beta^{(1)}(\omega) = 1/\varepsilon(\omega)$, where $\varepsilon(\omega)$ is the usual frequency-dependent dielectric function, Eq. (11.38) for $\omega_+$ is approximately the same as

$$\omega = \frac{k}{\sqrt{\varepsilon(\omega)}}, \qquad (11.39)$$

which is the relation between $\omega$ and $k$ we would expect for a wave travelling in a linear dielectric medium. This leaves us with the question of how to interpret $\omega_-$ and $b_k$.

An examination of Eq. (11.33) shows us that

$$\lambda_k = \frac{1}{A_k \sqrt{2}} (a_k + b_k^\dagger), \qquad (11.40)$$

while Eq. (11.36) implies that $b_k \sim e^{-i\omega_- t}$. Now $\omega_-$ is not too far from $\omega_\nu$, which implies that the $b_k^\dagger$ term in $\lambda_k$ has a time dependence given approximately by $e^{i\omega_\nu t}$. This places it outside the bandwidth for the field $\Lambda^\nu$. In order to be consistent we must assume that all of the $b_k$ modes are in the vacuum state and, thereby, drop these operators from the theory. This implies that the Hamiltonian for the full theory (as opposed to just the linear part) is

$$H = \sum_k \hbar \omega_+(k) a_k^\dagger a_k + \frac{1}{4} \beta^{(3)} A \int dx (\partial_x (\Lambda^\nu + \Lambda^{\nu\dagger}))^4, \qquad (11.41)$$

where

$$\Lambda = \sqrt{\frac{\hbar}{2V}} \sum_k \frac{1}{A_k} (a_k e^{ikx} + a_k^\dagger e^{-ikx}). \qquad (11.42)$$

We can express the quantity $A_k$ in terms of the group velocity, $v_k = d\omega_+/dk$. As we saw, $\omega_+$ is a solution to the equation $k^2 \beta^{(1)}(\omega_+) = \mu_0 \omega_+^2$, where $\beta^{(1)}$ is given by Eq. (11.6). Differentiating both sides with respect to $\omega_+$ gives

$$k \beta^{(1)}(\omega_+) = \left( \mu_0 \omega_+ - \frac{1}{2} k^2 \frac{d\beta^{(1)}}{d\omega_+} \right) \frac{d\omega_+}{dk}. \qquad (11.43)$$

Now, making use of Eqs. (11.37) and (11.35), we find that

$$\mu_0 \omega_+ - \frac{1}{2} k^2 \frac{d\beta^{(1)}}{d\omega_+} = \omega_+ (\mu_0 - \frac{1}{2} k^2 \beta''_\nu) - \frac{1}{2} k^2 \beta'_\nu = A_k^2. \qquad (11.44)$$

This, then, gives us

$$A_k = \sqrt{\frac{k\beta^{(1)}(\omega_+)}{v_k}}. \tag{11.45}$$

Finally, we have for the fields

$$\begin{aligned}
\mathbf{\Lambda}(x,t) &= \sum_k \left[\frac{\hbar\varepsilon(\omega_+)v_k}{2Vk}\right]^{1/2} (a_k e^{ikx} + a_k^\dagger e^{-ikx})\hat{\mathbf{y}} \\
\mathbf{D}(x,t) &= \sum_k i\left[\frac{\hbar k\varepsilon(\omega_+)v_k}{2V}\right]^{1/2} (a_k e^{ikx} - a_k^\dagger e^{-ikx})\hat{\mathbf{z}}.
\end{aligned} \tag{11.46}$$

We now have a theory which is capable of describing the propagation of quantum fields in nonlinear, dispersive media. Carter and Drummond have applied this theory to describe fields propagating through a fiber with a $\chi^{(3)}$ nonlinearity [15], and we shall briefly give the theory in the form in which they use it. We shall assume that the field has wave number components near $k_1$. First, we define the field

$$\Psi(x,t) = \frac{1}{\sqrt{L}} \sum_k e^{i[(k-k_1)x+\omega_1 t]} a_k, \tag{11.47}$$

where $\omega_1 = \omega(k_1)$. This field has equal time commutation relations given by

$$[\Psi(x,t), \Psi^\dagger(x',t)] = \tilde{\delta}(x-x'), \tag{11.48}$$

where $\tilde{\delta}$, because it has a band-limited Fourier transform, is a smoothed version of the Dirac delta function. We shall assume, however, that this smearing effect is not pronounced so that we can treat $\tilde{\delta}(x-x')$ as a delta function. Inverting the relation between $\Psi(x,t)$ and $a_k$, we find that

$$a_k = \frac{1}{\sqrt{L}} \int dx\, e^{-i[(k-k_1)x+\omega_1 t]} \Psi(x,t). \tag{11.49}$$

We can use this expression to express the Hamiltonian, Eq. (11.41), in terms of $\Psi(x,t)$. We first note that

$$\sum_k \omega(k) e^{i(k-k_1)(x-x')} = \frac{1}{L}\int dx \int dx' \left(\sum_k \omega(k) e^{i(k-k_1)(x-x')}\right) \Psi(x,t)^\dagger \Psi(x',t). \tag{11.50}$$

Expanding $\omega(k)$ around $k_1$, we have that

$$\omega(k) = \omega_1 + (k-k_1)v_1 + \frac{1}{2}(k-k_1)^2 \omega_1'' + \ldots \tag{11.51}$$

Here, $v_1$ is the group velocity at $k_1$ and $\omega_1''$ is the second derivative of $\omega(k)$ evaluated at $k_1$. This then gives us

$$\sum_k \omega(k) e^{i(k-k_1)(x-x')} \cong \sum_k [\omega_1 + \frac{i}{2}v_1(\partial_{x'} - \partial_x)$$

$$\begin{aligned} &+ \frac{1}{2}\omega_1''\partial_x\partial_{x'}]e^{i(k-k_1)(x-x')} \\ &\cong [\omega_1 + \frac{i}{2}v_1(\partial_{x'} - \partial_x) \\ &+ \frac{1}{2}\omega_1''\partial_x\partial_{x'}]\delta(x-x'). \end{aligned} \qquad (11.52)$$

This gives, for the linear part of the Hamiltonian, $H_{lin}$,

$$\begin{aligned} H_{lin} &= \hbar \int dx \left[ \omega_1 \Psi^\dagger \Psi + \frac{i}{2}v_1 \left( \frac{\partial \Psi^\dagger}{\partial x}\Psi - \Psi^\dagger \frac{\partial \Psi}{\partial x} \right) \right. \\ &\quad \left. + \frac{1}{2}\omega_1'' \frac{\partial \Psi^\dagger}{\partial x}\frac{\partial \Psi}{\partial x} \right]. \end{aligned} \qquad (11.53)$$

The nonlinear part of the Hamiltonian, $H_{nlin}$, is given by

$$\begin{aligned} \frac{1}{4}\beta^{(3)} A \int dx D^4 &\cong \frac{\beta^{(3)}}{4} A \left[ \left( \frac{\hbar}{2V} \right) k_1 v_1 \epsilon_1 L \right]^2 \int dx [e^{i(k_1 x - \omega_1 t)}\Psi(x,t) \\ &\quad - e^{-i(k_1 x - \omega_1 t)}\Psi(x,t)^\dagger]^4, \end{aligned} \qquad (11.54)$$

where $\beta^{(3)} = -\chi^3/\epsilon_1^3$. Keeping only the slowly varying terms, we obtain

$$H_{nlin} \cong \frac{3\beta^{(3)}}{8A}(\hbar k_1 v_1 \epsilon_1)^2 \int dx (\Psi^\dagger)^2 \Psi^2. \qquad (11.55)$$

The total Hamiltonian is, of course, just the sum of the two terms, $H = H_{lin} + H_{nlin}$.

## 12   Quantum solitons

As we shall shortly see, the above Hamiltonian gives rise to quantum solitons. A classical soliton in a dispersive, nonlinear medium is a pulse that propagates without changing shape. They were first observed in water waves [55] and more recently, and of more relevance to us, they have been observed in optical fibers [56]. Besides their fundamental interest, they could be of use in optical communications. When these waves are quantized, new effects emerge. We want to use the formalism we have developed to study these quantized solitons.

The theory of quantum solitons in optical fibers was developed by Carter and Drummond [57]. Among other effects, they predicted that quantum solitons are squeezed, which was subsequently confirmed experimentally [58]. This theory, and its subsequent elaborations [59] is comprehensive. It takes into account losses, Brillouin scattering, and Raman processes and its agreement with experiment is excellent. Here we only wish to present some of the basic features of quantum solitons, and to do so we will make us of a simpler, and less realistic, theory due to Lai and Haus [60]. It is based on the time-dependent Hartree approximation.

As a first step we shall go into a kind of interaction picture. Let

$$H_0 = \hbar \int dx \omega_1 \Psi^\dagger \Psi, \tag{12.1}$$

and define operators

$$\Psi_I = e^{-iH_0 t/\hbar} \Psi e^{iH_0 t/\hbar} \quad H_I = e^{-iH_0 t/\hbar} H e^{iH_0 t/\hbar}. \tag{12.2}$$

We then find that

$$H_I = \frac{\hbar}{2} \int dx \left[ iv_1 \left( \frac{\partial \Psi_I^\dagger}{\partial x} \Psi_I - \Psi_I^\dagger \frac{\partial \Psi_I}{\partial x} \right) + \frac{1}{2}\omega_1'' \frac{\partial \Psi_I^\dagger}{\partial x} \frac{\partial \Psi_I}{\partial x} \right.$$
$$\left. + \ 2g_3(\Psi_I^\dagger)^2 \Psi_I^2 \right], \tag{12.3}$$

where we have set

$$g_3 = \frac{3\beta^{(3)}}{8A} \hbar (k_1 v_1 \epsilon_1)^2. \tag{12.4}$$

This gives us the equation of motion

$$i\left(\frac{\partial}{\partial t} + v\frac{\partial}{\partial x}\right)\Psi_I = -\frac{\omega''}{2}\frac{\partial^2 \Psi_I}{\partial x^2} + 2g_3 \Psi_I^\dagger \Psi_I^2. \tag{12.5}$$

From here on, we shall drop the subscript $I$ with the understanding that all of the operators are in our interaction picture. Next, we go into a moving frame by considering $\Psi$ to be a function of $x_v = x - vt$ and $t$ rather than $x$ and $t$. In terms of the new coordinates, the equation of motion for $\Psi$ becomes

$$i\frac{\partial \Psi}{\partial t} = -\frac{\omega''}{2}\frac{\partial^2 \Psi}{\partial x_v^2} + 2g_3 \Psi^\dagger \Psi^2. \tag{12.6}$$

The above equation is an operator version of a nonlinear Schrödinger equation. The classical version of this equation gives rise to solitons. The operator version describes particles interacting by means of a delta-function potential, and the solutions to this problem are known. Let us make a short detour to describe the interacting-particle system, and then use the results to examine the properties of our solitons in a nonlinear medium.

Suppose that we have particles of mass $m$ in one dimension interacting via a potential $V(x - x')$. The second-quantized Hamiltonian for this system is

$$H = \int dx \frac{\hbar}{2m} \frac{\partial \phi^\dagger}{\partial x} \frac{\partial \phi}{\partial x} + \int dx \int dx' V(x - x') \phi^\dagger(x, t) \phi^\dagger(x', t) \phi(x', t) \phi(x, t), \quad (12.7)$$

where $\phi(x, t)^\dagger$ creates a particle at postion $x$, and

$$[\phi(x, t), \phi(x', t)^\dagger] = \delta(x - x'). \quad (12.8)$$

If the potential is given by $V(x - x') = \hbar g \delta(x - x')$, then we obtain a Hamiltonian that is the same as the one describing the propagation of the field in a nonlinear fiber. The fact that the number operator commutes with the Hamiltonian implies that the eigenstates will be states of well-defined particle number, and hence be of the form

$$|\psi_n\rangle = \frac{1}{\sqrt{n!}} \int dx_1 \ldots dx_n f_n(x_1, \ldots x_n) \phi(x_1, 0)^\dagger \ldots \phi(x_n, 0)^\dagger |0\rangle. \quad (12.9)$$

The function $f_n$ satisfies the equation

$$E_n f_n = \left[ -\frac{\hbar^2}{2m} \sum_{j=1}^n \frac{\partial^2}{\partial x_j^2} + 2\hbar g \sum_{1 \leq j < k \leq n} \delta(x_k - x_j) \right] f_n. \quad (12.10)$$

If $g < 0$, then there are bound states, and for these $f_n$ is proportional to [61,62]

$$f_n(x_1, \ldots x_n) \propto \exp[ip \sum_{j=1}^n x_j + (mg/\hbar) \sum_{1 \leq j < k \leq n} |x_k - x_j|]. \quad (12.11)$$

In the case of photons in a nonlinear fiber, this solution would be a bound state of $n$ photons, and, therefore, be a state of definite photon number. Because the solitons that propagate in nonlinear fibers are a result of light from a laser propagating through the fiber, and laser light, being close to a coherent state, does not have a definite photon number, the states we are looking for are superpositions of many of these states with different values of $n$. Rather than trying to find these soliton solutions by superposing the exact solutions, we shall make use of an approximate method.

We shall employ the time-dependent Hartree approximation. We start from the time dependent equation for $f_n$, which is just an $n$-particle Schrödinger equation,

$$i\hbar \frac{\partial f_n}{\partial t} = \left[ -\frac{\hbar^2}{2m} \sum_{j=1}^n \frac{\partial^2}{\partial x_j^2} + 2\hbar g \sum_{1 \leq j < k \leq n} \delta(x_k - x_j) \right] f_n, \quad (12.12)$$

and we assume that $f_n(x_1, \ldots x_n) = \Pi_{j=1}^n h_n(x_j, t)$. The effective potential felt by one of the particles is the $n$-particle potential averaged over all of the other particles

$$V_{eff}(x, t) = 2\hbar g \sum_{j=2}^n \int dx_j |h_n(x_j, t)|^2 \delta(x - x_j) = 2\hbar g (n - 1) |h_n(x, t)|^2. \quad (12.13)$$

We assume that $h_n(x,t)$ satisfies the one-particle Schrödinger equation with the effective potential

$$i\hbar \frac{\partial h_n}{\partial t} = -\frac{\hbar^2}{2m}\frac{\partial^2 h_n}{\partial x^2} + 2\hbar g(n-1)|h_n(x,t)|^2 h_n(x,t). \tag{12.14}$$

For $g < 0$ this equation has the soliton solution

$$h_n(x,t) = \frac{2\eta}{|g(n-1)|^{1/2}}\left(\frac{2m}{\hbar}\right)^{1/2} \exp\left[-4i(\xi^2 - \eta^2)t - 2i\xi\sqrt{\frac{2m}{\hbar}}(x - x_0)\right]$$

$$\operatorname{sech}\left[2\eta\left(\sqrt{\frac{2m}{\hbar}}(x - x_0) + 4\xi t\right)\right], \tag{12.15}$$

where $\xi$, $\eta$, and $x_0$ are arbitrary parameters. The normalization condition,

$$\int dx |h_n(x,t)|^2 = 1, \tag{12.16}$$

however, fixes the value of $\eta$

$$\eta = \frac{|g|(n-1)}{4}. \tag{12.17}$$

We now want to apply what we have learned about interacting particles to a field propagating in a nonlinear fiber. In order to do so it is useful to look at the field propagation problem in the Schrödinger picture. For the state $|\Phi\rangle$, we have

$$i\hbar \frac{\partial}{\partial t}|\Phi\rangle = (H_{lin} + H_{nlin})|\Phi\rangle. \tag{12.18}$$

Because the Hamiltonian conserves photon number, we can assume that

$$|\Phi\rangle = \frac{1}{\sqrt{n!}} \int dx_1 \ldots dx_n f_n(x_1, \ldots x_n, t) \Psi^\dagger(x_1, 0) \ldots \Psi^\dagger(x_n, 0)|0\rangle, \tag{12.19}$$

where

$$\int dx_1 [\ldots dx_n |f_n(x_1, \ldots x_n)|^2 = 1. \tag{12.20}$$

Inserting $|\Phi\rangle$ into the Schrödinger equation, we find that $f_n$ satisfies

$$i\frac{\partial f_n}{\partial t} = n\omega_1 f_n - iv_1 \sum_{j=1}^n \frac{\partial f_n}{\partial x_j} - \frac{1}{2}\omega_1'' \sum_{j=1}^n \frac{\partial^2 f_n}{\partial x_j^2}$$

$$+ 2g_3 \sum_{j=1}^n \sum_{k=1}^{j-1} \delta(x_j - x_k) f_n. \tag{12.21}$$

If we now define $\tilde{f}_n = \exp(in\omega_1 t) f_n$ and consider $\tilde{f}_n$ to be a function of $\{x_{vj} = x_j - v_1 t | j = 1,\ldots n\}$ and $t$ rather than of $\{x_j | j = 1,\ldots n\}$ and $t$, then we have that

$$i\frac{\partial \tilde{f}_n}{\partial t} = -\frac{1}{2}\omega_1'' \sum_{j=1}^n \frac{\partial^2 \tilde{f}_n}{\partial x_{vj}^2} + 2g_3 \sum_{j=1}^n \sum_{k=1}^{j-1} \delta(x_{vj} - x_{vk}) f_n, \tag{12.22}$$

which has the same form as the Schrödinger equation for $n$ particles interacting via a delta function potential. If we assume that $\omega_1'' > 0$ and $g_3 < 0$, we can make use of the solutions we obtained for the delta-function-interacting system. In particular, we can use the solutions we obtained from the Hartree approximation and set

$$f_n(x_1, \ldots x_n, t) = \prod_{j=1}^{n} \tilde{h}_n(x_j, t), \tag{12.23}$$

where

$$\tilde{h}_n(x_j, t) = e^{i\omega_1 t} h_n(x_j - v_1 t, t), \tag{12.24}$$

and in the explicit expresssion for $h_n$ we have made the replacements

$$\frac{\hbar}{2m} \to \frac{\omega_1''}{2} \quad g \to g_3. \tag{12.25}$$

As was mentioned, we are interested in solutions that are superpositions of states with different photon numbers, and, in particular, solutions that resemble coherent states. Defining

$$\Psi^\dagger[\tilde{h}_n] = \int dx \tilde{h}_n(x, t) \Psi^\dagger(x, 0), \tag{12.26}$$

we note that $\left[\Psi(x,0), \Psi^\dagger[\tilde{h}_n]\right] = \tilde{h}_n(x,t)$. Now consider the superposition of approximate solutions

$$|\Phi\rangle = e^{-|\alpha|^2/2} \sum_{n=0}^{\infty} \frac{\alpha^n}{n!} (\Psi^\dagger[\tilde{h}_n])^n |0\rangle. \tag{12.27}$$

This state has the property that

$$\langle \Phi | \Psi(x,0) | \Phi \rangle = \alpha e^{-|\alpha|^2} \sum_{n=0}^{\infty} \frac{|\alpha|^{2n}}{n!} \tilde{h}_{n+1}(x,t) (\langle \tilde{h}_n | \tilde{h}_{n+1} \rangle)^n. \tag{12.28}$$

Now let us assume that $n_0 = |\alpha|^2 \gg 1$, which implies that the average photon number in the state is large. The terms in the sum that will contribute most are those for which $n_0 - \sqrt{n_0} \leq n \leq n_0 + \sqrt{n_0}$. In this range we have that

$$(\langle \tilde{h}_n | \tilde{h}_{n+1} \rangle)^n \cong e^{itg_3^2 n(2n-1)/4} \tag{12.29}$$

and that the dominant $n$ dependence in $\tilde{h}_{n+1}(x,t)$ is the exponential factor $\exp(in^2 g_3^2 t/4)$. Therefore, if $g_3^2 t n_0 \sqrt{n_0} \ll 1$, the $n$ dependence of the terms in the sum in the important range will be weak, and we can replace them by their values at $n_0$, so that

$$\langle \Phi | \Psi(x,0) | \Phi \rangle = \alpha \tilde{h}_{n_0}(x,t). \tag{12.30}$$

This is just a propagating soliton, so that for sufficiently short times the average value of the field is just that of a classical soliton. For longer times, however, the $n$ dependence of the different terms in the sum comes into play, and these terms are no longer in phase. The phase of what had been a coherent state starts to diffuse, and this will cause the average value of the field to decay. This phase diffusion is a quantum effect; classically, the soliton would propagate with no change to its shape (ignoring losses). This phase diffusion can lead to squeezing, and it has been observed experimentally [58].

## 13 Conclusion

What has been presented here is merely an introduction to the quantum theory of nonlinear optics, and there are many aspects of it that we have not covered. For example, some of the mathematical methods that are useful in treating these systems, such as quasiprobability distributions and the stochastic differential equations to which they lead have either just been touched upon or have not been discussed. We have also not treated the input-output formalism that is employed to find the field emitted by a cavity containing a nonlinear medium, which is driven by input fields. These topics are covered elsewhere, once source being the book *Quantum Squeezing* [63].

Another aspect we have not covered is a second approach to nonlinear optics with quantized fields. We have characterized the medium by its susceptibilities. One can instead construct a model for the medium, and then quantize the entire system, fields and matter. This approach was pioneered for linear media by Hopfield [64]. The elementary excitations in this theory are mixed matter-field modes known as polaritons. When the medium is nonlinear, the polaritons interact. The Hamiltonian for this system is one that describes interacting polariton fields [65, 66]. This is consistent with the treatment presented here, because the creation and annihilation operators we have discussed are expressed in terms of the dual potential and the displacement field, which contains both matter and electromagnetic fields. Therefore, the excitations created by the creation operators are, in fact, mixed matter-field excitations.

Finally, it would be useful to have a formalism that describes fields entering a nonlinear medium, interacting within it, and propagating out into free space. This would, in fact, be a scattering theory approach, and it accords with what is done in the laboratory. The fields originate outside the medium, propagate through it, and are measured in free space. This is yet to be accomplished in the full multimode case. We expect that a quantum field theoretic scattering treatment would be a useful way to approach this problem.